\documentclass[a4paper,traditabstract,longauth]{aa}  

\usepackage{graphicx}
\usepackage{rotating,lscape}
\usepackage{longtable,multirow,tabu}
\usepackage{color,psfrag}
\usepackage{afterpage}
\usepackage{layouts}
\usepackage{url}

\usepackage[breaklinks, colorlinks, citecolor=blue]{hyperref}
\usepackage{breakurl}
\usepackage{natbib}
\usepackage{amsmath,amsfonts,amssymb}
\usepackage{bm}
\usepackage{txfonts}
\usepackage{subfig}

\def\threefigures#1#2#3{
\centerline{\includegraphics[width=0.3\linewidth]{#1}
            \quad
            \includegraphics[width=0.3\linewidth]{#2}
            \quad
            \includegraphics[width=0.3\linewidth]{#3}
\bigskip}}

\newcommand{\Ytot}{$Y_{\rm tot}$} 

\begin{document}

\graphicspath{{figs/EPS/}{figs/}}

\author{\small
Y.~C.~Perrott\inst{2}\thanks{Corresponding author: Y.~C.~Perrott, ycp21@mrao.cam.ac.uk}
\and
M.~Olamaie\inst{2}
\and
C.~Rumsey\inst{2}
\and
M.~L.~Brown\inst{15}
\and
F.~Feroz\inst{2}
\and
K.~J.~B.~Grainge\inst{2, 16, 15}
\and
M.~P.~Hobson\inst{2}
\and
A.~N.~Lasenby\inst{2, 16}
\and
C.~J.~MacTavish\inst{16}
\and
G.~G.~Pooley\inst{2}
\and
R.~D.~E.~Saunders\inst{2, 16}
\and
M.~P.~Schammel\inst{2, 10}
\and
P.~F.~Scott\inst{2}
\and
T.~W.~Shimwell\inst{2, 5}
\and
D.~J.~Titterington\inst{2}
\and
E.~M.~Waldram\inst{2}
\and
N.~Aghanim\inst{11}
\and
M.~Arnaud\inst{17}
\and
M.~Ashdown\inst{16, 2}
\and
H.~Aussel\inst{17}
\and
R.~Barrena\inst{14, 8}
\and
I.~Bikmaev\inst{7, 1}
\and
H.~B\"{o}hringer\inst{19}
\and
R.~Burenin\inst{21, 20}
\and
P.~Carvalho\inst{12, 16}
\and
G.~Chon\inst{19}
\and
B.~Comis\inst{18}
\and
H.~Dahle\inst{13}
\and
J.~Democles\inst{17}
\and
M.~Douspis\inst{11}
\and
D.~Harrison\inst{12, 16}
\and
A.~Hempel\inst{14, 8, 23}
\and
G.~Hurier\inst{11}
\and
I.~Khamitov\inst{22, 7}
\and
R.~Kneissl\inst{9, 3}
\and
J.~F.~Mac\'{\i}as-P\'{e}rez\inst{18}
\and
J.-B.~Melin\inst{6}
\and
E.~Pointecouteau\inst{24, 4}
\and
G.~W.~Pratt\inst{17}
\and
J.~A.~Rubi\~{n}o-Mart\'{\i}n\inst{14, 8}
\and
V.~Stolyarov\inst{2, 16, 21}
\and
D.~Sutton\inst{12, 16}
}

\institute{\small
Academy of Sciences of Tatarstan, Bauman Str., 20, Kazan, 420111, Republic of Tatarstan, Russia\\
\and
Astrophysics Group, Cavendish Laboratory, University of Cambridge, J J Thomson Avenue, Cambridge CB3 0HE, U.K.\\
\and
Atacama Large Millimeter/submillimeter Array, ALMA Santiago Central Offices, Alonso de Cordova 3107, Vitacura, Casilla 763 0355, Santiago, Chile\\
\and
CNRS, IRAP, 9 Av. colonel Roche, BP 44346, F-31028 Toulouse cedex 4, France\\
\and
CSIRO Astronomy \& Space Science, Australia Telescope National Facility, PO Box 76, Epping, NSW 1710, Australia\\
\and
DSM/Irfu/SPP, CEA-Saclay, F-91191 Gif-sur-Yvette Cedex, France\\
\and
Department of Astronomy and Geodesy, Kazan Federal University, Kremlevskaya Str., 18, Kazan, 420008, Russia\\
\and
Dpto. Astrof\'{\i}sica, Universidad de La Laguna (ULL), E-38206 La Laguna, Tenerife, Spain\\
\and
European Southern Observatory, ESO Vitacura, Alonso de Cordova 3107, Vitacura, Casilla 19001, Santiago, Chile\\
\and
INAF - Osservatorio Astronomico di Roma, via di Frascati 33, Monte Porzio Catone, Italy\\
\and
Institut d'Astrophysique Spatiale, CNRS (UMR8617) Universit\'{e} Paris-Sud 11, B\^{a}timent 121, Orsay, France\\
\and
Institute of Astronomy, University of Cambridge, Madingley Road, Cambridge CB3 0HA, U.K.\\
\and
Institute of Theoretical Astrophysics, University of Oslo, Blindern, Oslo, Norway\\
\and
Instituto de Astrof\'{\i}sica de Canarias, C/V\'{\i}a L\'{a}ctea s/n, La Laguna,Tenerife, Spain\\
\and
Jodrell Bank Centre for Astrophysics, Alan Turing Building, School of Physics and Astronomy, The University of Manchester, Oxford Road, Manchester, M13 9PL, U.K.\\
\and
Kavli Institute for Cosmology Cambridge, Madingley Road, Cambridge, CB3 0HA, U.K.\\
\and
Laboratoire AIM, IRFU/Service d'Astrophysique - CEA/DSM - CNRS - Universit\'{e} Paris Diderot, B\^{a}t. 709, CEA-Saclay, F-91191 Gif-sur-Yvette Cedex, France\\
\and
Laboratoire de Physique Subatomique et de Cosmologie, Universit\'{e} Joseph Fourier Grenoble I, CNRS/IN2P3, Institut National Polytechnique de Grenoble, 53 rue des Martyrs, 38026 Grenoble cedex, France\\
\and
Max-Planck-Institut f\"{u}r Extraterrestrische Physik, Giessenbachstra{\ss}e, 85748 Garching, Germany\\
\and
Moscow Institute of Physics and Technology, Dolgoprudny, Institutsky per., 9, 141700, Russia\\
\and
Space Research Institute (IKI), Russian Academy of Sciences, Profsoyuznaya Str, 84/32, Moscow, 117997, Russia\\
\and
T\"{U}B\.{I}TAK National Observatory, Akdeniz University Campus, 07058, Antalya, Turkey\\
\and
Universidad Andr\'{e}s Bello, Dpto. de Ciencias F\'{i}sicas, Facultad de Ciencias Exactas, Astronom\'{i}a, Campus Casona de Las Condes, Fern\'{a}ndez Concha 700, 7591538 Santiago / Las Condes, Chile\\
\and
Universit\'{e} de Toulouse, UPS-OMP, IRAP, F-31028 Toulouse cedex 4, France\\
}

\title{Comparison of Sunyaev-Zel'dovich measurements from \textit{Planck} and from the Arcminute Microkelvin Imager for 99 galaxy clusters} 

\date{Received ; Accepted } 

\abstract{We present observations and analysis of a sample of 123 galaxy clusters from the 2013 \emph{Planck} catalogue of Sunyaev-Zel'dovich sources with the Arcminute Microkelvin Imager (AMI), a ground-based radio interferometer.  AMI provides an independent measurement with higher angular resolution, 3\,arcmin compared to the \emph{Planck} beams of 5--10\,arcmin.  The AMI observations thus provide validation of the cluster detections, improved positional estimates, and a consistency check on the fitted `size' ($\theta_{s}$) and `flux' (\Ytot) parameters in the Generalised Navarro, Frenk and White (GNFW) model.  We detect 99 of the clusters.  We use the AMI positional estimates to check the positional estimates and error-bars produced by the \emph{Planck} algorithms PowellSnakes and MMF3.  We find that \Ytot \:values as measured by AMI are biased downwards with respect to the \emph{Planck} constraints, especially for high \emph{Planck}-SNR clusters.  We perform simulations to show that this can be explained by deviation from the `universal' pressure profile shape used to model the clusters.  We show that AMI data can constrain the $\alpha$ and $\beta$ parameters describing the shape of the profile in the GNFW model for individual clusters provided careful attention is paid to the degeneracies between parameters, but one requires information on a wider range of angular scales than are present in AMI data alone to correctly constrain all parameters simultaneously.}


\authorrunning{Y.~C.~Perrott et al.} 

\titlerunning{\emph{Planck} and AMI SZ measurements for 99 galaxy clusters}

\maketitle
 

\section{Introduction}

The \emph{Planck} satellite data-release of 2013 included a catalogue of 1227 galaxy clusters detected via the Sunyaev-Zel'dovich (SZ, \citealt{sunyaev72}) effect \citep{2013arXiv1303.5089P}.  This is the deepest all-sky cluster catalogue in SZ to date, consisting of clusters spanning redshifts up to $\approx$\,1, and masses of around $10^{14} \rm{M}_{\odot}$ to $10^{15} \rm{M}_{\odot}$.  SZ-selected samples have the advantage of a clean, and much less redshift-dependent (above $z$\,$\approx$\,0.3) selection function in mass than, for example, X-ray-selected samples \citep{2013arXiv1303.5080P}; in addition, simulations predict that the SZ `flux' correlates more tightly with mass than, for example, X-ray or optical observable quantities (e.g.\ \citealt{daSilva04}, \citealt{2005ApJ...623L..63M}, \citealt{nagai06}, \citealt{2009A&A...496..637A}, \citealt{2012MNRAS.426.2046A}, \citealt{2012MNRAS.422.1999K}).  The \emph{Planck} SZ catalogue is therefore a potentially very powerful tool for investigating the growth of structure in the Universe; clusters in the catalogue are being followed up with optical, radio and X-ray telescopes in order to provide multi-wavelength information to understand fully their properties.

The Arcminute Microkelvin Imager (AMI; \citealt{2008MNRAS.391.1545Z}) is a dual-array interferometer designed for SZ studies, which is situated near Cambridge, UK.  AMI consists of two arrays: the Small Array (SA), optimised for viewing arcminute-scale features, having an angular resolution of $\approx$\,3\,arcmin and sensitivity to structures up to $\approx$\,10\,arcmin in scale; and the Large Array (LA), with angular resolution of $\approx$\,30\,arcsec, which is insensitive to the arcminute-scale emission due to clusters and is used to characterise and subtract confusing radio sources.  Both arrays operate at a central frequency of $\approx$\,15\,GHz with a bandwidth of $\approx$\,4.5\,GHz, divided into six channels.  For further details of the instrument, see \citet{2008MNRAS.391.1545Z}.

In a previous paper, (\citealt{planck2012-II}, from here on AP2013) a sample of 11 clusters selected from the \emph{Planck} Early Release Catalogue was followed up with AMI in order to check the consistency of the cluster parameters as measured by the two telescopes, finding the SZ signals as measured by AMI to be, on average, fainter and of smaller angular size.   We have used AMI to observe all of the clusters in the \emph{Planck} 2013 SZ catalogue that are at declinations easily observable with AMI (excluding those at very low redshift).  This serves two purposes: (a) to investigate the discrepancies found in AP2013 further, and (b) to provide validation of, improved positional estimates for, and higher-resolution SZ maps of a large number of \emph{Planck} cluster detections.  We here present these observations and our analysis of them.

The paper is organised as follows.  In Section~\ref{sec:cluster_sample} we describe the selection of the cluster sample.  In Section~\ref{sec:amidata} we describe the AMI observations and data reduction, and in Section~\ref{sec:SZsignal} we outline the model used to describe the SZ signal.  In Section~\ref{sec:planck_data_analysis} we briefly describe the \emph{Planck} data analysis and describe in more detail the analysis of the AMI data in Section~\ref{sec:AMI_data_analysis}, including our detection criteria.  Section~\ref{sec:resultsreal} contains some representative examples of the results, and Sections~\ref{S:pos_comp} and \ref{S:Ytot_theta_comp} compare the cluster parameter estimates produced by AMI to those produced by \emph{Planck}.  In Section~\ref{sec:profiles} we use simulations to investigate the issue of variation from the `universal' model described in Section~\ref{sec:SZsignal}, and in Section~\ref{S:real_data_vary_ab} we present results from reanalysing the real data allowing the shape parameters in the model to vary.  Finally, we conclude in Section~\ref{sec:conclusions}.


\section{Selection of the cluster sample}
\label{sec:cluster_sample}

An initial selection cut of $20^{\circ} \le \delta < 87^{\circ}$ was applied to satisfy AMI's `easy' observing limits; although AMI can observe to lower declinations, increased interference due to geostationary satellites makes observing large samples below $\delta = 20^{\circ}$ currently difficult.  In addition, clusters with known redshifts of $z \le 0.100$ were excluded since these have large angular sizes and will be largely resolved out by AMI; although the brightest of these will still be detectable, it will be difficult to constrain their properties using AMI data.  These initial cuts resulted in an initial sample size of 337 with \emph{Planck} SNR values ranging from 4.5 -- 20.  In this paper, we present results for the subset of the sample with SNR $\ge 5$; this reduces the sample to 195.  Results for the remaining clusters with $4.5 \le \mathrm{SNR} < 5$ will be released at a later date.

As in the optical, where confusion due to a bright star or a crowded field can affect the detection likelihood, a benign radio point source environment is important for AMI, but the requisite benignness is difficult to quantify.  In practice, the effect of the source environment on the detection potential of a cluster depends on many factors including the number, location and orientation of the sources with respect to each other and to the sidelobes of the primary and synthesised beams.  Non-trivial source environments can create complex and overlapping sidelobe patterns which can create spurious sources or reduce the flux density of real sources.  In turn, the synthesised beam depends on $uv$-coverage, which changes for different $\delta$ and hour-angle coverage of observations of a given cluster.  The primary beam is a function of frequency so the effect of a source at a given offset from the pointing centre also depends on its spectrum.  These effects are almost impossible to quantify in a systematic way.  In order to apply at least consistent criteria across the whole sample, the following criteria were applied based on LA observations:  clusters were discarded if there were radio sources of peak flux density $S_{\rm peak} > 5$\,mJy within 3\,arcmin of the pointing centre, of $S_{\rm peak} > 20$\,mJy within 10\,arcmin of the pointing centre, or extended emission with fitted (deconvolved) major-axis size $> 2$\,arcmin and integrated flux density $S_{\rm int} > 2$\,mJy anywhere on the map; experience suggests that observation of the SZ signal in such clusters with AMI is unreliable.  Clusters were discarded for source environment based either on existing observations or, for clusters that had not been previously observed with AMI, based on a short pre-screening observation carried out with the LA.  It should be noted that some clusters which have been previously observed and detected by AMI are excluded by these cuts; some of the new clusters discarded by this process may also be observable.

In addition, clusters were visually inspected at various stages of the follow-up and analysis process, and some were rejected at later stages due to extra source environment problems such as extended emission not visible on the LA map, or very bright sources just outside the LA detection radius which affect the SA map due to the larger primary beam.  Here we present results for the so obtained final sub-sample, which we will refer to as the SZ sample, consisting of 123 clusters.  A breakdown of the numbers of clusters rejected for various reasons is shown in Table~\ref{T:cluster_count}.

\begin{table}
\centering
\caption{Numbers of clusters in the $20^{\circ} \le \delta < 87^{\circ}$, \emph{Planck} SNR $\ge 5$ sub-sample in various categories.}\label{T:cluster_count}
\begin{tabular}{lr}
\hline
Category & Number of clusters \\ \hline
Total & 229 \\
$z \le 0.100$ & 34 \\
Automatic radio-source environment rejection & 52 \\
Manual radio-source environment rejection & 20 \\
Included in sample & 123 \\
\hline
\end{tabular}
\end{table}

The full list of clusters within the AMI observational bounds and their reason for rejection, if not part of the SZ sample, is given in Appendix~\ref{sec:results_table}.  In addition, as a service to the community for each cluster we provide information on the 15\,GHz radio point source environment (available online at \url{http://www.astro.phy.cam.ac.uk/surveys/ami-planck/}).

\section{Description of AMI data}
\label{sec:amidata}

\begin{figure}
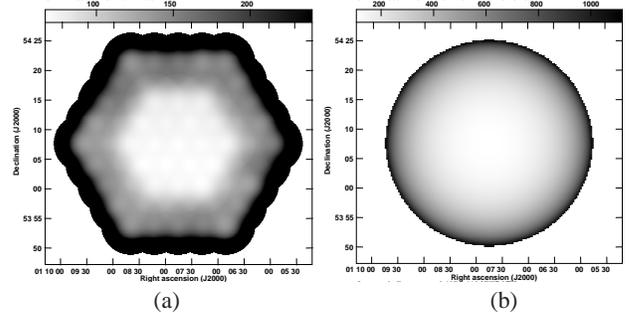

  \begin{center}
  \includegraphics[bb=47 148 570 632, clip=, width=0.45\linewidth]{CAJ0107+5407_LA_aips_noise_a.ps}
  \includegraphics[bb=45 146 570 632, clip=, width=0.45\linewidth]{CAJ0107+5407_SA_aips_noise_a.ps}
  \medskip
  \centerline{\hskip 0.05\linewidth(a) \hskip 0.45\linewidth (b)}
  \caption{Noise maps for a typical cluster observation at $\delta$\,$\approx$\,$54^{\circ}$ on the AMI-LA (a) and SA (b).  The grey-scales are in $\mu$Jy\,beam$^{-1}$ and on (a) the grey-scale is truncated to show the range of noise levels.  (b) is cut off at the 10\% power point of the primary beam.}
  \label{Fi:CAJ0107_noise}
  \end{center}
\end{figure}

\begin{figure*}
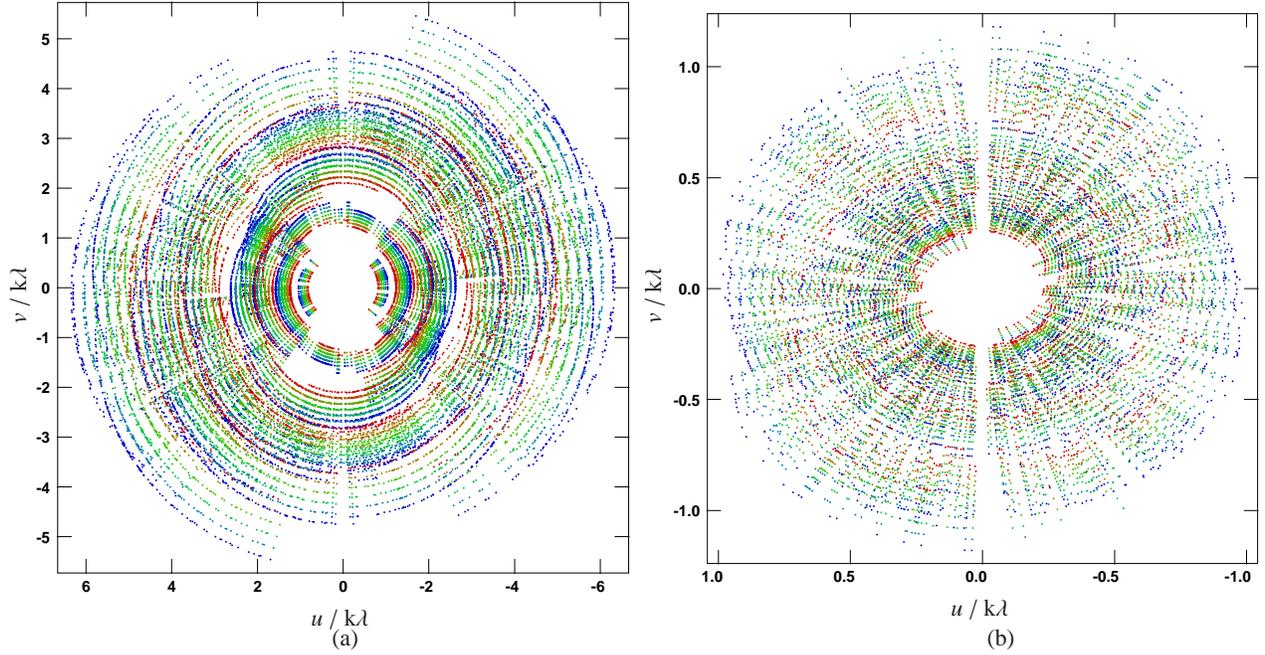

  \begin{center}
  \begin{psfrags}%
  \psfragscanon%
  \psfrag{x01}[t][t]{\hspace{25pt}\footnotesize{$u\:/\:{\rm k}\lambda$}}%
  \psfrag{y01}[b][b]{\hspace{22pt}\footnotesize{$v\:/\:{\rm k}\lambda$}}%
  \raisebox{-0.5\height}{\includegraphics[bb=48 114 570 631, clip=, width=0.45\linewidth]{CAJ0107+5407_LA_uvplt_uv.ps}}
  \end{psfrags}
  \begin{psfrags}%
  \psfragscanon%
  \psfrag{x01}[t][t]{\hspace{25pt}\footnotesize{$u\:/\:{\rm k}\lambda$}}%
  \psfrag{y01}[b][b]{\hspace{22pt}\footnotesize{$v\:/\:{\rm k}\lambda$}}%
  \raisebox{-0.5\height}{\includegraphics[bb=43 121 570 625, clip=, width=0.45\linewidth]{CAJ0107+5407_SA_uvplt_uv.ps}}
  \end{psfrags}
  \centerline{\mbox{}}
  \centerline{\hskip 0.05\linewidth (a) \hskip 0.45\linewidth (b)}
  \caption{\textit{uv}-coverages for a typical cluster observation at $\delta$\,$\approx$\,$54^{\circ}$, for the AMI-LA (a) and SA (b).  The colours indicate different channels.  Note the different axis scales; the short baselines of the SA are designed for sensitivity to arcminute-scale cluster emission, while the longer baselines of the LA are insensitive to emission on this scale and are used to characterise and subtract the foreground radio sources.}
  \label{Fi:CAJ0107_uv}
  \end{center}
\end{figure*}

Clusters are observed using a single pointing centre on the SA, which has a primary beam of size $\approx$\,20\,arcmin FWHM, to noise levels of $\lessapprox$\,120\,$\mu$Jy\,beam$^{-1}$.  To cover the same area with the LA, which has a primary beam of size $\approx$\,6\,arcmin FWHM, the cluster field is observed as a 61-point hexagonal raster.  The noise level of the raster is $\lessapprox$\,100\,$\mu$Jy\,beam$^{-1}$ in the central 19 pointings, and slightly higher in the outer regions.  Typical noise maps and \textit{uv}-coverages are displayed for both arrays in Figs.~\ref{Fi:CAJ0107_noise} and ~\ref{Fi:CAJ0107_uv}.  The average observation time for a cluster is $\approx$\,30 hours on both arrays.

Data on both arrays are flagged for interference and calibrated using the AMI in-house software package \textsc{reduce}.  Flux calibration is applied using contemporaneous observations of the primary calibration sources 3C286, 3C48, and 3C147.  The assumed flux densities for 3C286 were converted from Very Large Array total-intensity measurements \citep{2013ApJS..204...19P}, and are consistent with the \cite{1987Icar...71..159R} model of Mars transferred on to absolute scale, using results from the \emph{Wilkinson Microwave Anisotropy Probe}.  The assumed flux densities for 3C48 and 3C147 are based on long-term monitoring with the SA using 3C286 for flux calibration (see Table~\ref{tab:Fluxes-of-3C286}).  Phase calibration is applied using interleaved observations of a nearby bright source selected from the VLBA Calibrator survey \citep{2008AJ....136..580P}; in the case of the LA, a secondary amplitude calibration is also applied using contemporaneous observations of the phase-calibration source on the SA.

\begin{table}
\centering
\caption{Assumed I~+~Q flux densities of 3C286, 3C48 and 3C147.} \label{tab:Fluxes-of-3C286}
\bigskip
\begin{tabular}{ccccc}
\hline
 Channel & $\bar{\nu}$/GHz & $S^{\rm 3C286}$/Jy & $S^{\rm 3C48}$/Jy & $S^{\rm 3C147}$/Jy \phantom{$S^{\rm{X^{X}}}$} \\ \hline 
 3 & 13.88 & 3.74 & 1.89 & 2.72 \\
 4 & 14.63 & 3.60 & 1.78 & 2.58 \\
 5 & 15.38 & 3.47 & 1.68 & 2.45 \\
 6 & 16.13 & 3.35 & 1.60 & 2.34 \\
 7 & 16.88 & 3.24 & 1.52 & 2.23 \\
 8 & 17.63 & 3.14 & 1.45 & 2.13 \\ 
\hline 
\end{tabular}
\end{table}

Maps of the SA and LA data are made using \textsc{aips}\footnote{\url{http://aips.nrao.edu/}}, \textsc{clean}ing in an automated manner.  Source-finding is carried out at 4$\sigma$ on the LA continuum map, as described in \citet{davies11} and \citet{franzen11}, and sources that are detected at $\ge 3\sigma$ on at least three channel maps and are not extended have a spectral index $\alpha$ fitted across the AMI band.  SA data are binned on a grid in $uv$-space in order to reduce the memory required for subsequent analysis.

\section{Analysing the SZ signal}
\label{sec:SZsignal}

\subsection{Cluster model}

For consistency with the \emph{Planck} catalogue, in this paper we assume the electron pressure profile $P_{\rm e}(r)$ of each cluster follows a generalised Navarro-Frenk-White (NFW, \citealt{navarro97}) model, which is given by (assuming spherical geometry)

\begin{equation}
P_{\rm e}(r) = P_{0} \left ( \frac{r}{r_{\rm s}} \right )^{-\gamma} \left [ 1+\left(\frac{r}{r_{\rm s}} \right)^{\alpha} \right ]^{(\gamma - \beta)/\alpha},
\end{equation}
where $P_{0}$ is a normalisation coefficient, $r$ is the physical radius, $r_{\rm s}$ is a characteristic scale radius, and the parameters $(\gamma, \alpha, \beta)$ describe the slopes of the pressure profile at radii $r \ll r_{\rm s}$, $r \approx r_{\rm s}$, and $r \gg r_{\rm s}$ respectively \citep{nagai07}.  Following \citet{arnaud10}, we fix the slope parameters to their `universal' values, $\gamma = 0.3081, \alpha = 1.0510, \beta = 5.4905$ derived from the REXCESS sample \citep{bohringer07}.  They are also fixed to these values in the \emph{Planck} analysis.

Given this model, the integrated SZ surface brightness, or integrated Compton-$y$ parameter, for a cluster is given by

\begin{equation}
Y_{\rm sph}(r) = \frac{\sigma_{\rm T}}{m_{\rm e}c^2} \int_{0}^{r} P_{\rm e}(r')4\pi r'^2 {\rm d} r',
\end{equation}
where $\sigma_{\rm T}$ is the Thomson scattering cross-section, $m_{\rm e}$ is the electron mass, and $c$ is the speed of light.  This has an analytical solution as $r \rightarrow \infty$, giving the total integrated Compton-$y$ parameter $Y_{\rm tot,phys}$ as

\begin{equation}
Y_{\rm tot,phys} = \frac{4 \pi \sigma_{\rm T}}{m_{\rm e} c^{2}} P_{0} r_{\rm s}^{3} \frac{\Gamma \left ( \frac{3-\gamma}{\alpha} \right ) \Gamma \left ( \frac{\beta - 3}{\alpha} \right )}{\alpha \Gamma \left ( \frac{\beta-\gamma}{\alpha} \right )}.
\end{equation}\label{Eq:Ytot}

With $(\gamma, \alpha, \beta)$ fixed, a cluster's appearance on the sky may be described using four (observational) parameters only: $(x_{0}, y_{0}, \theta_{\rm s}, Y_{\rm tot})$, where $x_{0}$ and $y_{0}$ are the positional coordinates for the cluster, $\theta_{\rm s} = r_{\rm s}/D_{\rm A}$ is the characteristic angular scale of the cluster on the sky ($D_{\rm A}$ is the angular diameter distance to the cluster), and $Y_{\rm tot} = Y_{\rm tot,phys}/D_{\rm A}^2$ is the SZ surface brightness integrated over the cluster's extent on the sky.

This model does not require any redshift information; physical quantities such as $r_{\rm s}$ and $Y_{\rm tot,phys}$ can be recovered from $\theta_{\rm s}$ and \Ytot \:given a redshift.  Alternatively, $r_{X}$ and $M_{X}$ for some overdensity radius $X$ can be recovered given a redshift, a concentration parameter $c_{X} \equiv r_{X}/r_{\rm s}$ and some model or scaling relationship for translating $Y$ into mass (e.g.\ \citealt{2013arXiv1303.5080P}, \citealt{2012MNRAS.423.1534O}).  Physical modelling will not be addressed in this paper.

Note that in the \emph{Planck} analysis, in order to impose a finite integration extent, $Y_{5R_{500}}$ (the SZ surface brightness integrated to $5 \times R_{500}$) is estimated rather than {\Ytot}.  For the `universal' GNFW parameter values, (with $c_{500} = 1.177$), the two quantities are equivalent to within 5\%.

\subsection{Analysis of \emph{Planck} data}
\label{sec:planck_data_analysis}

The \emph{Planck} SZ catalogue is the union of the catalogues produced by three detection algorithms: MMF1 and MMF3, which are multi-frequency matched-filter detection methods, and PowellSnakes (PwS), which is a Bayesian detection method.  Full details of these algorithms are provided in \citet{melin2006}, \citet{carvalho2009}, \citet{carvalho11} and \citet{2012A&A...548A..51M}.  Since the PwS analysis methodology most closely matches the Bayesian analysis procedures used to analyse AMI data, we take cluster parameters produced by PwS as our preferred `\emph{Planck}' values, followed by MMF3, and finally MMF1 values where a particular cluster is not detected by all algorithms.

\subsection{Analysis of AMI data}
\label{sec:AMI_data_analysis}

The model attempting to describe the AMI data is produced by a combination of the cluster model described above, the radio source environment as measured by the LA and a generalised Gaussian noise component comprising instrumental noise, confusion noise from radio sources below the detection threshold, and contamination from primordial CMB anisotropies.

Each foreground radio source is modelled by the parameters $(x_{S}, y_{S}, S_{0}, \alpha$).  Positions $(x_{S}, y_{S})$ and initial estimates of the flux density at a central frequency ($S_{0}$) are produced from the LA channel-averaged maps; for sources detected at $\ge 3\sigma$ on at least three of the individual channel maps, a spectral index $\alpha$ is also fitted to the channel flux densities.  The flux density and spectral index of sources which are detected at $\ge 4\sigma$ on the SA map are modelled simultaneously with the cluster; this accounts for possible source variability (although we attempt to observe clusters close in time on the two arrays, this is not always possible due to different demands on the observing time of the arrays) and inter-array calibration uncertainty.  Flux densities are given a Gaussian prior with $\sigma = 40$\%; where $\alpha$ has been fitted from the LA data, a Gaussian prior with width corresponding to the fitting uncertainty is applied, otherwise a prior based on the 10C survey is applied \citep{davies11}.  Sources detected at $<4\sigma$ on the SA map are subtracted directly based on the LA values of $S_{0}$ and $\alpha$ (or the median of the 10C prior where $\alpha$ has not been fitted) initially.  If the cluster position output from the analysis has directly-subtracted sources within 3\,arcmin, the analysis is repeated with those sources also modelled.  The positions of the sources are always fixed to their LA values as the LA has higher positional precision.

In the cluster model, $x_{0}$ and $y_{0}$ are the offsets in RA and $\delta$ from the pointing centre of the SA observation; for previously-known clusters with existing AMI data, the pointing centre is the X-ray position of the cluster, while for new clusters it is the \emph{Planck} position.  Gaussian priors are used on $x_{0}$ and $y_{0}$, centred on the \emph{Planck} position (i.e.\ offset from the pointing/phase reference centre, if the pointing centre is the X-ray position) and with width given by the \emph{Planck} positional uncertainty up to a maximum of 5\,arcmin; larger priors allow the detection algorithm to fix on noise features toward the edges of the SA primary beam, which has a FWHM of $\approx$\,20\,arcmin.  In practice, no PwS positional errors in the sample are greater than 5\,arcmin.  MMF1 does not give positional error estimates, so clusters detected only by MMF1 are given the maximum 5\,arcmin error; some clusters detected by MMF3 (but not PwS) have positional errors $> 5$\,arcmin, but as will be shown in Section~\ref{S:pos_comp}, MMF3 positional errors tend to be over-estimated.

Model parameter estimation is performed in a fully Bayesian manner using the AMI in-house software package \textsc{McADAM}, in $uv$-space (see, e.g.\ \citealt{feroz09b} for more details).  Bayes' theorem states that

\begin{equation} 
\Pr(\mathbf{\Theta}|\mathbf{D}, H) =\frac{\Pr(\mathbf{D}|\,\mathbf{\Theta},H)\Pr(\mathbf{\Theta}|H)}{\Pr(\mathbf{D}|H)},
\end{equation}
where $\mathbf{\Theta}$ is a set of parameters for a model, $H$, and $\mathbf{D}$ is the data.  Thus, the posterior probability distribution, $\Pr(\mathbf{\Theta}|\mathbf{D}, H)$, is proportional to the likelihood, $\Pr(\mathbf{D}|\,\mathbf{\Theta},H)$, multiplied by the prior, $\Pr(\mathbf{\Theta}|H)$.  The normalising factor is the evidence, $\Pr(\mathbf{D}|H) \equiv \mathcal{Z}$.  \textsc{McADAM} uses the nested sampler \textsc{MultiNEST} (\citealt{feroz08}, \citealt{feroz09a}) to obtain the posterior distribution for all parameters, which can be marginalised to provide two- and one-dimensional parameter constraints.

\textsc{MultiNEST} also calculates the evidence, which can be ignored for parameter estimation but is important for model selection, since it represents the probability of the data given a model and a prior, marginalised over the the model's parameter space:

\begin{equation}
\mathcal{Z} = \int{\Pr(\mathbf{D}|\,\mathbf{\Theta},H) \Pr(\mathbf{\Theta}|H)}d^D\mathbf{\Theta},
\end{equation}
where $D$ is the dimensionality of the parameter space.  The probability of two different models given the data can be compared using their evidence ratio:

\begin{equation}
\frac{\Pr(H_{1}|\mathbf{D})}{\Pr(H_{0}|\mathbf{D})}
=\frac{\Pr(\mathbf{D}|H_{1})\Pr(H_{1})}{\Pr(\mathbf{D}|
H_{0})\Pr(H_{0})}
=\frac{\mathcal{Z}_1}{\mathcal{Z}_0}\frac{\Pr(H_{1})}{\Pr(H_{0})},
\end{equation}
where $\Pr(H_{1})/\Pr(H_{0})$ is the \emph{a priori} probability ratio for the two models.  To assess the detection significance of a cluster, we therefore perform two parameter estimation runs -- one with the full cluster + radio source environment model $(H_{1})$, and one with only the radio source environment model (the `null' run, $H_{0}$).  We set $\Pr(H_{1})/\Pr(H_{0}) = 1$ so that $\mathcal{Z}_{1}/\mathcal{Z}_{0}$ is a measure of the detection significance for the cluster.  This ratio takes into account the various sources of noise as well as the goodness of fit of the radio source and cluster models.

\begin{figure}
  \begin{center}
  \centerline{
%
%
\begin{psfrags}%
\psfragscanon%
%
\psfrag{s03}[t][t]{\color[rgb]{0,0,0}\setlength{\tabcolsep}{0pt}\begin{tabular}{c}$\Delta \ln (\mathcal{Z})$\end{tabular}}%
\psfrag{s04}[b][b]{\color[rgb]{0,0,0}\setlength{\tabcolsep}{0pt}\begin{tabular}{c}Number of clusters\end{tabular}}%
%
\psfrag{x01}[t][t]{0}%
\psfrag{x02}[t][t]{20}%
\psfrag{x03}[t][t]{40}%
\psfrag{x04}[t][t]{60}%
\psfrag{x05}[t][t]{80}%
\psfrag{x06}[t][t]{100}%
\psfrag{x07}[t][t]{120}%
%
\psfrag{v01}[r][r]{0}%
\psfrag{v02}[r][r]{5}%
\psfrag{v03}[r][r]{10}%
\psfrag{v04}[r][r]{15}%
\psfrag{v05}[r][r]{20}%
%
\resizebox{8.9cm}{!}{\includegraphics{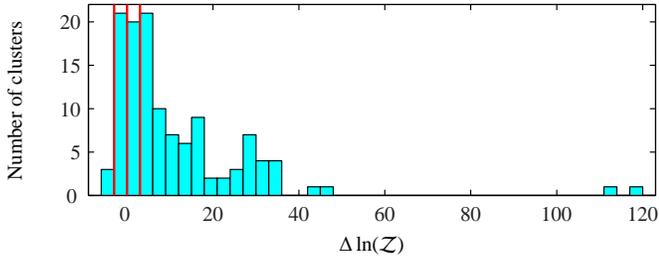}}%
\end{psfrags}%
%
}
  \caption{The distribution of evidence ratio values in the SZ sample, with the division into detection categories given in Table~\ref{T:ev_cats} indicated by red vertical lines.}
  \label{Fi:evidence_hist}
  \end{center}
\end{figure}

Fig.~\ref{Fi:evidence_hist} shows the distribution of $\Delta \ln (\mathcal{Z})$ values in the SZ sample.  It is also useful to define discrete `detection' and `non-detection' categories based on the continuous evidence ratio values.  We follow \citet{Jeffreys61} in taking $\Delta \ln (\mathcal{Z}) = 0$ as the boundary between detections and non-detections.  We also define an additional boundary $\Delta \ln(\mathcal{Z}) = 3$ between `moderate' and `clear' detections, where `moderate' detections are cases where the data are more consistent with the presence of a cluster than not, but there is not enough information in the data to constrain the model parameters well.  For symmetry, we also define a boundary at $\Delta \ln(\mathcal{Z}) = -3$ to indicate cases where the cluster model is strongly rejected by the data.  These boundaries were chosen empirically, by inspecting final maps and posterior distributions.  The four categories are listed in Table~\ref{T:ev_cats}.

\begin{table}[hbt]
\centering
\caption{The evidence difference ($\Delta \ln(\mathcal{Z})$) boundaries used for categorising clusters as clear detections, moderate detections, non-detections and clear non-detections, and the number of clusters in each category in the SZ sample.}\label{T:ev_cats}
\begin{tabular}{lcr}
\hline
Category & $\Delta \ln(\mathcal{Z})$ boundaries & Number \\ \hline
Clear detection (Y) & $\Delta \ln(\mathcal{Z}) \geq 3$ & 79 \\
Moderate detection (M) & $0 \leq \Delta \ln(\mathcal{Z}) < 3$ & 20 \\
Non-detection (N) & $-3 \leq \Delta \ln(\mathcal{Z}) < 0$ & 21 \\
Clear non-detection (NN) & $\Delta \ln(\mathcal{Z}) < -3$ & 3 \\
\hline
\end{tabular}
\end{table}

\subsubsection{Prior on \Ytot \:and $\theta_{\rm s}$}\label{sec:2D_prior}

 The priors assigned to \Ytot \:and $\theta_{s}$ in AP2013 and used for the \emph{Planck} \textsc{PwS} analysis are based on marginalised distributions of \Ytot \:and $\theta_{s}$ in a simulated population of clusters generated according to the Jenkins mass function \citep{2001MNRAS.321..372J}, as described in \citet{carvalho11}.  The parameterisation functions for these priors are listed in Table~\ref{T:cluster_priors}.  These priors ignore, however, the correlation between \Ytot \:and $\theta_{\rm s}$; in addition, they take into account the \emph{Planck} selection function only in assuming minimum and maximum cutoffs in each parameter.  

To produce a better approximation to the true distribution of clusters expected to be detected by \emph{Planck}, we used the results of the \emph{Planck} completeness simulation (\citealt{2013arXiv1303.5089P}, Section~3.1 and 3.2, Fig.~9).  This simulation was produced by drawing a cluster population from the Tinker mass function \citep{2008ApJ...688..709T}, and converting the redshifts and masses to $Y_{500}$ and $\theta_{500}$ observable quantities using the scaling relations in \citet{planck2011-5.2a}.  This cluster population was injected into the real \emph{Planck} data assuming GNFW pressure profiles with the shape parameters varying according to results from \citet{planck2012-V} and a simulated union catalogue was created by running the \emph{Planck} detection pipelines on the simulated dataset in the usual manner; see \citet{2013arXiv1303.5089P} for more details.

\begin{figure}
  \begin{center}
  \centerline{
%
%
\begin{psfrags}%
\psfragscanon%
%
\psfrag{s04}[t][t]{\color[rgb]{0,0,0}\setlength{\tabcolsep}{0pt}\begin{tabular}{c}$\theta_{s}$ / arcmin\end{tabular}}%
\psfrag{s05}[b][b]{\color[rgb]{0,0,0}\setlength{\tabcolsep}{0pt}\begin{tabular}{c}$Y_{\rm tot}$ / arcmin$^{2}$\end{tabular}}%
\psfrag{s08}[][]{\color[rgb]{0,0,0}\setlength{\tabcolsep}{0pt}\begin{tabular}{c} \end{tabular}}%
\psfrag{s09}[][]{\color[rgb]{0,0,0}\setlength{\tabcolsep}{0pt}\begin{tabular}{c} \end{tabular}}%
\psfrag{s10}[b][b]{\color[rgb]{0,0,0}\setlength{\tabcolsep}{0pt}\begin{tabular}{c}Number of clusters\end{tabular}}%
\psfrag{s13}[t][t]{\color[rgb]{0,0,0}\setlength{\tabcolsep}{0pt}\begin{tabular}{c}$\theta_{s}$ / arcmin\end{tabular}}%
\psfrag{s18}[][]{\color[rgb]{0,0,0}\setlength{\tabcolsep}{0pt}\begin{tabular}{c} \end{tabular}}%
\psfrag{s19}[][]{\color[rgb]{0,0,0}\setlength{\tabcolsep}{0pt}\begin{tabular}{c} \end{tabular}}%
\psfrag{s20}[b][b]{\color[rgb]{0,0,0}\setlength{\tabcolsep}{0pt}\begin{tabular}{c}Number of clusters\end{tabular}}%
%
\psfrag{x01}[B][B]{-20}%
\psfrag{x02}[B][B]{-10}%
\psfrag{x03}[B][B]{0}%
\psfrag{x04}[B][B]{10}%
\psfrag{x05}[B][B]{20}%
\psfrag{x06}[t][t]{2}%
\psfrag{x07}[t][t]{5}%
\psfrag{x08}[t][t]{10}%
\psfrag{x09}[t][t]{20}%
\psfrag{x10}[t][t]{40}%
\psfrag{x11}[B][B]{0}%
\psfrag{x12}[B][B]{20}%
\psfrag{x13}[B][B]{40}%
\psfrag{x14}[t][t]{2}%
\psfrag{x15}[t][t]{5}%
\psfrag{x16}[t][t]{10}%
\psfrag{x17}[t][t]{20}%
\psfrag{x18}[t][t]{40}%
%
\psfrag{v01}[r][r]{0}%
\psfrag{v02}[r][r]{0.5}%
\psfrag{v03}[r][r]{1}%
\psfrag{v04}[r][r]{0.001}%
\psfrag{v05}[r][r]{0.01}%
\psfrag{v06}[r][r]{0.1}%
\psfrag{v07}[r][r]{0}%
\psfrag{v08}[r][r]{0.5}%
\psfrag{v09}[r][r]{1}%
\psfrag{v10}[r][r]{0.001}%
\psfrag{v11}[r][r]{0.01}%
\psfrag{v12}[r][r]{0.1}%
%
\resizebox{8.9cm}{!}{\includegraphics{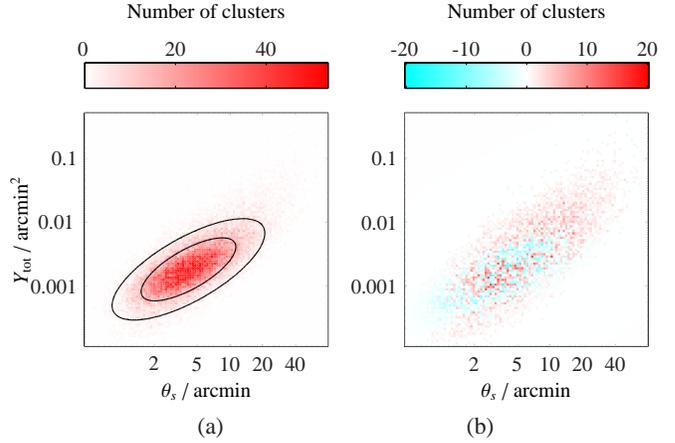}}%
\end{psfrags}%
%
}\medskip
  \centerline{(a) \hskip 0.35\linewidth (b)}
  \caption{(a) shows the sampled distribution (red histogram), and the two-dimensional elliptical Gaussian fit to the \Ytot \:vs $\theta_{s}$ distribution in log-space (black lines, enclosing 68\% and 95\% of the probability).  (b) shows the residuals with respect to the simulated distribution.  Note that the colour-axis scales are different.}
  \label{Fi:2d_gaussian}
  \end{center}
\end{figure}

We noted that the resulting two-dimensional distribution in $\theta_{\rm s}$ and \Ytot \:in log-space was elliptical in shape with roughly Gaussian distribution along the principal axes and performed a two-dimensional Gaussian fit to the distribution, parameterised by width and offset in $x = \log_{10}(\theta_{\rm s})$, width and offset in $y = \log_{10}(Y_{\rm tot})$, and angle $\phi$ measured clockwise from the $y$-axis.  The best-fit parameters are listed in Table~\ref{T:cluster_priors}, and the fit and residuals with respect to the simulated population are shown in Fig.~\ref{Fi:2d_gaussian}.  We use this fit to the simulated population as our prior on $\theta_{s}$ and \Ytot.

\begin{table*}
\begin{center}
\caption{Priors used on profile fit parameters}\label{T:cluster_priors}
\begin{tabular}{lccc}
\hline
Parameter & Prior type & Parameters & Limits \\ \hline
$x_{0}$, $y_{0}$ & Gaussian, $e^{-x^{2}/2\sigma^{2\phantom{^{2}}}}$ & $\sigma = \max(5\,\mathrm{arcmin}, \sigma_{\mathrm{Planck}})$ & - \\[5pt]
\Ytot \:(old) & Power-law, $x^{-a}$ & $a = 1.6$ & $0.0005 < x < 0.2$ \\[5pt]
$\theta_{s}$ (old) & Exponential, $\lambda e^{-\lambda x}$ & $\lambda = 0.2$ & $1.3 < x < 45$ \\[5pt]
\multirow{3}{*}{\Ytot, $\theta_{s}$ (new)} & 2D elliptical Gaussian & $x_{0} = 0.6171, \sigma_{x} = 0.1153,$  & \multirow{3}{*}{$1.3 < \theta_{s}$} \\
& in $x = \log_{10}(\theta_{s}),$ & $y_{0} = -2.743, \sigma_{y} = 0.2856,$ &  \\
& $y = \log_{10}(Y_{\rm tot})$ & $\phi = 40.17^{\circ}$ & \\
\hline
\end{tabular}
\end{center}
\end{table*}

\subsection{Results}
\label{sec:resultsreal}

In the SZ sample, 79 are clear detections, 20 are moderate detections, 21 are non-detections and 3 are clear non-detections.  A summary of the results for each cluster in the sample is presented in Appendix~\ref{sec:results_table}.

Some representative examples from each category are discussed in the following.  In each case, two foreground-source-subtracted maps are shown; both are produced using natural weighting, and the second also has a Gaussian weighting function with the 30\% point at 600\,$\lambda$ applied (the `$uv$-tapered' map).  This taper downweights the longer baselines, which are only sensitive to small-angular-scale features, making the extended cluster more visible.  The symbols $\times$ and $+$ show the positions of subtracted sources, respectively either modelled in \textsc{McAdam} or directly subtracted based on LA values. $\Box$ shows the AMI (\textsc{McAdam}-determined) position of the cluster, and the $1\times\sigma_{Planck}$ positional error radius is shown as a circle.  Contours are plotted at $\pm (2, 3, 4, ..., 10)\times$ the r.m.s.\ noise level (measured using the \textsc{aips} task \textsc{imean}), and dashed contours are negative.  The synthesised beam is shown in the bottom left-hand corner.  We emphasise that these maps are only shown for visual inspection and to assess the residual foreground contamination; all parameter estimation is done in $uv$-space.

Posterior distributions for position offset, cluster model parameters and the flux densities of the closest radio sources to the cluster centre are also shown; in these plots the units are arcsec on the sky for offset in RA ($x_{0}$) and $\delta$ ($y_{0}$), arcmin$^{2}$ for \Ytot, arcmin for $\theta_{s}$ and mJy for radio source flux densities.  The blue (pink) areas correspond to regions of higher (lower) probability density.  The \Ytot-$\theta_{s}$ posterior distribution is shown separately with solid black lines for the AMI constraints overlaid with that obtained by \textsc{PwS} using \emph{Planck} data for the cluster in red, as well as the AMI prior (black dashed lines).  The joint constraint is shown in yellow where appropriate.  In all cases, the contours mark the 68\% and 95\% confidence limits in the posterior or prior probability distributions.  Similar maps and posterior distribution plots for the entire sample are available online at \url{http://www.astro.phy.cam.ac.uk/surveys/ami-planck/}.

\subsubsection{Clear detections}

\begin{figure}
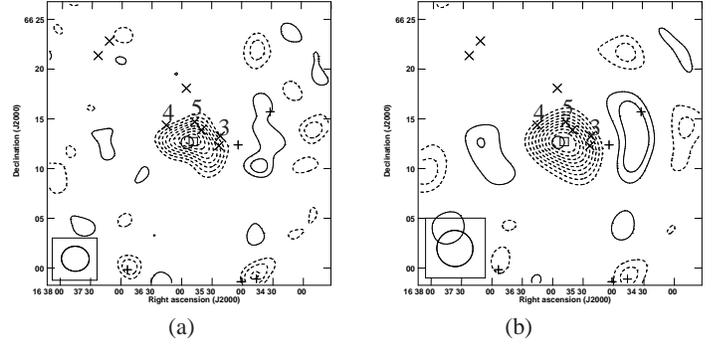

  \begin{psfrags}%
  \psfragscanon%
  \small
  \psfrag{s03}[c][c]{3}%
  \psfrag{s04}[c][c]{4}%
  \psfrag{s05}[c][c]{5}%
  \centerline{\includegraphics[bb=47 156 570 648, clip=, height=0.45\linewidth]{A2218_nw.ps}\qquad\includegraphics[bb=47 156 570 648, clip=, clip=, height=0.45\linewidth]{A2218_uvtap.ps}}
  \end{psfrags}
  \centerline{\hskip 0.05\linewidth (a) \hskip 0.45\linewidth (b)\phantom{$S^{\rm{X^{X^{X}}}}$}}
  \caption{SA source-subtracted map of A2218 with (a) natural weighting and (b) a $uv$-taper.  The r.m.s.\ noise levels are 131 and 163\,$\mu$Jy\,beam$^{-1}$ respectively.  The numbered sources have posterior distributions for their flux densities plotted in Fig.~\ref{Fi:A2218_post}.  See Section~\ref{sec:resultsreal} for more details on the plots.}\label{Fi:A2218_maps}
\end{figure}

\begin{figure}
  \fbox{
%
%
\begin{psfrags}%
\psfragscanon%
%
\psfrag{s29}[t][t]{\color[rgb]{0,0,0}\setlength{\tabcolsep}{0pt}\begin{tabular}{c}$S_{5}$\end{tabular}}%
\psfrag{s33}[t][t]{\color[rgb]{0,0,0}\setlength{\tabcolsep}{0pt}\begin{tabular}{c}$y_0$\end{tabular}}%
\psfrag{s37}[t][t]{\color[rgb]{0,0,0}\setlength{\tabcolsep}{0pt}\begin{tabular}{c}$\theta_s$\end{tabular}}%
\psfrag{s41}[t][t]{\color[rgb]{0,0,0}\setlength{\tabcolsep}{0pt}\begin{tabular}{c}$Y_{\rm tot}$\\$\times 10^3$\end{tabular}}%
\psfrag{s45}[t][t]{\color[rgb]{0,0,0}\setlength{\tabcolsep}{0pt}\begin{tabular}{c}$S_{3}$\end{tabular}}%
\psfrag{s49}[t][t]{\color[rgb]{0,0,0}\setlength{\tabcolsep}{0pt}\begin{tabular}{c}$S_{4}$\end{tabular}}%
\psfrag{s53}[t][t]{\color[rgb]{0,0,0}\setlength{\tabcolsep}{0pt}\begin{tabular}{c}$x_0$\end{tabular}}%
\psfrag{s54}[t][t]{\color[rgb]{0,0,0}\setlength{\tabcolsep}{0pt}\begin{tabular}{c}$S_{5}$\end{tabular}}%
\psfrag{s73}[t][t]{\color[rgb]{0,0,0}\setlength{\tabcolsep}{0pt}\begin{tabular}{c}$y_0$\end{tabular}}%
\psfrag{s89}[t][t]{\color[rgb]{0,0,0}\setlength{\tabcolsep}{0pt}\begin{tabular}{c}$\theta_s$\end{tabular}}%
\psfrag{s01}[t][t]{\color[rgb]{0,0,0}\setlength{\tabcolsep}{0pt}\begin{tabular}{c}$Y_{\rm tot} \times 10^3$\end{tabular}}%
\psfrag{s02}[t][t]{\color[rgb]{0,0,0}\setlength{\tabcolsep}{0pt}\begin{tabular}{c}$S_{3}$\end{tabular}}%
\psfrag{s03}[t][t]{\color[rgb]{0,0,0}\setlength{\tabcolsep}{0pt}\begin{tabular}{c}$S_{4}$\end{tabular}}%
%
\psfrag{x12}[t][t]{1.5}%
\psfrag{x13}[t][t]{2}%
\psfrag{x14}[t][t]{5.5}%
\psfrag{x15}[t][t]{6}%
\psfrag{x20}[t][t]{6}%
\psfrag{x21}[t][t]{6.2}%
\psfrag{x22}[t][t]{5}%
\psfrag{x23}[t][t]{15}%
\psfrag{x24}[t][t]{5}%
\psfrag{x25}[t][t]{10}%
\psfrag{x26}[t][t]{15}%
\psfrag{x27}[t][t]{5}%
\psfrag{x28}[t][t]{10}%
\psfrag{x29}[t][t]{15}%
\psfrag{x30}[t][t]{4}%
\psfrag{x31}[t][t]{8}%
\psfrag{x32}[t][t]{12}%
\psfrag{x48}[t][t]{-30}%
\psfrag{x49}[t][t]{0}%
\psfrag{x50}[t][t]{30}%
\psfrag{x71}[t][t]{-20}%
\psfrag{x72}[t][t]{20}%
\psfrag{x73}[t][t]{60}%
\psfrag{x99}[t][t]{2.5}%
\psfrag{x01}[t][t]{3}%
%
\psfrag{v91}[r][r]{2.5}%
\psfrag{v92}[r][r]{3}%
\psfrag{v93}[r][r]{1.5}%
\psfrag{v94}[r][r]{2}%
\psfrag{v95}[r][r]{5.5}%
\psfrag{v96}[r][r]{6}%
\psfrag{v97}[r][r]{5}%
\psfrag{v98}[r][r]{15}%
\psfrag{v99}[r][r]{5}%
\psfrag{v01}[r][r]{10}%
\psfrag{v02}[r][r]{-30}%
\psfrag{v03}[r][r]{0}%
\psfrag{v04}[r][r]{30}%
%
\resizebox{0.95\linewidth}{!}{\includegraphics{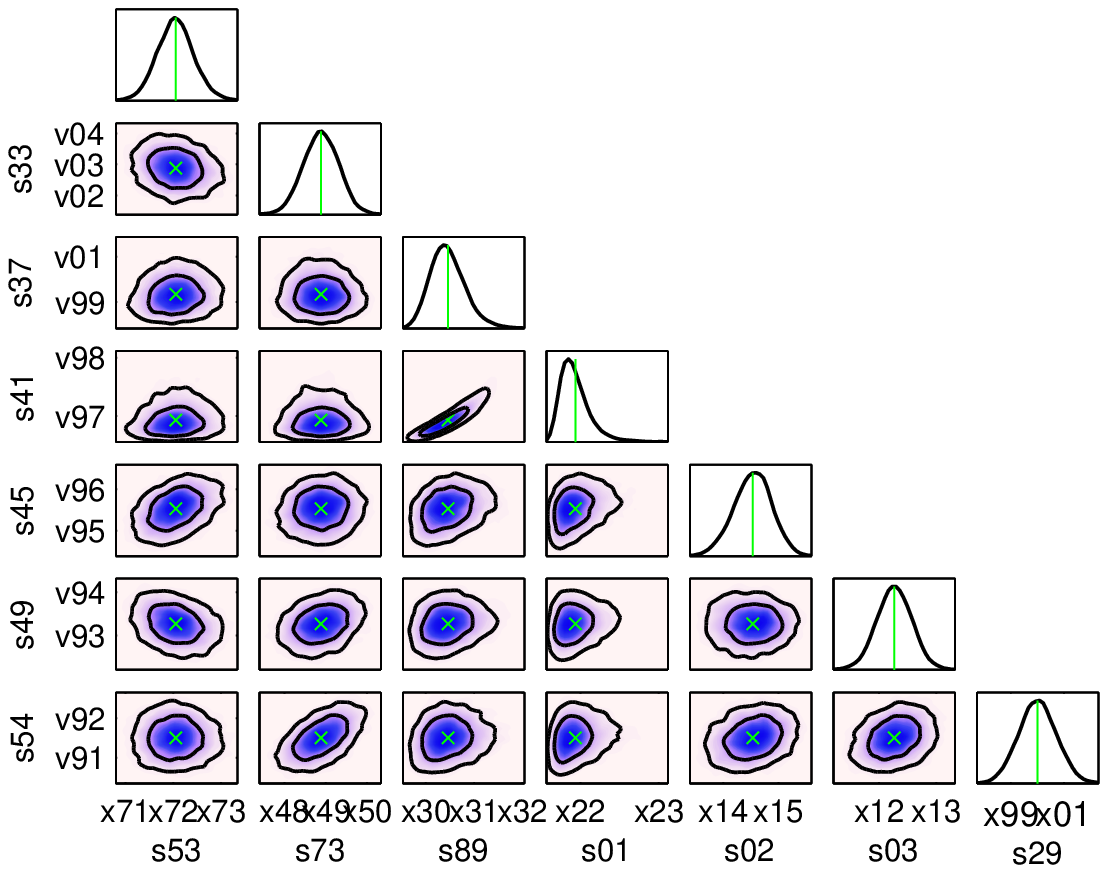}}%
\end{psfrags}%
%
}\hspace{-0.42\linewidth}
  \fbox{\raisebox{\height}{\includegraphics[width=0.4\linewidth]{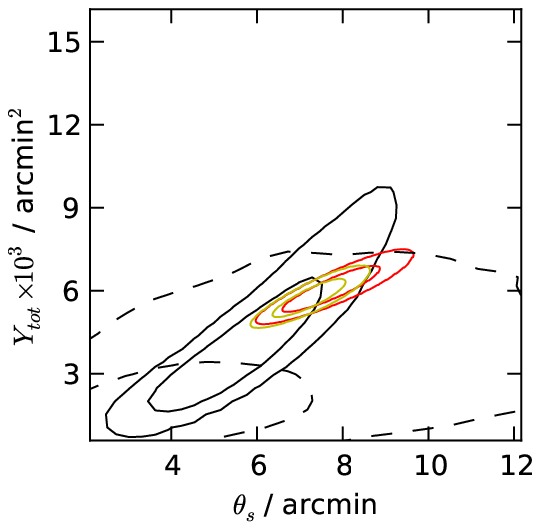}}}
  \caption{AMI posterior distributions for A2218 and the \Ytot-$\theta_{s}$ posterior overlaid with that obtained by \emph{Planck} in red, and the prior as a black dotted line (upper right-hand corner).  The joint constraint is shown in yellow.  See Section~\ref{sec:resultsreal} for more details on the plots.}\label{Fi:A2218_post}
\end{figure}

\paragraph{Abell 2218 (PSZ1~G097.72+38.13)} \mbox{}\\
\noindent Abell 2218 \citep{1958ApJS....3..211A} is an extremely well-known cluster and one of the earliest SZ detections (e.g. \citealt{1978Natur.275...40B}, \citealt{1984Natur.309...34B}, \citealt{1993Natur.365..320J}).  It lies at redshift $z = 0.171$ \citep{1978ApJ...221..383K}.  It has been observed by AMI previously as part of the LoCuSS sample \citep{2012MNRAS.425..162L} and was also in AP2013.  It has the highest \emph{Planck} SNR in the final subsample and is also well-detected by AMI with $\Delta \ln(\mathcal{Z}) = 34$.  Fig.~\ref{Fi:A2218_maps} shows that the cluster is resolved by AMI as the depth of the decrement increases in the $uv$-tapered map, and structure can be clearly seen in the naturally-weighted map.  The posterior distributions (Fig.~\ref{Fi:A2218_post}) show good constraints in both position and the cluster model parameters.  The two-dimensional posterior distributions for the flux densities of the three most significant nearby sources are included in the plot; it can be seen that there is some correlation between the flux densities of the sources and \Ytot, i.e.\ lower values of the flux densities allow lower values of \Ytot, but this does not affect the parameter constraints significantly.  There is also some correlation between the flux densities of the sources and the cluster position.  The remaining two sources near the cluster centre are fainter and were not modelled in the initial analysis since they appear at $<4\sigma$ on the SA map; there is no evidence for degeneracy between the flux densities of these sources and the cluster parameters.  As in AP2013 (see their Fig.~5), the \textsc{PwS} \Ytot-$\theta_{s}$ posterior overlaps with the AMI posterior, but AMI finds the cluster to be smaller and fainter than \emph{Planck} (at low significance for this particular cluster).

\Ytot \:is the total SZ signal of the cluster and corresponds to the zero-spacing flux, which is not measured by an interferometer; the constraints produced by AMI on \Ytot \:therefore rely on extrapolating the signal on the angular scales that AMI does measure ($\approx$\,200 to 1200$\lambda$, corresponding to $\approx$\,15 to 3\,arcmin) to $0\lambda$ assuming a fixed profile.  Since this is a relatively nearby, large-angular-size cluster (i.e.\ $\theta_{500}$ inferred from the X-ray luminosity is 6.4\,arcmin (\citealt{bohringer00}, \citealt{2011A&A...534A.109P}) corresponding to $\theta_{s} = 5.4$\,arcmin for the `universal' value of $c_{500} = 1.177$, in agreement with the AMI constraint and slightly smaller than the preferred \emph{Planck} value), much of the flux of the cluster exists on scales that are not measured by AMI.  \Ytot \:is therefore not well constrained and the \Ytot-$\theta_{s}$ degeneracy is large compared to that produced by \emph{Planck}, which measures \Ytot \:directly.  Nonetheless, the different degeneracy direction means that combining the two posteriors results in a tighter constraint (assuming no systematic difference between the two instruments, which will be discussed in Section~\ref{S:Ytot_theta_comp}).

\setlength\fboxrule{0pt}

\begin{figure}
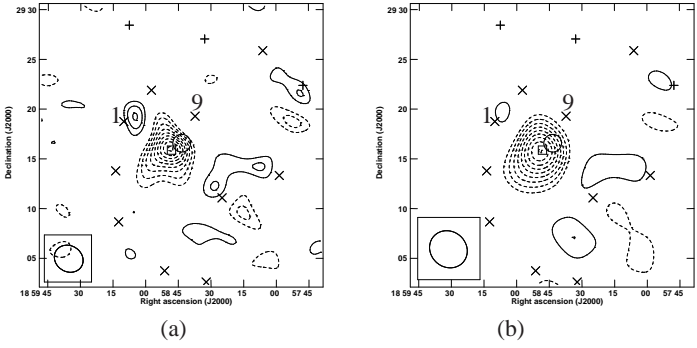

  \begin{psfrags}%
  \psfragscanon%
  \small
  \psfrag{s01}[c][c]{1}%
  \psfrag{s09}[c][c]{9}%
  \centerline{\includegraphics[bb=47 156 570 648, clip=, clip=, height=0.45\linewidth]{CAJ1858_nw.ps}\qquad\includegraphics[bb=47 156 570 648, clip=, clip=, height=0.45\linewidth]{CAJ1858_uvtap.ps}}
  \end{psfrags}
  \centerline{\hskip 0.05\linewidth (a) \hskip 0.45\linewidth (b)\phantom{$S^{\rm{X^{X^{X}}}}$}}
  \caption{SA source-subtracted map of PSZ1 G060.12+11.42 with (a) natural weighting and (b) a $uv$-taper.  The r.m.s.\ noise levels are 96 and 131\,$\mu$Jy\,beam$^{-1}$ respectively.  The numbered sources have posterior distributions for their flux densities plotted in Fig.~\ref{Fi:CAJ1858_post}.  See Section~\ref{sec:resultsreal} for more details on the plots.}\label{Fi:CAJ1858_maps}
\end{figure}

\begin{figure}
  \fbox{
%
%
\begin{psfrags}%
\psfragscanon%
\small
%
\psfrag{s01}[t][t]{\color[rgb]{0,0,0}\setlength{\tabcolsep}{0pt}\begin{tabular}{c}$S_{9}$\end{tabular}}%
\psfrag{s02}[b][b]{\color[rgb]{0,0,0}\setlength{\tabcolsep}{0pt}\begin{tabular}{c}$y_0$\end{tabular}}%
\psfrag{s03}[b][b]{\color[rgb]{0,0,0}\setlength{\tabcolsep}{0pt}\begin{tabular}{c}$\theta_s$\end{tabular}}%
\psfrag{s04}[b][b]{\color[rgb]{0,0,0}\setlength{\tabcolsep}{0pt}\begin{tabular}{c}$Y_{tot}$\end{tabular}}%
\psfrag{s05}[b][b]{\color[rgb]{0,0,0}\setlength{\tabcolsep}{0pt}\begin{tabular}{c}$S_{1}$\end{tabular}}%
\psfrag{s06}[t][t]{\color[rgb]{0,0,0}\setlength{\tabcolsep}{0pt}\begin{tabular}{c}$x_0$\end{tabular}}%
\psfrag{s07}[b][b]{\color[rgb]{0,0,0}\setlength{\tabcolsep}{0pt}\begin{tabular}{c}$S_{9}$\end{tabular}}%
\psfrag{s08}[t][t]{\color[rgb]{0,0,0}\setlength{\tabcolsep}{0pt}\begin{tabular}{c}$y_0$\end{tabular}}%
\psfrag{s09}[t][t]{\color[rgb]{0,0,0}\setlength{\tabcolsep}{0pt}\begin{tabular}{c}$\theta_s$\end{tabular}}%
\psfrag{s10}[t][t]{\color[rgb]{0,0,0}\setlength{\tabcolsep}{0pt}\begin{tabular}{c}$Y_{tot}$\end{tabular}}%
\psfrag{s11}[t][t]{\color[rgb]{0,0,0}\setlength{\tabcolsep}{0pt}\begin{tabular}{c}$S_{1}$\end{tabular}}%
%
\psfrag{x01}[t][t]{0}%
\psfrag{x02}[t][t]{0.1}%
\psfrag{x03}[t][t]{0.2}%
\psfrag{x04}[t][t]{0.3}%
\psfrag{x05}[t][t]{0.4}%
\psfrag{x06}[t][t]{0.5}%
\psfrag{x07}[t][t]{0.6}%
\psfrag{x08}[t][t]{0.7}%
\psfrag{x09}[t][t]{0.8}%
\psfrag{x10}[t][t]{0.9}%
\psfrag{x11}[t][t]{1}%
\psfrag{x12}[t][t]{4}%
\psfrag{x13}[t][t]{4.5}%
\psfrag{x14}[t][t]{5}%
\psfrag{x15}[t][t]{0}%
\psfrag{x16}[t][t]{4}%
\psfrag{x17}[t][t]{8}%
\psfrag{x18}[t][t]{0}%
\psfrag{x19}[t][t]{5}%
\psfrag{x20}[t][t]{0}%
\psfrag{x21}[t][t]{5}%
\psfrag{x22}[t][t]{10}%
\psfrag{x23}[t][t]{0}%
\psfrag{x24}[t][t]{5}%
\psfrag{x25}[t][t]{10}%
\psfrag{x26}[t][t]{0}%
\psfrag{x27}[t][t]{5}%
\psfrag{x28}[t][t]{10}%
\psfrag{x29}[t][t]{-100}%
\psfrag{x30}[t][t]{-50}%
\psfrag{x31}[t][t]{-100}%
\psfrag{x32}[t][t]{-80}%
\psfrag{x33}[t][t]{-60}%
\psfrag{x34}[t][t]{-40}%
\psfrag{x35}[t][t]{-20}%
\psfrag{x36}[t][t]{-100}%
\psfrag{x37}[t][t]{-80}%
\psfrag{x38}[t][t]{-60}%
\psfrag{x39}[t][t]{-40}%
\psfrag{x40}[t][t]{-20}%
\psfrag{x41}[t][t]{-100}%
\psfrag{x42}[t][t]{-80}%
\psfrag{x43}[t][t]{-60}%
\psfrag{x44}[t][t]{-40}%
\psfrag{x45}[t][t]{-20}%
\psfrag{x46}[t][t]{-100}%
\psfrag{x47}[t][t]{-50}%
\psfrag{x48}[t][t]{-100}%
\psfrag{x49}[t][t]{-50}%
\psfrag{x50}[t][t]{-100}%
\psfrag{x51}[t][t]{-50}%
\psfrag{x52}[t][t]{-100}%
\psfrag{x53}[t][t]{-50}%
\psfrag{x54}[t][t]{-100}%
\psfrag{x55}[t][t]{-50}%
\psfrag{x56}[t][t]{0}%
\psfrag{x57}[t][t]{0.5}%
\psfrag{x58}[t][t]{1}%
\psfrag{x59}[t][t]{1.5}%
\psfrag{x60}[t][t]{4}%
\psfrag{x61}[t][t]{4.5}%
\psfrag{x62}[t][t]{5}%
\psfrag{x63}[t][t]{0}%
\psfrag{x64}[t][t]{5}%
\psfrag{x65}[t][t]{0}%
\psfrag{x66}[t][t]{5}%
\psfrag{x67}[t][t]{10}%
\psfrag{x68}[t][t]{-100}%
\psfrag{x69}[t][t]{-80}%
\psfrag{x70}[t][t]{-60}%
\psfrag{x71}[t][t]{-40}%
\psfrag{x72}[t][t]{-20}%
\psfrag{x73}[t][t]{-100}%
\psfrag{x74}[t][t]{-50}%
%
\psfrag{v01}[r][r]{0}%
\psfrag{v02}[r][r]{0.1}%
\psfrag{v03}[r][r]{0.2}%
\psfrag{v04}[r][r]{0.3}%
\psfrag{v05}[r][r]{0.4}%
\psfrag{v06}[r][r]{0.5}%
\psfrag{v07}[r][r]{0.6}%
\psfrag{v08}[r][r]{0.7}%
\psfrag{v09}[r][r]{0.8}%
\psfrag{v10}[r][r]{0.9}%
\psfrag{v11}[r][r]{1}%
\psfrag{v12}[r][r]{1}%
\psfrag{v13}[r][r]{1.5}%
\psfrag{v14}[r][r]{1}%
\psfrag{v15}[r][r]{1.5}%
\psfrag{v16}[r][r]{4}%
\psfrag{v17}[r][r]{4.5}%
\psfrag{v18}[r][r]{5}%
\psfrag{v19}[r][r]{1}%
\psfrag{v20}[r][r]{1.5}%
\psfrag{v21}[r][r]{4}%
\psfrag{v22}[r][r]{4.5}%
\psfrag{v23}[r][r]{5}%
\psfrag{v24}[r][r]{0}%
\psfrag{v25}[r][r]{5}%
\psfrag{v26}[r][r]{1}%
\psfrag{v27}[r][r]{1.5}%
\psfrag{v28}[r][r]{4}%
\psfrag{v29}[r][r]{4.5}%
\psfrag{v30}[r][r]{5}%
\psfrag{v31}[r][r]{0}%
\psfrag{v32}[r][r]{5}%
\psfrag{v33}[r][r]{0}%
\psfrag{v34}[r][r]{5}%
\psfrag{v35}[r][r]{10}%
\psfrag{v36}[r][r]{0}%
\psfrag{v37}[r][r]{0.5}%
\psfrag{v38}[r][r]{1}%
\psfrag{v39}[r][r]{1.5}%
\psfrag{v40}[r][r]{4}%
\psfrag{v41}[r][r]{4.5}%
\psfrag{v42}[r][r]{5}%
\psfrag{v43}[r][r]{0}%
\psfrag{v44}[r][r]{4}%
\psfrag{v45}[r][r]{8}%
\psfrag{v46}[r][r]{0}%
\psfrag{v47}[r][r]{5}%
\psfrag{v48}[r][r]{10}%
\psfrag{v49}[r][r]{-100}%
\psfrag{v50}[r][r]{-50}%
\psfrag{v51}[r][r]{0}%
\psfrag{v52}[r][r]{0.5}%
\psfrag{v53}[r][r]{1}%
\psfrag{v54}[r][r]{0}%
\psfrag{v55}[r][r]{0.5}%
\psfrag{v56}[r][r]{1}%
\psfrag{v57}[r][r]{0}%
\psfrag{v58}[r][r]{0.5}%
\psfrag{v59}[r][r]{1}%
\psfrag{v60}[r][r]{0}%
\psfrag{v61}[r][r]{0.5}%
\psfrag{v62}[r][r]{1}%
\psfrag{v63}[r][r]{0}%
\psfrag{v64}[r][r]{0.5}%
\psfrag{v65}[r][r]{1}%
\psfrag{v66}[r][r]{0}%
%
\resizebox{0.95\linewidth}{!}{\includegraphics{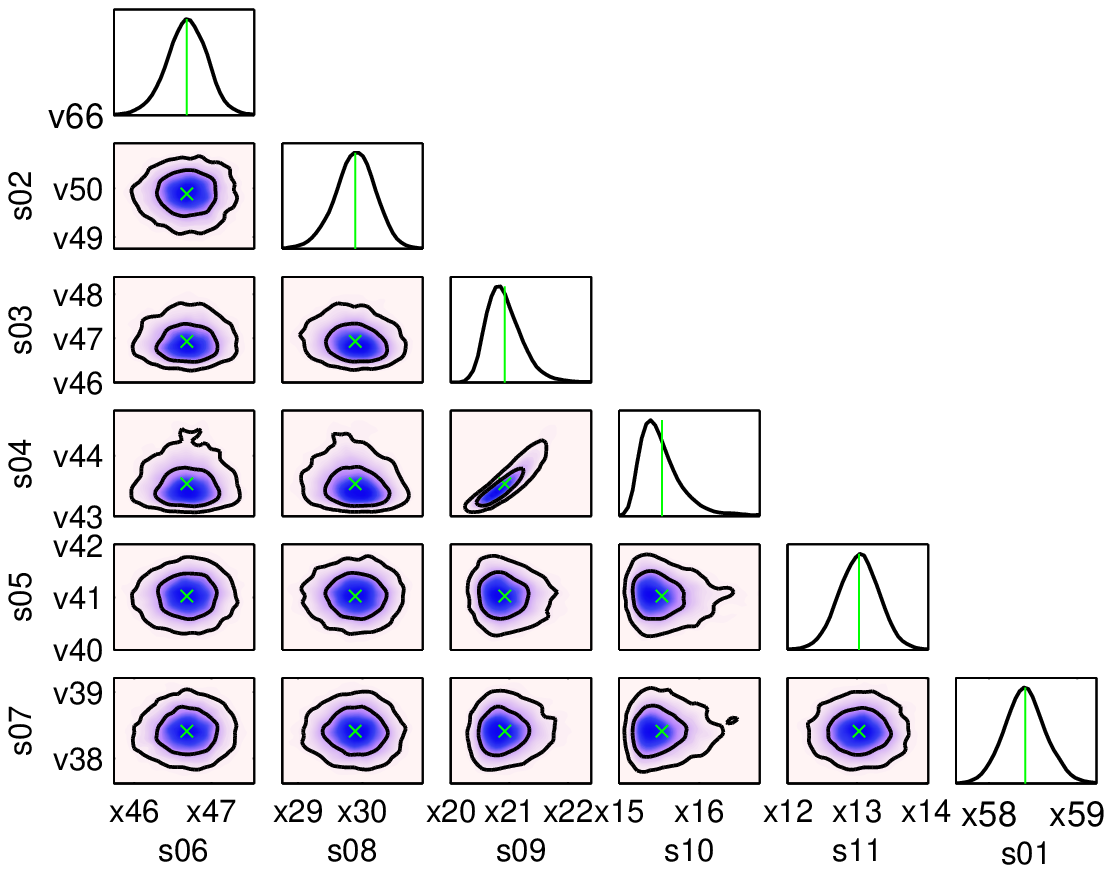}}%
\end{psfrags}%
%
}\hspace{-0.42\linewidth}
  \fbox{\raisebox{\height}{\includegraphics[width=0.4\linewidth]{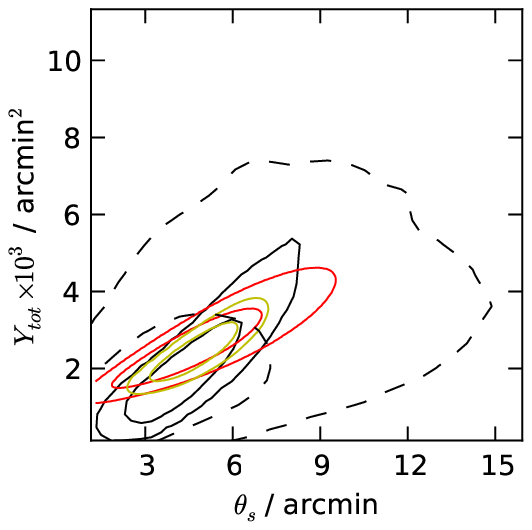}}}
  \caption{AMI posterior distributions for PSZ1 G060.12+11.42 and the \Ytot-$\theta_{s}$ posterior overlaid with that obtained by \emph{Planck} (upper right-hand corner).  The joint constraint is shown in yellow.  See Section~\ref{sec:resultsreal} for more details on the plots.}\label{Fi:CAJ1858_post}
\end{figure}

\paragraph{PSZ1 G060.12+11.42}\mbox{}\\
\noindent This is a new, previously unconfirmed (at the time the catalogue was published) cluster discovered by \emph{Planck} at high SNR (7.2) and clearly detected by AMI with $\Delta \ln(\mathcal{Z}) = 16$.  The source-subtracted maps for the cluster are shown in Fig.~\ref{Fi:CAJ1858_maps}, and the posterior distributions in Fig.~\ref{Fi:CAJ1858_post}.  Again, it is clear that AMI resolves the cluster.  The source flux densities of the two nearest sources are shown in the posterior distributions; there is no apparent degeneracy between the source flux densities and any of the parameters.  In this case, the posterior distributions for $\theta_{s}$ and \Ytot \:are very consistent with the \textsc{PwS} posteriors.  The AMI and \textsc{PwS} degeneracies are in different directions, meaning that the joint constraints produced by combining the two are considerably tighter.

\subsubsection{Moderate detections}

\begin{figure}
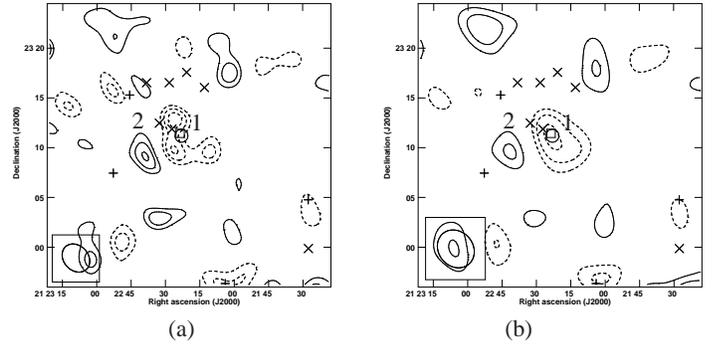

  \begin{psfrags}%
  \psfragscanon%
  \small
  \psfrag{s01}[c][c]{1}%
  \psfrag{s02}[c][c]{2}%
  \centerline{\includegraphics[bb=47 156 570 648, clip=, clip=, height=0.45\linewidth]{ZW8503_nw.ps}\qquad\includegraphics[bb=47 156 570 648, clip=, clip=, height=0.45\linewidth]{ZW8503_uvtap.ps}}
  \end{psfrags}
  \centerline{\hskip 0.05\linewidth (a) \hskip 0.45\linewidth (b)\phantom{$S^{\rm{X^{X^{X}}}}$}}
  \caption{SA source-subtracted map of ZW8503 with (a) natural weighting and (b) a $uv$-taper.  The r.m.s.\ noise levels are 90 and 122\,$\mu$Jy\,beam$^{-1}$ respectively.  The numbered sources have posterior distributions for their flux densities plotted in Fig.~\ref{Fi:ZW8503_post}.  See Section~\ref{sec:resultsreal} for more details on the plots.}\label{Fi:ZW8503_maps}
\end{figure}

\begin{figure}
  \fbox{
%
%
\begin{psfrags}%
\psfragscanon%
%
\psfrag{s20}[t][t]{\color[rgb]{0,0,0}\setlength{\tabcolsep}{0pt}\begin{tabular}{c}$S_{2}$\end{tabular}}%
\psfrag{s24}[b][b]{\color[rgb]{0,0,0}\setlength{\tabcolsep}{0pt}\begin{tabular}{c}$y_0$\end{tabular}}%
\psfrag{s28}[b][b]{\color[rgb]{0,0,0}\setlength{\tabcolsep}{0pt}\begin{tabular}{c}$\theta_s$\end{tabular}}%
\psfrag{s32}[b][b]{\color[rgb]{0,0,0}\setlength{\tabcolsep}{0pt}\begin{tabular}{c}$Y_{tot}$\end{tabular}}%
\psfrag{s36}[b][b]{\color[rgb]{0,0,0}\setlength{\tabcolsep}{0pt}\begin{tabular}{c}$S_{1}$\end{tabular}}%
\psfrag{s40}[t][t]{\color[rgb]{0,0,0}\setlength{\tabcolsep}{0pt}\begin{tabular}{c}$x_0$\end{tabular}}%
\psfrag{s41}[b][b]{\color[rgb]{0,0,0}\setlength{\tabcolsep}{0pt}\begin{tabular}{c}$S_{2}$\end{tabular}}%
\psfrag{s56}[t][t]{\color[rgb]{0,0,0}\setlength{\tabcolsep}{0pt}\begin{tabular}{c}$y_0$\end{tabular}}%
\psfrag{s68}[t][t]{\color[rgb]{0,0,0}\setlength{\tabcolsep}{0pt}\begin{tabular}{c}$\theta_s$\end{tabular}}%
\psfrag{s76}[t][t]{\color[rgb]{0,0,0}\setlength{\tabcolsep}{0pt}\begin{tabular}{c}$Y_{tot}$\end{tabular}}%
\psfrag{s80}[t][t]{\color[rgb]{0,0,0}\setlength{\tabcolsep}{0pt}\begin{tabular}{c}$S_{1}$\end{tabular}}%

\small
%
\psfrag{x01}[t][t]{0}%
\psfrag{x02}[t][t]{0.1}%
\psfrag{x03}[t][t]{0.2}%
\psfrag{x04}[t][t]{0.3}%
\psfrag{x05}[t][t]{0.4}%
\psfrag{x06}[t][t]{0.5}%
\psfrag{x07}[t][t]{0.6}%
\psfrag{x08}[t][t]{0.7}%
\psfrag{x09}[t][t]{0.8}%
\psfrag{x10}[t][t]{0.9}%
\psfrag{x11}[t][t]{1}%
\psfrag{x12}[t][t]{4.5}%
\psfrag{x13}[t][t]{5}%
\psfrag{x14}[t][t]{2}%
\psfrag{x15}[t][t]{6}%
\psfrag{x16}[t][t]{10}%
\psfrag{x17}[t][t]{0}%
\psfrag{x18}[t][t]{0.5}%
\psfrag{x19}[t][t]{1}%
\psfrag{x20}[t][t]{10}%
\psfrag{x21}[t][t]{30}%
\psfrag{x22}[t][t]{0}%
\psfrag{x23}[t][t]{0.5}%
\psfrag{x24}[t][t]{1}%
\psfrag{x25}[t][t]{0}%
\psfrag{x26}[t][t]{0.5}%
\psfrag{x27}[t][t]{1}%
\psfrag{x28}[t][t]{0}%
\psfrag{x29}[t][t]{100}%
\psfrag{x30}[t][t]{0}%
\psfrag{x31}[t][t]{0.5}%
\psfrag{x32}[t][t]{1}%
\psfrag{x33}[t][t]{0}%
\psfrag{x34}[t][t]{0.5}%
\psfrag{x35}[t][t]{1}%
\psfrag{x36}[t][t]{0}%
\psfrag{x37}[t][t]{0.5}%
\psfrag{x38}[t][t]{1}%
\psfrag{x39}[t][t]{-100}%
\psfrag{x40}[t][t]{0}%
\psfrag{x41}[t][t]{0}%
\psfrag{x42}[t][t]{0.5}%
\psfrag{x43}[t][t]{1}%
\psfrag{x44}[t][t]{0}%
\psfrag{x45}[t][t]{0.5}%
\psfrag{x46}[t][t]{1}%
\psfrag{x47}[t][t]{0}%
\psfrag{x48}[t][t]{0.5}%
\psfrag{x49}[t][t]{1}%
\psfrag{x50}[t][t]{0}%
\psfrag{x51}[t][t]{0.5}%
\psfrag{x52}[t][t]{1}%
\psfrag{x53}[t][t]{1}%
\psfrag{x54}[t][t]{1.5}%
\psfrag{x55}[t][t]{0}%
\psfrag{x56}[t][t]{0.5}%
\psfrag{x57}[t][t]{1}%
\psfrag{x58}[t][t]{0}%
\psfrag{x59}[t][t]{0.5}%
\psfrag{x60}[t][t]{1}%
\psfrag{x61}[t][t]{0}%
\psfrag{x62}[t][t]{0.5}%
\psfrag{x63}[t][t]{1}%
\psfrag{x64}[t][t]{0}%
\psfrag{x65}[t][t]{0.5}%
\psfrag{x66}[t][t]{1}%
\psfrag{x67}[t][t]{0}%
\psfrag{x68}[t][t]{0.5}%
\psfrag{x69}[t][t]{1}%
%
\psfrag{v01}[r][r]{0}%
\psfrag{v02}[r][r]{0.1}%
\psfrag{v03}[r][r]{0.2}%
\psfrag{v04}[r][r]{0.3}%
\psfrag{v05}[r][r]{0.4}%
\psfrag{v06}[r][r]{0.5}%
\psfrag{v07}[r][r]{0.6}%
\psfrag{v08}[r][r]{0.7}%
\psfrag{v09}[r][r]{0.8}%
\psfrag{v10}[r][r]{0.9}%
\psfrag{v11}[r][r]{1}%
\psfrag{v12}[r][r]{0}%
\psfrag{v13}[r][r]{0.5}%
\psfrag{v14}[r][r]{1}%
\psfrag{v15}[r][r]{0}%
\psfrag{v16}[r][r]{0.5}%
\psfrag{v17}[r][r]{1}%
\psfrag{v18}[r][r]{0}%
\psfrag{v19}[r][r]{0.5}%
\psfrag{v20}[r][r]{1}%
\psfrag{v21}[r][r]{0}%
\psfrag{v22}[r][r]{0.5}%
\psfrag{v23}[r][r]{1}%
\psfrag{v24}[r][r]{0}%
\psfrag{v25}[r][r]{0.5}%
\psfrag{v26}[r][r]{1}%
\psfrag{v27}[r][r]{0}%
\psfrag{v28}[r][r]{0.5}%
\psfrag{v29}[r][r]{1}%
\psfrag{v30}[r][r]{0}%
\psfrag{v31}[r][r]{0.5}%
\psfrag{v32}[r][r]{1}%
\psfrag{v33}[r][r]{0}%
\psfrag{v34}[r][r]{0.5}%
\psfrag{v35}[r][r]{1}%
\psfrag{v36}[r][r]{0}%
\psfrag{v37}[r][r]{0.5}%
\psfrag{v38}[r][r]{1}%
\psfrag{v39}[r][r]{0}%
\psfrag{v40}[r][r]{0.5}%
\psfrag{v41}[r][r]{1}%
\psfrag{v42}[r][r]{1}%
\psfrag{v43}[r][r]{1.5}%
\psfrag{v44}[r][r]{4.5}%
\psfrag{v45}[r][r]{5}%
\psfrag{v46}[r][r]{2}%
\psfrag{v47}[r][r]{6}%
\psfrag{v48}[r][r]{10}%
\psfrag{v49}[r][r]{10}%
\psfrag{v50}[r][r]{30}%
\psfrag{v51}[r][r]{0}%
\psfrag{v52}[r][r]{100}%
\psfrag{v53}[r][r]{0}%
\psfrag{v54}[r][r]{0.5}%
\psfrag{v55}[r][r]{1}%
\psfrag{v56}[r][r]{0}%
\psfrag{v57}[r][r]{0.5}%
\psfrag{v58}[r][r]{1}%
\psfrag{v59}[r][r]{0}%
\psfrag{v60}[r][r]{0.5}%
\psfrag{v61}[r][r]{1}%
\psfrag{v62}[r][r]{0}%
\psfrag{v63}[r][r]{0.5}%
\psfrag{v64}[r][r]{1}%
\psfrag{v65}[r][r]{0}%
\psfrag{v66}[r][r]{0.5}%
\psfrag{v67}[r][r]{1}%
\psfrag{v68}[r][r]{0}%
\psfrag{v69}[r][r]{0.5}%
\psfrag{v70}[r][r]{1}%
%
\resizebox{0.95\linewidth}{!}{\includegraphics{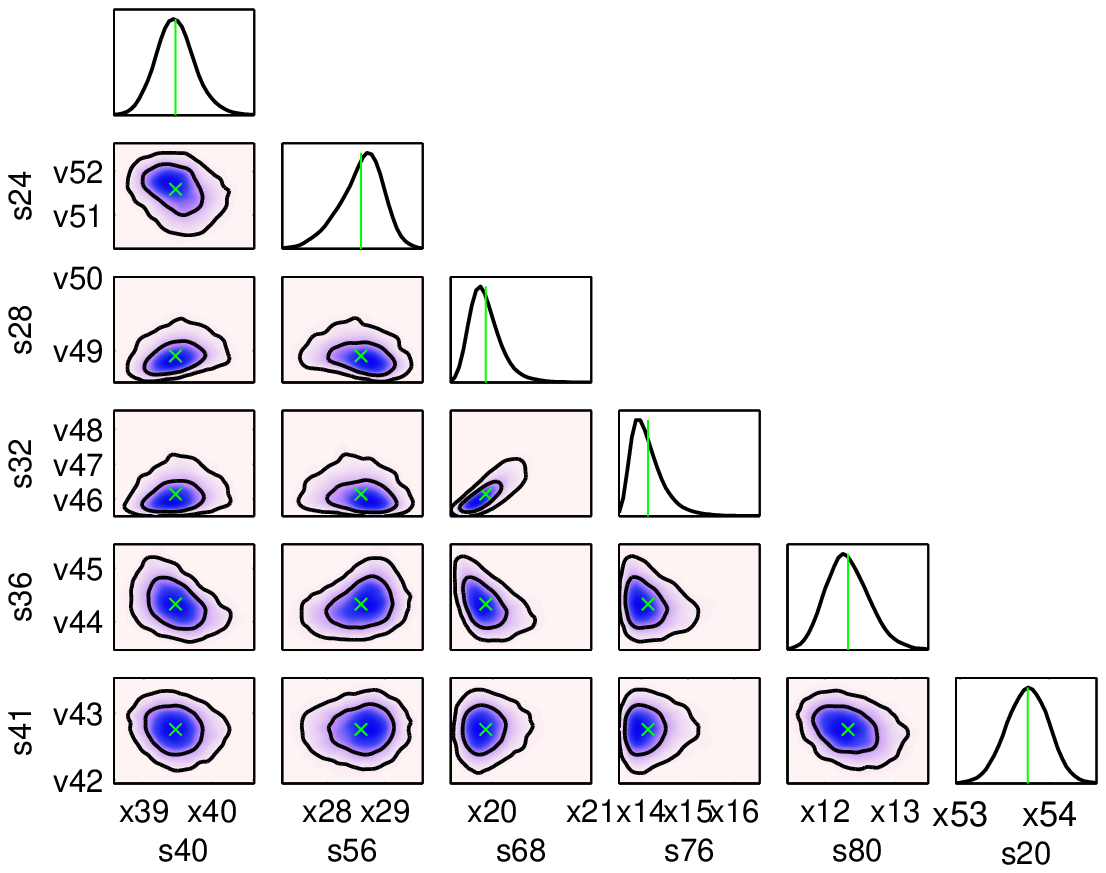}}%
\end{psfrags}%
%
}\hspace{-0.42\linewidth}
  \fbox{\raisebox{1.0\height}{\includegraphics[width=0.4\linewidth]{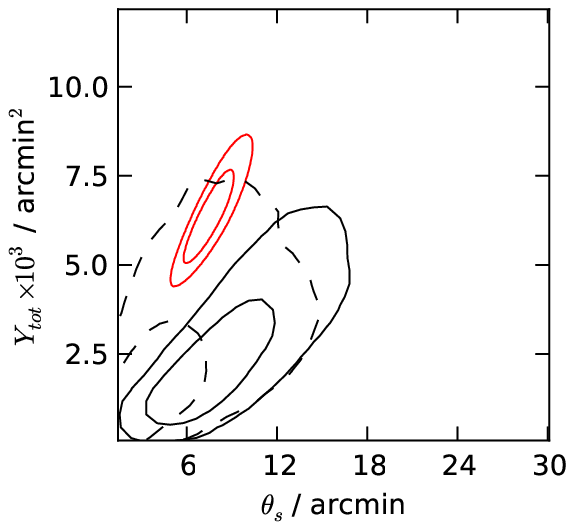}}}
  \caption{AMI posterior distributions for ZW8503 and the \Ytot-$\theta_{s}$ posterior overlaid with that obtained by \emph{Planck} (upper right hand corner).  See Section~\ref{sec:resultsreal} for more details on the plots.}\label{Fi:ZW8503_post}
\end{figure}

\begin{figure}
  \begin{center}
  \includegraphics[bb=84 125 529 670, clip=, height=0.45\linewidth, angle=270]{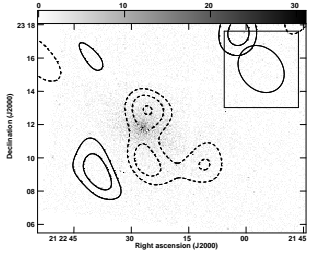}\qquad
  \caption{A \emph{Chandra} X-ray map of ZW8503\protect\footnotemark \:with AMI-SA contours at $\pm (2, 3, 4) \times 100\,\mu$Jy overlaid to show the substructure.  The grey-scale is in units of counts per pixel and is truncated at the peak value in the centre of the cluster.  The AMI synthesised beam is shown in the top right-hand corner.  Note that the axis scale is different to Fig.~\ref{Fi:ZW8503_maps}}\label{Fi:ZW8503_Xray}
  \end{center}
\end{figure}

\paragraph{ZW8503 (PSZ1~G072.78-18.70)}\mbox{}\\
ZW8503 is a well-known cluster at $z = 0.143$ \citep{1992MNRAS.259...67A} with a large angular size ($\theta_{s}$\,$\approx$\,8\,arcmin as measured by \emph{Planck}); it is therefore not too surprising that AMI does not detect it well.  A decrement at the phase centre is visible in the source-subtracted maps (Fig.~\ref{Fi:ZW8503_maps}), and a model with a cluster is favoured over one without by $\Delta \ln(\mathcal{Z}) = 1.8$, but Fig.~\ref{Fi:ZW8503_post} shows that there is not enough information in the AMI data to constrain the cluster parameters well, and the \Ytot-$\theta_{s}$ posterior distribution is strongly influenced by the prior (plotted as a black dotted line for comparison).  There is also significant degeneracy between the cluster parameters ($x_{0}, y_{0}, \theta_{s}, $\Ytot) and the flux densities of the closest sources.  The parameter space indicated by the \emph{Planck} posterior is completely ruled out by the AMI posterior distribution.  The AMI map shows a good positional coincidence with the X-ray emission (Fig.~\ref{Fi:ZW8503_Xray}) and also shows some substructure within the cluster; if this is real, the spherical cluster model with the `universal' pressure profile (derived from fits to relaxed clusters) may not provide a good fit and the extrapolated {\Ytot} result may be biased.

\footnotetext{Courtesy of the \emph{Chandra} X-ray Observatory Center and the \emph{Chandra} Data Archive, \url{http://cxc.cfa.harvard.edu/cda/} (ivo://ADS/Sa.CXO\#obs/13379)}

\subsubsection{Non-detections}

\begin{figure}
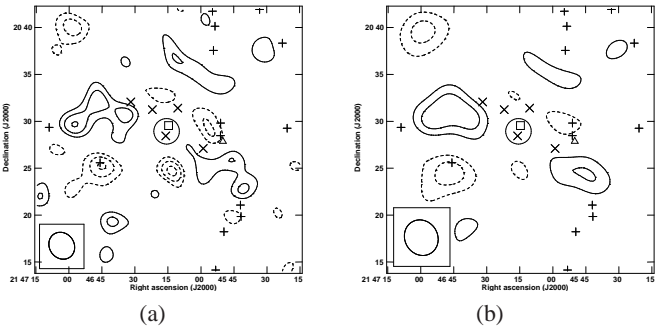

  \begin{center}
  \includegraphics[bb=48 157 570 648, clip=, width=0.45\linewidth]{CAJ2146_nw.ps}\qquad
  \includegraphics[bb=48 157 570 648, clip=, width=0.45\linewidth]{CAJ2146_uvtap.ps}
  \centerline{\hskip 0.05\linewidth (a) \hskip 0.45\linewidth (b)\phantom{$S^{\rm{X^{X^{X}}}}$}}
  \caption{SA source-subtracted map of PSZ1 G074.75-24.59 with (a) natural weighting and (b) a $uv$-taper.  The r.m.s.\ noise levels are 105 and 166\,$\mu$Jy\,beam$^{-1}$ respectively.  The position of ZwCl 2143.5+2014 is shown as a triangle \citep{1968cgcg.bookR....Z}.  See Section~\ref{sec:resultsreal} for more details on the plots.}\label{Fi:CAJ2146_maps}
  \end{center}
\end{figure}

\paragraph{PSZ1 G074.75-24.59}\mbox{}\\
\noindent PSZ1 G074.75-24.59 is associated in the \emph{Planck} catalogue with ZwCl 2143.5+2014.  Despite having an SNR of 6.1 and being detected by all three of the \emph{Planck} detection algorithms, it is not detected by AMI, with an evidence difference of $\Delta \ln (\mathcal{Z}) = -2.6$.  Although there is some negative flux visible on the map, it is ruled out by the \emph{Planck} positional prior (Fig.~\ref{Fi:CAJ2146_maps}).

\begin{figure}
  \fbox{
%
%
\begin{psfrags}%
\psfragscanon%
%
\psfrag{s18}[t][t]{\color[rgb]{0,0,0}\setlength{\tabcolsep}{0pt}\begin{tabular}{c}$Y_{\rm tot} \times 10^3$\end{tabular}}%
\psfrag{s22}[b][b]{\color[rgb]{0,0,0}\setlength{\tabcolsep}{0pt}\begin{tabular}{c}$y_0$\end{tabular}}%
\psfrag{s25}[b][b]{\color[rgb]{0,0,0}\setlength{\tabcolsep}{0pt}\begin{tabular}{c}$\theta_s$\end{tabular}}%
\psfrag{s29}[t][t]{\color[rgb]{0,0,0}\setlength{\tabcolsep}{0pt}\begin{tabular}{c}$x_0$\end{tabular}}%
\psfrag{s30}[b][b]{\color[rgb]{0,0,0}\setlength{\tabcolsep}{0pt}\begin{tabular}{c}$Y_{\rm tot}$ \\ $\times 10^3$\end{tabular}}%
\psfrag{s37}[t][t]{\color[rgb]{0,0,0}\setlength{\tabcolsep}{0pt}\begin{tabular}{c}$y_0$\end{tabular}}%
\psfrag{s41}[t][t]{\color[rgb]{0,0,0}\setlength{\tabcolsep}{0pt}\begin{tabular}{c}$\theta_s$\end{tabular}}%
%
\psfrag{x01}[t][t]{0}%
\psfrag{x02}[t][t]{0.1}%
\psfrag{x03}[t][t]{0.2}%
\psfrag{x04}[t][t]{0.3}%
\psfrag{x05}[t][t]{0.4}%
\psfrag{x06}[t][t]{0.5}%
\psfrag{x07}[t][t]{0.6}%
\psfrag{x08}[t][t]{0.7}%
\psfrag{x09}[t][t]{0.8}%
\psfrag{x10}[t][t]{0.9}%
\psfrag{x11}[t][t]{1}%
\psfrag{x12}[t][t]{20}%
\psfrag{x13}[t][t]{40}%
\psfrag{x14}[t][t]{60}%
\psfrag{x15}[t][t]{-150}%
\psfrag{x16}[t][t]{0}%
\psfrag{x17}[t][t]{150}%
\psfrag{x18}[t][t]{-100}%
\psfrag{x19}[t][t]{0}%
\psfrag{x20}[t][t]{100}%
\psfrag{x21}[t][t]{200}%
\psfrag{x22}[t][t]{-100}%
\psfrag{x23}[t][t]{0}%
\psfrag{x24}[t][t]{100}%
\psfrag{x25}[t][t]{-100}%
\psfrag{x26}[t][t]{0}%
\psfrag{x27}[t][t]{100}%
\psfrag{x28}[t][t]{-100}%
\psfrag{x29}[t][t]{0}%
\psfrag{x30}[t][t]{100}%
\psfrag{x31}[t][t]{0}%
\psfrag{x32}[t][t]{5}%
\psfrag{x33}[t][t]{10}%
\psfrag{x34}[t][t]{20}%
\psfrag{x35}[t][t]{40}%
\psfrag{x36}[t][t]{60}%
\psfrag{x37}[t][t]{-100}%
\psfrag{x38}[t][t]{0}%
\psfrag{x39}[t][t]{100}%
\psfrag{x40}[t][t]{200}%
\psfrag{x41}[t][t]{-100}%
\psfrag{x42}[t][t]{0}%
\psfrag{x43}[t][t]{100}%
%
\psfrag{v01}[r][r]{0}%
\psfrag{v02}[r][r]{0.1}%
\psfrag{v03}[r][r]{0.2}%
\psfrag{v04}[r][r]{0.3}%
\psfrag{v05}[r][r]{0.4}%
\psfrag{v06}[r][r]{0.5}%
\psfrag{v07}[r][r]{0.6}%
\psfrag{v08}[r][r]{0.7}%
\psfrag{v09}[r][r]{0.8}%
\psfrag{v10}[r][r]{0.9}%
\psfrag{v11}[r][r]{1}%
\psfrag{v12}[r][r]{0}%
\psfrag{v13}[r][r]{5}%
\psfrag{v14}[r][r]{10}%
\psfrag{v15}[r][r]{0}%
\psfrag{v16}[r][r]{5}%
\psfrag{v17}[r][r]{10}%
\psfrag{v18}[r][r]{20}%
\psfrag{v19}[r][r]{40}%
\psfrag{v20}[r][r]{60}%
\psfrag{v21}[r][r]{0}%
\psfrag{v22}[r][r]{5}%
\psfrag{v23}[r][r]{10}%
\psfrag{v24}[r][r]{20}%
\psfrag{v25}[r][r]{40}%
\psfrag{v26}[r][r]{60}%
\psfrag{v27}[r][r]{-150}%
\psfrag{v28}[r][r]{0}%
\psfrag{v29}[r][r]{150}%
\psfrag{v30}[r][r]{0}%
\psfrag{v31}[r][r]{0.5}%
\psfrag{v32}[r][r]{1}%
\psfrag{v33}[r][r]{0}%
\psfrag{v34}[r][r]{0.5}%
\psfrag{v35}[r][r]{1}%
\psfrag{v36}[r][r]{0}%
\psfrag{v37}[r][r]{0.5}%
\psfrag{v38}[r][r]{1}%
\psfrag{v39}[r][r]{0}%
\psfrag{v40}[r][r]{0.5}%
\psfrag{v41}[r][r]{1}%
%
\resizebox{0.95\linewidth}{!}{\includegraphics{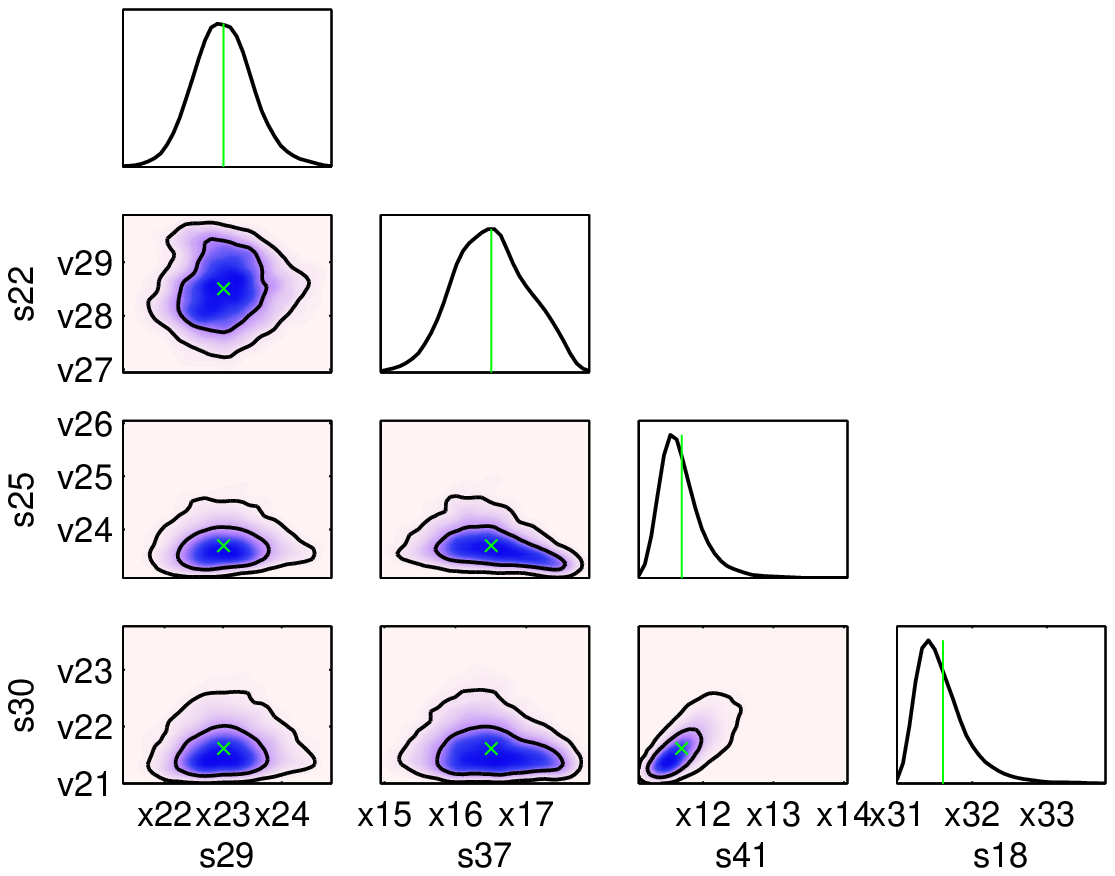}}%
\end{psfrags}%
%
}\hspace{-0.42\linewidth}
  \fbox{\raisebox{\height}{\includegraphics[width=0.4\linewidth]{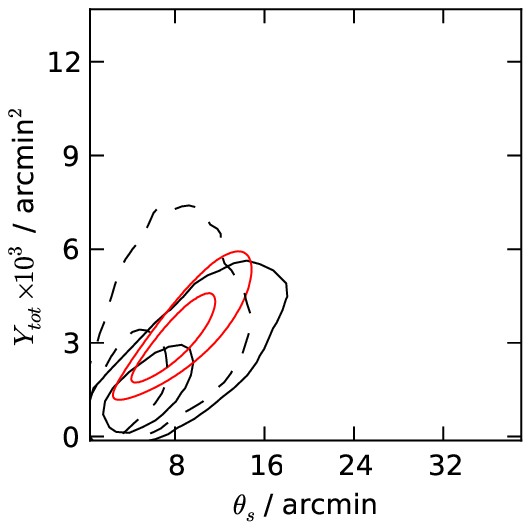}}}
  \caption{AMI posterior distributions for PSZ1 G074.75-24.59 and the \Ytot-$\theta_{s}$ posterior overlaid with that obtained by \emph{Planck} (upper right hand corner).  See Section~\ref{sec:resultsreal} for more details on the plots.}\label{Fi:CAJ2146_post}
\end{figure}

A simulated cluster using the \textsc{PwS} maximum a-posteriori values for $\theta_{s}$ and \Ytot, `observed' using the same visibilities and noise levels as those in the real AMI observation, shows that this cluster should be detected at a SNR of $\approx$\,8 in the naturally-weighted map, and $\approx$\,9 in the $uv$-tapered map.  However, the posterior distributions (Fig.~\ref{Fi:CAJ2146_post}) show that the $\theta_s$/{\Ytot} parameter space preferred by \emph{Planck} cannot be ruled out by the AMI observations, so the cluster could be more extended than the \emph{Planck} MAP estimate shows (although the redshift is given as 0.250 so this seems unlikely) and/or be significantly offset from its given position.

\subsubsection{Clear non-detections}\label{sec:clear_non}

\begin{figure}
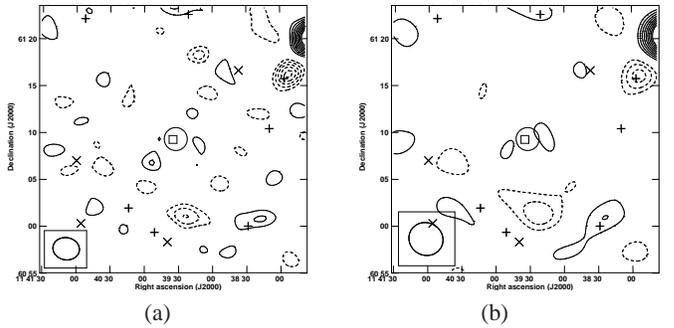

  \begin{center}
  \includegraphics[bb=48 157 570 648, clip=, width=0.45\linewidth]{CAJ1139_nw.ps}\qquad
  \includegraphics[bb=48 157 570 648, clip=, width=0.45\linewidth]{CAJ1139_uvtap.ps}
  \centerline{\hskip 0.05\linewidth (a) \hskip 0.45\linewidth (b)\phantom{$S^{\rm{X^{X^{X}}}}$}}
  \caption{SA source-subtracted map of PSZ1 G137.56+53.88 with (a) natural weighting and (b) a $uv$-taper.  The r.m.s.\ noise levels are 109 and 150\,$\mu$Jy\,beam$^{-1}$ respectively.  See Section~\ref{sec:resultsreal} for more details on the plots.}\label{Fi:CAJ1139_maps}
  \end{center}
\end{figure}

\begin{figure}
  \fbox{
%
%
\begin{psfrags}%
\psfragscanon%
%
\psfrag{s17}[t][t]{\color[rgb]{0,0,0}\setlength{\tabcolsep}{0pt}\begin{tabular}{c}$Y_{\rm tot} \times 10^3$\end{tabular}}%
\psfrag{s21}[t][t]{\color[rgb]{0,0,0}\setlength{\tabcolsep}{0pt}\begin{tabular}{c}$y_0$\end{tabular}}%
\psfrag{s25}[t][t]{\color[rgb]{0,0,0}\setlength{\tabcolsep}{0pt}\begin{tabular}{c}$\theta_s$\end{tabular}}%
\psfrag{s29}[t][t]{\color[rgb]{0,0,0}\setlength{\tabcolsep}{0pt}\begin{tabular}{c}$x_0$\end{tabular}}%
\psfrag{s30}[t][t]{\color[rgb]{0,0,0}\setlength{\tabcolsep}{0pt}\begin{tabular}{c}$Y_{\rm tot}$\\$\times 10^3$\end{tabular}}%
\psfrag{s37}[t][t]{\color[rgb]{0,0,0}\setlength{\tabcolsep}{0pt}\begin{tabular}{c}$y_0$\end{tabular}}%
\psfrag{s41}[t][t]{\color[rgb]{0,0,0}\setlength{\tabcolsep}{0pt}\begin{tabular}{c}$\theta_s$\end{tabular}}%
%
\psfrag{x01}[t][t]{0}%
\psfrag{x02}[t][t]{0.1}%
\psfrag{x03}[t][t]{0.2}%
\psfrag{x04}[t][t]{0.3}%
\psfrag{x05}[t][t]{0.4}%
\psfrag{x06}[t][t]{0.5}%
\psfrag{x07}[t][t]{0.6}%
\psfrag{x08}[t][t]{0.7}%
\psfrag{x09}[t][t]{0.8}%
\psfrag{x10}[t][t]{0.9}%
\psfrag{x11}[t][t]{1}%
\psfrag{x12}[t][t]{20}%
\psfrag{x13}[t][t]{40}%
\psfrag{x14}[t][t]{60}%
\psfrag{x15}[t][t]{-200}%
\psfrag{x16}[t][t]{0}%
\psfrag{x17}[t][t]{200}%
\psfrag{x18}[t][t]{-200}%
\psfrag{x19}[t][t]{0}%
\psfrag{x20}[t][t]{200}%
\psfrag{x21}[t][t]{-200}%
\psfrag{x22}[t][t]{0}%
\psfrag{x23}[t][t]{200}%
\psfrag{x24}[t][t]{-200}%
\psfrag{x25}[t][t]{0}%
\psfrag{x26}[t][t]{200}%
\psfrag{x27}[t][t]{-200}%
\psfrag{x28}[t][t]{0}%
\psfrag{x29}[t][t]{200}%
\psfrag{x30}[t][t]{0}%
\psfrag{x31}[t][t]{10}%
\psfrag{x32}[t][t]{20}%
\psfrag{x33}[t][t]{40}%
\psfrag{x34}[t][t]{60}%
\psfrag{x35}[t][t]{-200}%
\psfrag{x36}[t][t]{0}%
\psfrag{x37}[t][t]{200}%
\psfrag{x38}[t][t]{-200}%
\psfrag{x39}[t][t]{0}%
\psfrag{x40}[t][t]{200}%
%
\psfrag{v01}[r][r]{0}%
\psfrag{v02}[r][r]{0.1}%
\psfrag{v03}[r][r]{0.2}%
\psfrag{v04}[r][r]{0.3}%
\psfrag{v05}[r][r]{0.4}%
\psfrag{v06}[r][r]{0.5}%
\psfrag{v07}[r][r]{0.6}%
\psfrag{v08}[r][r]{0.7}%
\psfrag{v09}[r][r]{0.8}%
\psfrag{v10}[r][r]{0.9}%
\psfrag{v11}[r][r]{1}%
\psfrag{v12}[r][r]{0}%
\psfrag{v13}[r][r]{5}%
\psfrag{v14}[r][r]{10}%
\psfrag{v15}[r][r]{15}%
\psfrag{v16}[r][r]{0}%
\psfrag{v17}[r][r]{5}%
\psfrag{v18}[r][r]{10}%
\psfrag{v19}[r][r]{15}%
\psfrag{v20}[r][r]{20}%
\psfrag{v21}[r][r]{40}%
\psfrag{v22}[r][r]{60}%
\psfrag{v23}[r][r]{0}%
\psfrag{v24}[r][r]{10}%
\psfrag{v25}[r][r]{20}%
\psfrag{v26}[r][r]{40}%
\psfrag{v27}[r][r]{60}%
\psfrag{v28}[r][r]{-200}%
\psfrag{v29}[r][r]{0}%
\psfrag{v30}[r][r]{200}%
\psfrag{v31}[r][r]{0}%
\psfrag{v32}[r][r]{0.5}%
\psfrag{v33}[r][r]{1}%
\psfrag{v34}[r][r]{0}%
\psfrag{v35}[r][r]{0.5}%
\psfrag{v36}[r][r]{1}%
\psfrag{v37}[r][r]{0}%
\psfrag{v38}[r][r]{0.5}%
\psfrag{v39}[r][r]{1}%
\psfrag{v40}[r][r]{0}%
\psfrag{v41}[r][r]{0.5}%
\psfrag{v42}[r][r]{1}%
%
\resizebox{0.95\linewidth}{!}{\includegraphics{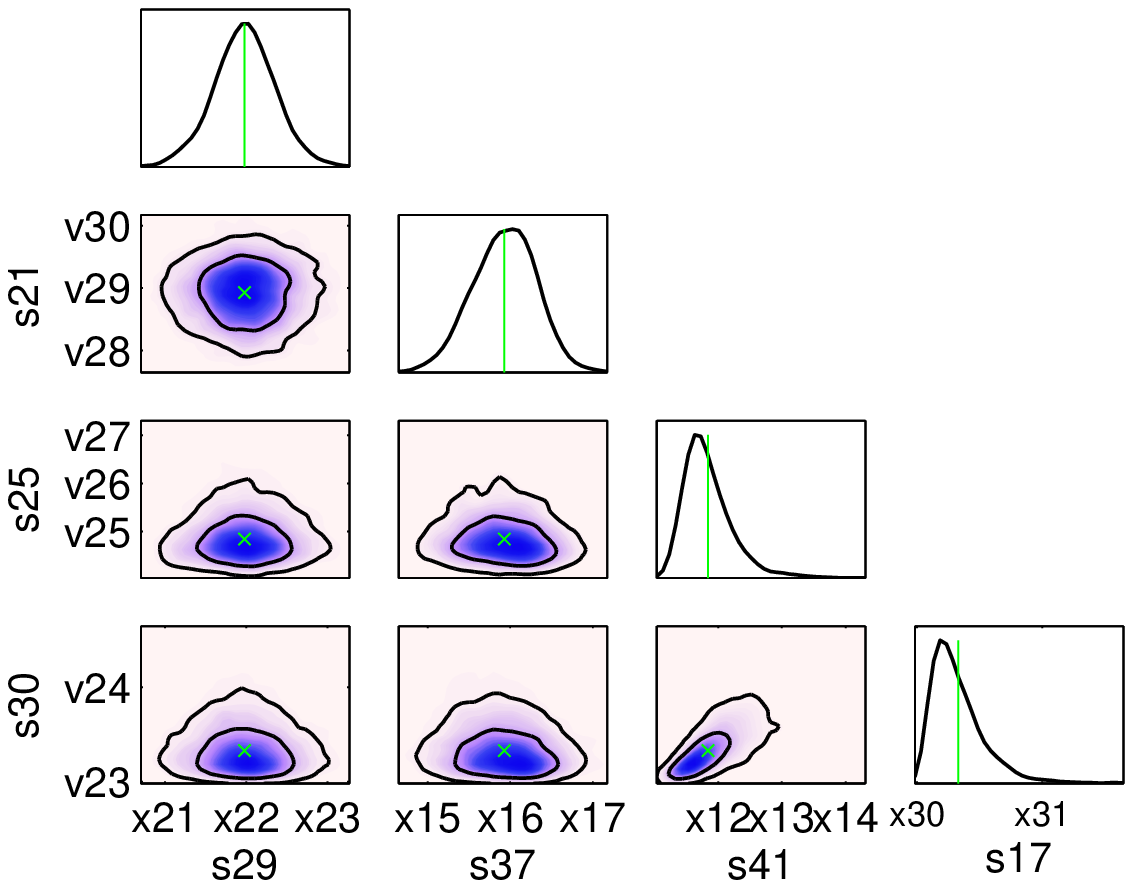}}%
\end{psfrags}%
%
}\hspace{-0.42\linewidth}
  \fbox{\raisebox{\height}{\includegraphics[width=0.4\linewidth]{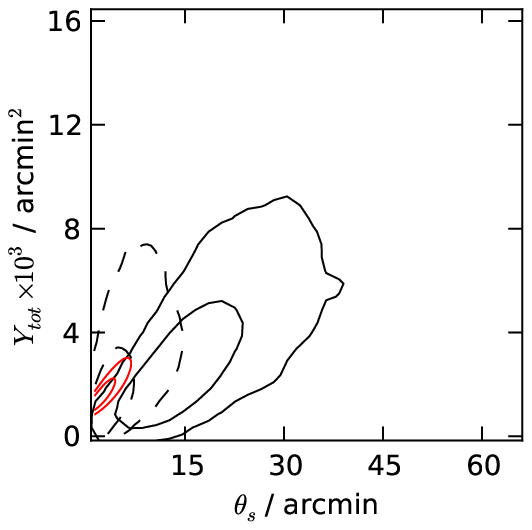}}}
  \caption{AMI posterior distributions for PSZ1 G137.56+53.88  and the \Ytot-$\theta_{s}$ posterior overlaid with that obtained by \emph{Planck} (upper right hand corner).  See Section~\ref{sec:resultsreal} for more details on the plots.}\label{Fi:CAJ1139_post}
\end{figure}

PSZ1~G137.56+53.88 is a clear non-detection with evidence ratio $\Delta \ln(\mathcal{Z}) = -4.1$.  There is no negative flux near the phase centre and no nearby point sources or positive extended emission to cause the non-detection of the cluster (Fig.~\ref{Fi:CAJ1139_maps}).  Simulations show the cluster should have a significance of $\approx$\,17 in both the naturally-weighted and $uv$-tapered maps.  The posterior distribution (Fig.~\ref{Fi:CAJ1139_post}) shows that very large values of $\theta_{s}$ are required to provide any kind of consistency with the data, so that nearly all of the cluster flux would be resolved out, in disagreement with the small value for $\theta_{s}$ indicated by PwS.  Noting also that although the cluster has an SNR of 5.7, it was detected by PwS only and not the other algorithms, we consider it likely to be a spurious detection.  

\subsubsection{Validation}

\paragraph{Detection of new clusters}\mbox{}\\

\noindent Of our SZ sample, 82 clusters are previously known (the `validation' flag in the \emph{Planck} catalogue is 20).  16 of the new clusters are already confirmed by other followup (`validation' = 10); of these, we re-confirm 14.

We detect 14 of the remaining 25 new clusters that have not been previously confirmed by other methods, at the time of publishing of the catalogue.  All of these are detected by at least two \emph{Planck} pipelines, and 8 are detected by all three.  For these clusters, the \emph{Planck} catalogue provides a quality assessment flag between 1 and 3 (1 being the most reliable); there are 6, 4 and 4 AMI detections in the 1, 2 and 3 categories respectively.

\paragraph{Discussion of AMI non-detections}\mbox{}\\

\noindent Across the whole sample, 75\% of the AMI non- and clear non-detections have less than three \emph{Planck} pipeline detections, compared to 18\% for the AMI clear and moderate detections; of the previously unconfirmed clusters, none of the AMI non- and clear non-detections has a quality flag value of 1.  Although it is difficult to rule out the presence of a cluster entirely using AMI data alone, these correlations indicate that an AMI non-detection is a useful indicator for a possible spurious \emph{Planck} detection.  Fig.~\ref{Fi:all_post2} shows $\theta_{s}$-\Ytot \:posteriors for all of the non-detections; the \emph{Planck} parameter space is often ruled out by the AMI posterior.

All of the three clear non-detections have $<$3 \emph{Planck} pipeline detections.  Two of these (PSZ1~G053.50+09.56 and PSZ1~G142.17+37.28) are within 5\,arcmin of thermal, compact sources at 545 and/or 857\,GHz, which are another indicator of a potentially spurious \emph{Planck} detection caused by contamination by dust emission.  The third has been addressed in Section~\ref{sec:clear_non}; we consider these three likely to be spurious.

The \emph{Planck} catalogue produced by the intersection of detections by the three algorithms is expected to be $\approx$\,99\% pure at SNR $\ge 5$ \citep{2013arXiv1303.5089P}.  Our SZ sample of 123 clusters contains 87 in the `intersection' catalogue, of which 81 are detected by AMI.  This leaves six non-detections.  Of these, three (PSZ1~G099.48+55.62, PSZ1~G107.32-31.51, and PSZ1~G084.84+35.04) are at known, low redshift and the posteriors in Fig.~\ref{Fi:all_post2} show that the region of $\theta_{s}$-\Ytot\ parameter space preferred by \emph{Planck} cannot be ruled out by the AMI observations; i.e.\ these clusters are likely to be too large in angular size (and not bright enough) to be seen by AMI.  Of the remaining four, PSZ1~G094.69+26.34 is predicted to have a low SNR of $\approx$\,4 in the AMI data based on the \emph{Planck} maximum a-posteriori values of $\theta_{s}$ and \Ytot, and could also be resolved out if the true values are toward the upper edge of the constraint.  Also, although PSZ1~G050.46+67.54 should be well-detected according to its \emph{Planck} size estimate of $\theta_{s}$\,$\approx$\,3\,arcmin, it is within 220\,arcsec of an MCXC cluster with size $\theta_{500} = 6.89$\,arcmin \citep{2011A&A...534A.109P}, corresponding to $\theta_{s} = 5.85$\,arcmin for $c_{500} = 1.177$ and may therefore also be resolved out if the \emph{Planck} size is an under-estimate.

This leaves one cluster only in the `intersection' catalogue, PSZ1~G074.75-24.59, which simulations based on the \emph{Planck} maximum a-posteriori parameter estimates predict should be well-detected by AMI; the AMI maps (Fig.~\ref{Fi:CAJ2146_maps}) show no source environment problems which could explain its non-detection.  More follow-up data will be required to definitively determine if this is a spurious detection, as the pressure profile of the cluster gas could deviate significantly from the `universal' pressure profile and/or the \emph{Planck} position estimates could be offset significantly from the true position, so that the simulations do not accurately predict the AMI detection significances.

\subsubsection{Positional comparison}\label{S:pos_comp}

\begin{figure*}
  \centerline{
%
%
\begin{psfrags}%
\psfragscanon%
%
\psfrag{s02}[b][b]{\color[rgb]{0,0,0}\setlength{\tabcolsep}{0pt}\begin{tabular}{c}$\Delta \delta$ / arcsec\end{tabular}}%
\psfrag{s05}[l][l]{\color[rgb]{0,0,0}\setlength{\tabcolsep}{0pt}\begin{tabular}{l}PwS\end{tabular}}%
\psfrag{s10}[l][l]{\color[rgb]{0,0,0}\setlength{\tabcolsep}{0pt}\begin{tabular}{l}MMF1\end{tabular}}%
\psfrag{s11}[t][t]{\color[rgb]{0,0,0}\setlength{\tabcolsep}{0pt}\begin{tabular}{c}$\Delta$ RA / arcsec\end{tabular}}%
\psfrag{s15}[l][l]{\color[rgb]{0,0,0}\setlength{\tabcolsep}{0pt}\begin{tabular}{l}MMF3\end{tabular}}%
%
\psfrag{x01}[t][t]{0}%
\psfrag{x02}[t][t]{0.1}%
\psfrag{x03}[t][t]{0.2}%
\psfrag{x04}[t][t]{0.3}%
\psfrag{x05}[t][t]{0.4}%
\psfrag{x06}[t][t]{0.5}%
\psfrag{x07}[t][t]{0.6}%
\psfrag{x08}[t][t]{0.7}%
\psfrag{x09}[t][t]{0.8}%
\psfrag{x10}[t][t]{0.9}%
\psfrag{x11}[t][t]{1}%
\psfrag{x12}[t][t]{-200}%
\psfrag{x13}[t][t]{0}%
\psfrag{x14}[t][t]{200}%
\psfrag{x15}[t][t]{-200}%
\psfrag{x16}[t][t]{0}%
\psfrag{x17}[t][t]{200}%
\psfrag{x18}[t][t]{-200}%
\psfrag{x19}[t][t]{0}%
\psfrag{x20}[t][t]{200}%
%
\psfrag{v01}[r][r]{0}%
\psfrag{v02}[r][r]{0.1}%
\psfrag{v03}[r][r]{0.2}%
\psfrag{v04}[r][r]{0.3}%
\psfrag{v05}[r][r]{0.4}%
\psfrag{v06}[r][r]{0.5}%
\psfrag{v07}[r][r]{0.6}%
\psfrag{v08}[r][r]{0.7}%
\psfrag{v09}[r][r]{0.8}%
\psfrag{v10}[r][r]{0.9}%
\psfrag{v11}[r][r]{1}%
\psfrag{v12}[r][r]{-200}%
\psfrag{v13}[r][r]{0}%
\psfrag{v14}[r][r]{200}%
\psfrag{v15}[r][r]{-200}%
\psfrag{v16}[r][r]{0}%
\psfrag{v17}[r][r]{200}%
\psfrag{v18}[r][r]{-200}%
\psfrag{v19}[r][r]{0}%
\psfrag{v20}[r][r]{200}%
%
\resizebox{14.5cm}{!}{\includegraphics{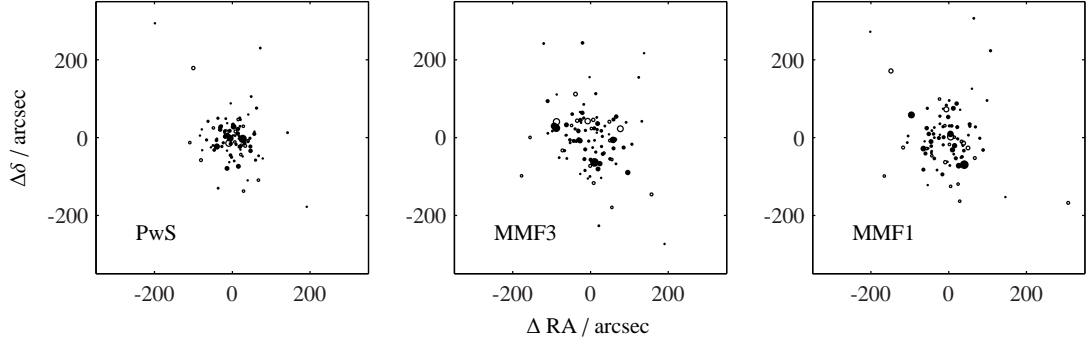}}%
\end{psfrags}%
%
}
  \bigskip
  \caption{Positional offset from AMI for the three \emph{Planck} detection algorithms.  The size of the points plotted increases with increasing \emph{Planck} SNR; clear detections are plotted as filled circles, and moderate detections as empty circles.}\label{Fi:pos_comp}
\end{figure*}

The higher angular resolution of AMI enables a more accurate positional estimate to be produced for the clusters (although in practice this depends on a variety of factors such as signal-to-noise over the angular scales observed by both telescopes, and how successful the decoupling of the signal from the foregrounds is).  This allows the accuracy of the \emph{Planck} positions and error estimates to be checked.  Fig.~\ref{Fi:pos_comp} compares positional offsets between AMI and the three \emph{Planck} detection algorithms.  The offsets for MMF1 and MMF3 are very similar.  The \textsc{PwS} offsets are slightly more clustered toward zero, and also show a greater correlation with the SNR (i.e. the highest SNR points are closer to zero than the low-SNR points).

\begin{figure}
  \centerline{
%
%
\begin{psfrags}%
\psfragscanon%
%
\psfrag{s06}[b][b]{\color[rgb]{0,0,0}\setlength{\tabcolsep}{0pt}\begin{tabular}{c}Probability density\end{tabular}}%
\psfrag{s07}[t][t]{\color[rgb]{0,0,0}\setlength{\tabcolsep}{0pt}\begin{tabular}{c}(AMI-PwS separation) / $\sigma_{\rm tot}$\end{tabular}}%
\psfrag{s08}[t][t]{\color[rgb]{0,0,0}\setlength{\tabcolsep}{0pt}\begin{tabular}{c}(AMI-MMF3 separation) / $\sigma_{\rm tot}$\end{tabular}}%
%
\psfrag{x01}[t][t]{0}%
\psfrag{x02}[t][t]{1}%
\psfrag{x03}[t][t]{2}%
\psfrag{x04}[t][t]{3}%
\psfrag{x05}[t][t]{4}%
\psfrag{x06}[t][t]{5}%
\psfrag{x07}[t][t]{0}%
\psfrag{x08}[t][t]{1}%
\psfrag{x09}[t][t]{2}%
\psfrag{x10}[t][t]{3}%
\psfrag{x11}[t][t]{4}%
\psfrag{x12}[t][t]{5}%
%
\psfrag{v01}[r][r]{0}%
\psfrag{v02}[r][r]{0.5}%
\psfrag{v03}[r][r]{1}%
\psfrag{v04}[r][r]{1.5}%
\psfrag{v05}[r][r]{2}%
\psfrag{v06}[r][r]{2.5}%
\psfrag{v07}[r][r]{0}%
\psfrag{v08}[r][r]{0.2}%
\psfrag{v09}[r][r]{0.4}%
\psfrag{v10}[r][r]{0.6}%
\psfrag{v11}[r][r]{0.8}%
\psfrag{v12}[r][r]{1}%
\psfrag{v13}[r][r]{1.2}%
%
\resizebox{8.9cm}{!}{\includegraphics{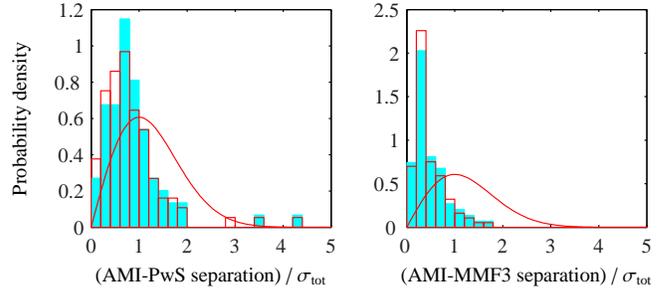}}%
\end{psfrags}%
%
}
  \bigskip
  \caption{Positional offset from AMI, normalised by total error $\sigma_{\rm tot} = \sqrt{\sigma_{\rm AMI}^{2}+\sigma_{Planck}^{2}}$, for PwS and MMF3.  The solid histogram shows the clear detections only, and the red outline shows clear and moderate detections together.  A Rayleigh distribution is plotted in red for comparison.}\label{Fi:pos_comp2}
\end{figure}

The MMF1 algorithm does not currently output positional errors, so Fig.~\ref{Fi:pos_comp2} shows the distribution of positional offsets normalised by the total error $\left ( \sqrt{\sigma_{\rm AMI}^{2}+\sigma_{Planck}^{2}} \right )$ for \textsc{PwS} and MMF3 only.  A Rayleigh distribution, $(x/\sigma^{2}) \exp(-x^2/2\sigma^2)$ with $\sigma = 1$, is plotted for comparison -- this is the expected distribution assuming the errors in RA and $\delta$ are uncorrelated and normally distributed.  The \textsc{PwS} distribution is a reasonable match, showing that the error estimates are a good representation of the true uncertainty in the positions.  In contrast, the MMF3 errors are generally overestimated in this version of the \emph{Planck} catalogue.

\begin{figure}
  \centerline{
%
%
\begin{psfrags}%
\psfragscanon%
%
\psfrag{s01}[t][t]{\color[rgb]{0,0,0}\setlength{\tabcolsep}{0pt}\begin{tabular}{c}(AMI-MMF3 separation) / $\sigma_{\rm tot, rescaled}$\end{tabular}}%
\psfrag{s02}[b][b]{\color[rgb]{0,0,0}\setlength{\tabcolsep}{0pt}\begin{tabular}{c}Probability density\end{tabular}}%
\psfrag{s05}[t][t]{\color[rgb]{0,0,0}\setlength{\tabcolsep}{0pt}\begin{tabular}{c}PwS (compatibility) SNR\end{tabular}}%
\psfrag{s06}[t][t]{\color[rgb]{0,0,0}\setlength{\tabcolsep}{0pt}\begin{tabular}{c}$\Delta_{\rm MMF3} / \Delta_{\rm PwS}$\end{tabular}}%
%
\psfrag{x01}[t][t]{0}%
\psfrag{x02}[t][t]{0.1}%
\psfrag{x03}[t][t]{0.2}%
\psfrag{x04}[t][t]{0.3}%
\psfrag{x05}[t][t]{0.4}%
\psfrag{x06}[t][t]{0.5}%
\psfrag{x07}[t][t]{0.6}%
\psfrag{x08}[t][t]{0.7}%
\psfrag{x09}[t][t]{0.8}%
\psfrag{x10}[t][t]{0.9}%
\psfrag{x11}[t][t]{1}%
\psfrag{x12}[t][t]{4}%
\psfrag{x13}[t][t]{8}%
\psfrag{x14}[t][t]{12}%
\psfrag{x15}[t][t]{16}%
\psfrag{x16}[t][t]{0}%
\psfrag{x17}[t][t]{1}%
\psfrag{x18}[t][t]{2}%
\psfrag{x19}[t][t]{3}%
\psfrag{x20}[t][t]{4}%
\psfrag{x21}[t][t]{5}%
%
\psfrag{v01}[r][r]{0}%
\psfrag{v02}[r][r]{0.1}%
\psfrag{v03}[r][r]{0.2}%
\psfrag{v04}[r][r]{0.3}%
\psfrag{v05}[r][r]{0.4}%
\psfrag{v06}[r][r]{0.5}%
\psfrag{v07}[r][r]{0.6}%
\psfrag{v08}[r][r]{0.7}%
\psfrag{v09}[r][r]{0.8}%
\psfrag{v10}[r][r]{0.9}%
\psfrag{v11}[r][r]{1}%
\psfrag{v12}[r][r]{$10^{0}$}%
\psfrag{v13}[r][r]{$10^{1}$}%
\psfrag{v14}[r][r]{0}%
\psfrag{v15}[r][r]{0.2}%
\psfrag{v16}[r][r]{0.4}%
\psfrag{v17}[r][r]{0.6}%
\psfrag{v18}[r][r]{0.8}%
\psfrag{v19}[r][r]{1}%
%
\resizebox{8.9cm}{!}{\includegraphics{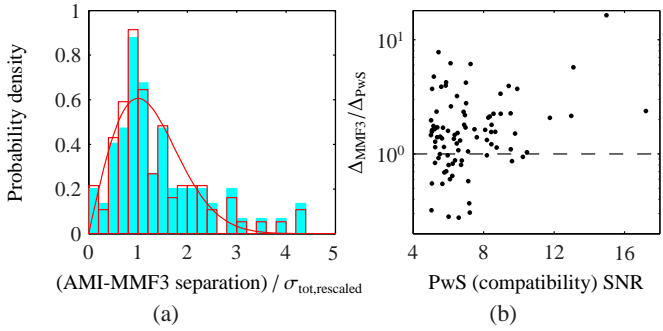}}%
\end{psfrags}%
%
}
  \bigskip
  \centerline{(a) \hskip 0.45\linewidth (b)}
  \bigskip
  \caption{(a) shows the MMF3 positional offset from AMI, normalised by rescaled total error $\sigma_{\rm tot} = \sqrt{\sigma_{\rm AMI}^{2}+(0.28 \times \sigma_{\rm MMF3})^{2}}$.  The solid histogram shows the clear detections only, and the red outline shows clear and moderate detections together.  A Rayleigh distribution is plotted in red for comparison. (b) shows the ratio between the absolute offsets ($\Delta$) between AMI and MMF3 and AMI and PwS as a function of SNR; as shown in Fig.~\ref{Fi:pos_comp}, PwS does better at high SNR.}\label{Fi:pos_comp3}
\end{figure}

We estimate a rescaling factor of 0.28 for the MMF3 errors, by minimising the Kolmogorov-Smirnov test statistic between the distribution and the Rayleigh distribution.  Fig.~\ref{Fi:pos_comp3} shows the rescaled histogram, which agrees much more closely with the Rayleigh distribution.  In contrast, the same procedure gives a rescaling factor of 0.51 for the PwS errors.  Fig.~\ref{Fi:pos_comp3} also shows a comparison between the absolute offsets between AMI and PwS and AMI and MMF3; confirming what is seen in Fig.~\ref{Fi:pos_comp}, the PwS offsets are generally smaller, especially at high SNR.

The MMF3 rescaling factor is in agreement with that estimated via internal \emph{Planck} quality assessment, and later versions of the catalogue have been corrected for this; see \url{http://wiki.cosmos.esa.int/planckpla/index.php/Catalogues\#The\_SZ\_catalogues} under `Caveats'.

\subsubsection{\Ytot-$\theta_{s}$ comparison}\label{S:Ytot_theta_comp}

A major conclusion of AP2013 was that the clusters were found overall to be smaller in angular size and fainter (lower \Ytot) by AMI than by \emph{Planck}.  The comparison for the larger sample shows a similar trend.  

To properly compare the quantities, it is necessary to look at the full, two-dimensional posteriors for \Ytot \:and $\theta_{s}$ since the quantities are correlated.  Fig.~\ref{Fi:all_post} shows the two-dimensional posteriors for $\theta_{s}$ and \Ytot \:as measured by both AMI and \emph{Planck}, and the joint constraints where appropriate, in descending \emph{Planck} SNR order.  It is clear that, especially at the high-SNR end, there are many cases where the constraints are inconsistent and in these cases the \emph{Planck} posteriors usually prefer higher values of $\theta_{s}$ and \Ytot.

\begin{figure*}
  \centerline{
%
%
\begin{psfrags}%
\psfragscanon%
%
\psfrag{s05}[t][t]{\color[rgb]{0,0,0}\setlength{\tabcolsep}{0pt}\begin{tabular}{c}$\theta_{s, \mathrm{AMI}}$ / arcmin\end{tabular}}%
\psfrag{s06}[b][b]{\color[rgb]{0,0,0}\setlength{\tabcolsep}{0pt}\begin{tabular}{c}$\theta_{s, \mathrm{PwS}}$ / arcmin\end{tabular}}%
\psfrag{s07}[t][t]{\color[rgb]{0,0,0}\setlength{\tabcolsep}{0pt}\begin{tabular}{c}$Y_{\rm tot, AMI}$ / arcmin$^2$\end{tabular}}%
\psfrag{s08}[b][b]{\color[rgb]{0,0,0}\setlength{\tabcolsep}{0pt}\begin{tabular}{c}$Y_{\rm tot, PwS}$ / arcmin$^2$\end{tabular}}%
%
\psfrag{x01}[t][t]{$10^{-3}$}%
\psfrag{x02}[t][t]{$10^{-2}$}%
\psfrag{x03}[t][t]{5}%
\psfrag{x04}[t][t]{10}%
\psfrag{x05}[t][t]{15}%
%
\psfrag{v01}[r][r]{$10^{-3}$}%
\psfrag{v02}[r][r]{$10^{-2}$}%
\psfrag{v03}[r][r]{5}%
\psfrag{v04}[r][r]{10}%
\psfrag{v05}[r][r]{15}%
%
\resizebox{17.8cm}{!}{\includegraphics{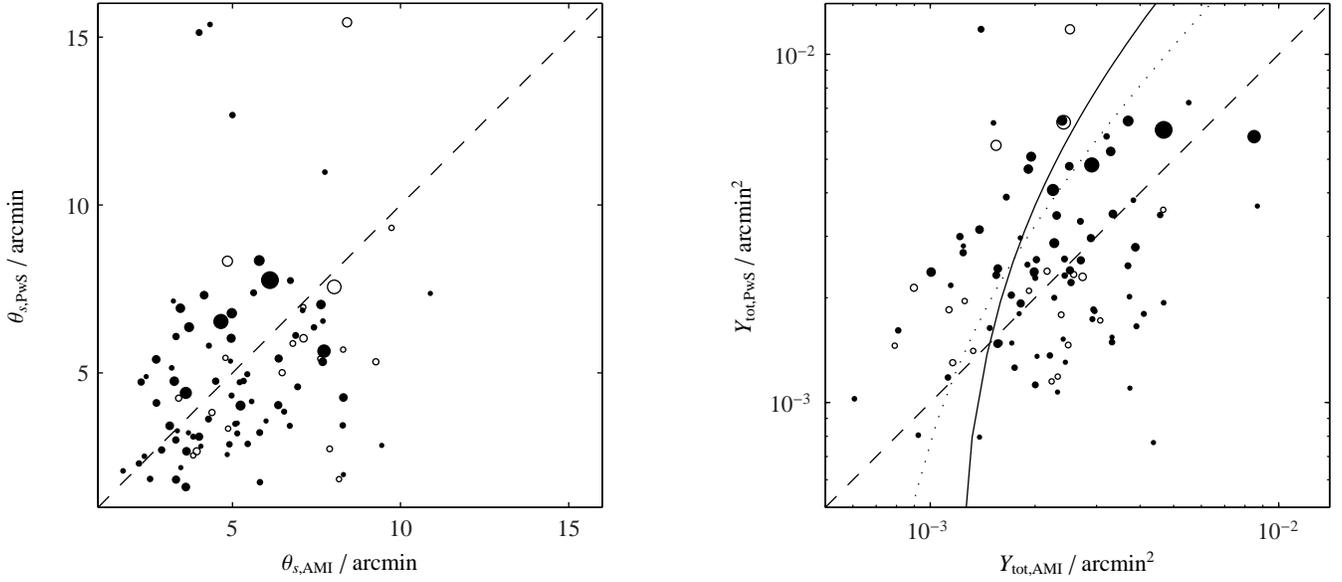}}%
\end{psfrags}%
%
}
  \bigskip
  \caption{Comparison between PwS and AMI mean \Ytot \:and $\theta_{s}$ values.  The size of the points plotted increases with increasing \emph{Planck} SNR; clear detections are plotted as filled circles, and moderate detections as empty circles.  Error bars are omitted for clarity and since the errors in \Ytot and $\theta_{s}$ are correlated.   The one-to-one relationship is plotted as a black dashed line.  The fitted linear relationship for all clusters (clear AMI detections only) with SNR greater than 7 is plotted as a black solid (black dotted) line.}\label{Fi:Y_theta_comp}
\end{figure*}

Fig.~\ref{Fi:Y_theta_comp} shows the comparison between the AMI and \textsc{PwS} mean values for the entire sample of clear and moderate detections.  Aside from some outliers, the $\theta_{s}$ values do not seem to be biased, but only correlate weakly, with a Pearson correlation coefficient of 0.25 (0.18) for all common AMI and PwS detections (clear AMI detections only).  However, the \Ytot \:values for the high-SNR clusters as measured by AMI are still lower overall than the \emph{Planck} values; for lower SNR clusters, the bias may be obscured by the noise.  Following \citet{2013arXiv1303.5080P} for the definition of `high-SNR', we make a cut at \emph{Planck} SNR of 7 and fit a linear model to the \emph{Planck} and AMI results for \Ytot, using the \textsc{SciPy} orthogonal distance regression function\footnote{\url{http://docs.scipy.org/doc/scipy/reference/odr.html}} to take into account errors in both the $x$ and $y$ direction.  The best fit slope for all clusters (clear AMI detections only) above SNR of 7 is $4.2 \pm 1.5$ ($2.45 \pm 0.72$); note that the slope for all clusters is driven by one very discrepant moderate detection.  The slope for clear AMI detections only is consistent with the slope found in AP2013 ($1.05 \pm 0.05$) at $<2\sigma$ significance; note however that this relationship was obtained by fixing the cluster size to the $\theta_{500}$ inferred from the X-ray luminosity for improved consistency.

The comparison between AMI values and the values produced by the MMF algorithms is very similar.

\begin{figure*}
  \centerline{
%
%
\begin{psfrags}%
\psfragscanon%
%
\psfrag{s05}[t][t]{\color[rgb]{0,0,0}\setlength{\tabcolsep}{0pt}\begin{tabular}{c}$\theta_s$ / arcmin\end{tabular}}%
\psfrag{s06}[b][b]{\color[rgb]{0,0,0}\setlength{\tabcolsep}{0pt}\begin{tabular}{c}$Y_{\rm tot, PwS} / Y_{\rm tot, AMI}$\end{tabular}}%
\psfrag{s07}[t][t]{\color[rgb]{0,0,0}\setlength{\tabcolsep}{0pt}\begin{tabular}{c}$\theta_{s}$ / arcmin\end{tabular}}%
\psfrag{s08}[b][b]{\color[rgb]{0,0,0}\setlength{\tabcolsep}{0pt}\begin{tabular}{c}$Y_{\rm tot} \times 10^3$ / arcmin$^{2}$\end{tabular}}%
%
\psfrag{x01}[t][t]{0}%
\psfrag{x02}[t][t]{5}%
\psfrag{x03}[t][t]{10}%
\psfrag{x04}[t][t]{15}%
\psfrag{x05}[t][t]{0}%
\psfrag{x06}[t][t]{5}%
\psfrag{x07}[t][t]{10}%
\psfrag{x08}[t][t]{15}%
%
\psfrag{v01}[r][r]{0}%
\psfrag{v02}[r][r]{4}%
\psfrag{v03}[r][r]{8}%
\psfrag{v04}[r][r]{12}%
\psfrag{v05}[r][r]{16}%
\psfrag{v06}[r][r]{$10^{-1}$}%
\psfrag{v07}[r][r]{$10^{0}$}%
\psfrag{v08}[r][r]{$10^{1}$}%
%
\resizebox{17.8cm}{!}{\includegraphics{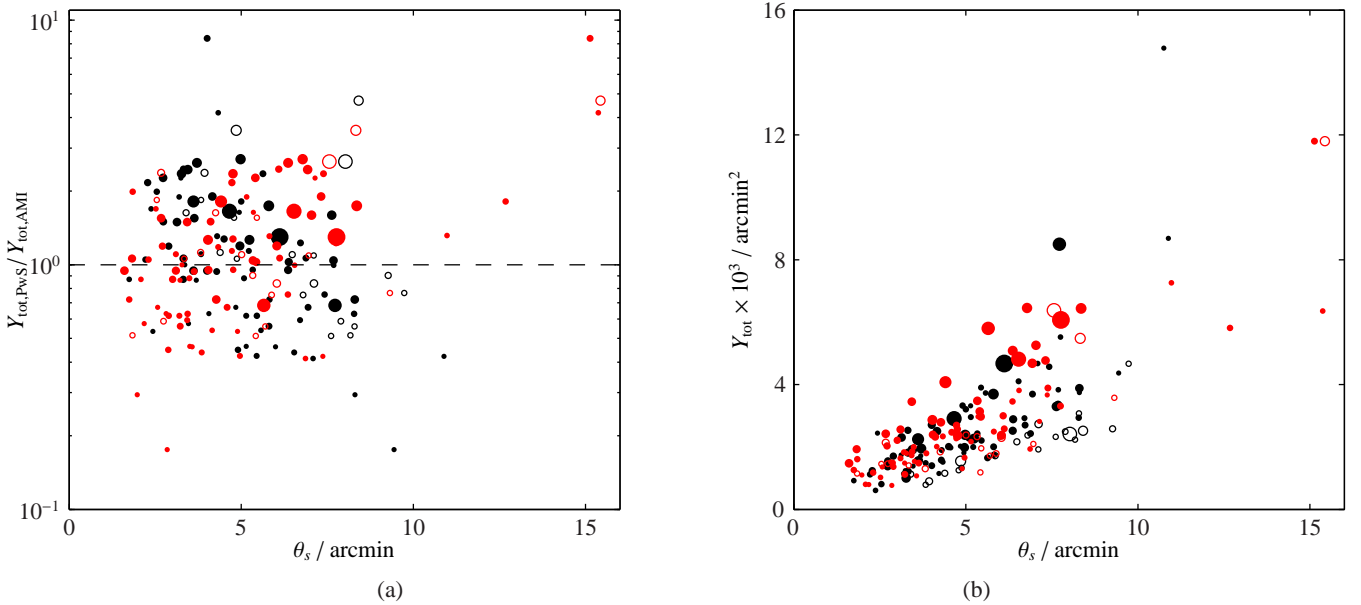}}%
\end{psfrags}%
%
}
  \medskip
  \centerline{(a) \hskip 0.4\linewidth (b)}
  \caption{(a) shows a comparison between \textsc{PwS} and AMI MAP \Ytot \:values as a function of AMI (PwS) $\theta_{s}$ values in black (red).  The one-to-one relationship is plotted as a black dashed line. (b) shows \Ytot \:as a function of $\theta_{s}$ as measured by AMI (black) and \textsc{PwS} (red) for all of the moderate and clear detections.  In both plots, the size of the points plotted increases with increasing \emph{Planck} SNR, clear detections are plotted as filled circles, and moderate detections as empty circles.  Error bars are omitted for clarity and since the errors in \Ytot and $\theta_{s}$ are correlated.}\label{Fi:Y_theta_comp2}
\end{figure*}

This inconsistency could be due to the fact that AMI does not measure \Ytot \:directly, since it is an interferometer and therefore resolves out the larger scales; as long as the cluster is resolved, the zero-spacing flux, and therefore \Ytot, is never measured directly.  In this case the discrepancy should be worse for larger angular-size clusters since more of an extrapolation is required to infer the zero-spacing flux.  In Fig.~\ref{Fi:Y_theta_comp2}(a), the ratio of the \Ytot \:values is plotted as a function of $\theta_{s}$ as measured by AMI and \emph{Planck}; the discrepancy does appear worse for larger values of $\theta_{s, \mathit{Planck}}$, but occurs across all values of $\theta_{s, \mathrm{AMI}}$.  In Fig.~\ref{Fi:Y_theta_comp2}(b) the correlation between $\theta_{s}$ and \Ytot \:is plotted as measured by AMI and \emph{Planck}, which also shows that the discrepancy occurs over the entire sample.

\paragraph{Potential origins of the discrepancy}\mbox{}\\

\begin{figure}
  \centerline{
%
%
\begin{psfrags}%
\psfragscanon%
%
\psfrag{s03}[t][t]{\color[rgb]{0,0,0}\setlength{\tabcolsep}{0pt}\begin{tabular}{c}Frequency / GHz\end{tabular}}%
\psfrag{s04}[b][b]{\color[rgb]{0,0,0}\setlength{\tabcolsep}{0pt}\begin{tabular}{c}Flux density / Jy\end{tabular}}%
\psfrag{s05}[l][l]{\color[rgb]{0,0,0}\setlength{\tabcolsep}{0pt}\begin{tabular}{l}3C286\end{tabular}}%
\psfrag{s06}[l][l]{\color[rgb]{0,0,0}\setlength{\tabcolsep}{0pt}\begin{tabular}{l}3C147\end{tabular}}%
%
\psfrag{x01}[t][t]{15}%
\psfrag{x02}[t][t]{30}%
\psfrag{x03}[t][t]{44}%
%
\psfrag{v01}[r][r]{1}%
\psfrag{v02}[r][r]{2}%
\psfrag{v03}[r][r]{4}%
%
\resizebox{7cm}{!}{\includegraphics{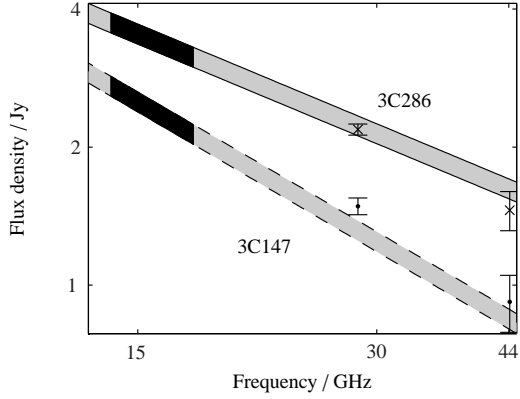}}%
\end{psfrags}%
%
}
  \caption{The power-law relationships used to calculate primary calibration flux densities for AMI for two calibrators, 3C286 and 3C147, are shown with $\pm 5$\% uncertainty limits as the grey filled bands.  The AMI frequency band is shown in black.  Flux densities for both sources at 30 and 44\,GHz taken from the \emph{Planck} Compact Source Catalogue \citep{2013arXiv1303.5088P} are shown as points with errorbars.}\label{Fi:calibration}
\end{figure}

\noindent To first eliminate the possibility that the discrepancy is caused by absolute calibration problems, we obtained flux densities for two of our primary calibration sources, 3C286 and 3C147, at 30 and 44\,GHz from the \emph{Planck} Compact Source Catalogue \citep{2013arXiv1303.5088P}.  These are shown in Fig.~\ref{Fi:calibration} with the power-law used to calculate the AMI primary calibration flux densities for comparison.  All flux densities are within 3$\sigma$ of the power-law, and there does not appear to be a systematic bias.  We therefore discard absolute calibration as a potential cause of the discrepancy.

Several potential origins of the discrepancy were investigated in AP2013, as follows.

\begin{enumerate}
\item{The possibility that a population of faint sources existed below the LA detection threshold and acted to `fill in' the decrement was investigated by obtaining very deep LA observations toward the central pointing of the raster for each cluster, obtaining r.m.s.\ noise levels $\lessapprox$\,30\,$\mu$Jy\,beam$^{-1}$, and re-extracting the cluster parameters, subtracting any extra sources detected.  In one case this shifted the \Ytot \:estimate upward by $\approx$\,1$\sigma$, but the parameters for the remaining 10 cases were not significantly changed.  This is clearly not the source of the discrepancy.}
\item{To eliminate any effects from differing centroid positions, the AMI and \emph{Planck} data were both analysed with the position of the cluster fixed to the best-fit position obtained from an initial AMI analysis where the central position was allowed to vary.  Fixing the position also had a negligible effect on the derived $\theta_{s}$ and \Ytot \:posterior distributions.}
\item{For five clusters with measured X-ray profiles, the cluster parameters were re-extracted using the appropriate X-ray-determined $\gamma$ and $\alpha$ parameters rather than the `universal' parameters.  This did not significantly improve the agreement.  Note that the parameter affecting the cluster outskirts, $\beta$, was not varied since the X-ray data do not extend to this region.  See AP2013 for more details.}
\end{enumerate}

\begin{figure}
  \centerline{
%
%
\begin{psfrags}%
\psfragscanon%
%
\psfrag{s05}[t][t]{\color[rgb]{0,0,0}\setlength{\tabcolsep}{0pt}\begin{tabular}{c}$\theta_{s}$ / arcmin\end{tabular}}%
\psfrag{s06}[b][b]{\color[rgb]{0,0,0}\setlength{\tabcolsep}{0pt}\begin{tabular}{c}$Y_{\rm tot, PwS} / Y_{\rm tot, AMI}$\end{tabular}}%
\psfrag{s07}[t][t]{\color[rgb]{0,0,0}\setlength{\tabcolsep}{0pt}\begin{tabular}{c}$\theta_{s}$ / arcmin\end{tabular}}%
\psfrag{s08}[b][b]{\color[rgb]{0,0,0}\setlength{\tabcolsep}{0pt}\begin{tabular}{c}$Y_{\rm tot} \times 10^3$ / arcmin$^{2}$\end{tabular}}%
%
\psfrag{x01}[t][t]{0}%
\psfrag{x02}[t][t]{5}%
\psfrag{x03}[t][t]{10}%
\psfrag{x04}[t][t]{15}%
\psfrag{x05}[t][t]{0}%
\psfrag{x06}[t][t]{5}%
\psfrag{x07}[t][t]{10}%
\psfrag{x08}[t][t]{15}%
%
\psfrag{v01}[r][r]{0}%
\psfrag{v02}[r][r]{4}%
\psfrag{v03}[r][r]{8}%
\psfrag{v04}[r][r]{12}%
\psfrag{v05}[r][r]{16}%
\psfrag{v06}[r][r]{$10^{-1}$}%
\psfrag{v07}[r][r]{$10^{0}$}%
\psfrag{v08}[r][r]{$10^{1}$}%
%
\resizebox{8.9cm}{!}{\includegraphics{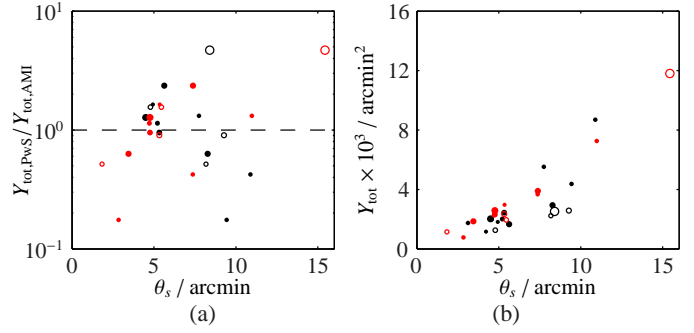}}%
\end{psfrags}%
%
}
  \medskip
  \centerline{(a) \hskip 0.4\linewidth (b)}
  \caption{Comparison between \textsc{PwS} and AMI $\theta_s$ and \Ytot \:MAP values, for clusters selected to have no radio point sources within 3\,arcmin of the cluster position.  In both plots, the black (red) points show the AMI (\textsc{PwS}) values, larger points have higher \emph{Planck} SNR values and filled (empty) circles represent AMI clear (moderate) detections.}\label{Fi:Y_theta_comp2_ns}
\end{figure}

When a point source very near the cluster centre is fitted simultaneously with the cluster model, there is often a correlation between the point source flux and the \Ytot \:value, i.e.\ the data can constrain the sum of the point source flux and the cluster flux well, but not separate the two components.  If this effect led to biases in the fitted \Ytot \:values, it would worsen for smaller angular-size clusters since it becomes more difficult to distinguish between the profiles in $uv$-space of a marginally-resolved cluster and an unresolved point source.  To test whether this could cause the discrepancy, we replotted Fig.~\ref{Fi:Y_theta_comp2} using only clusters with no fitted sources within 3\,arcmin of the cluster position.  This is shown in Fig.~\ref{Fi:Y_theta_comp2_ns}; although the number of clusters in the plot is much smaller, the discrepancy is clearly not resolved.  In addition, we conducted tests on simulations of clusters with point sources of varying flux densities and at varying distances from the cluster centres, and found that we were able to recover \Ytot \:values correctly.

Another potential problem is the mismatch between the spherical model and the real data; the higher resolution AMI data will be much more sensitive to this issue than the \emph{Planck} data (in some cases, also dependent on other factors as discussed in Section~\ref{S:pos_comp}).  Some of the clusters have clearly non-spherical shapes in the AMI maps, but modelling with an ellipsoidal GNFW profile does not change the constraints on \Ytot \:and $\theta_s$ significantly.

\section{Profile investigation}
\label{sec:profiles}

The outstanding issue to be considered is the use of the `universal' profile shape for all clusters.  AMI-SA data are not of high enough resolution to measure $\gamma$; the range of scales measured by the SA corresponds to $0.3 \lesssim \theta/\theta_{s} \lesssim 9$ for clusters with angular sizes $\theta_{s}$ in the range 2 to 10\,arcmin.  For the smallest (largest) clusters in the sample, $\alpha$ ($\beta$) will be the parameter most affecting AMI data; for most clusters, both will be important.

\subsection{Analysis of simulations}\label{S:simulation}

\begin{figure*}
    \centerline{\input{figs/rexcess_sim_CAJ0441.tex}}
    \centerline{
%
%
\begin{psfrags}%
\psfragscanon%
%
\psfrag{s09}[t][t]{}%
\psfrag{s10}[b][b]{\color[rgb]{0,0,0}\setlength{\tabcolsep}{0pt}\begin{tabular}{c}$Y_{tot} \times 10^3$ / arcmin$^2$\end{tabular}}%
\psfrag{s11}[t][t]{}%
\psfrag{s12}[b][b]{}%
%
\psfrag{x01}[t][t]{0}%
\psfrag{x02}[t][t]{0.1}%
\psfrag{x03}[t][t]{0.2}%
\psfrag{x04}[t][t]{0.3}%
\psfrag{x05}[t][t]{0.4}%
\psfrag{x06}[t][t]{0.5}%
\psfrag{x07}[t][t]{0.6}%
\psfrag{x08}[t][t]{0.7}%
\psfrag{x09}[t][t]{0.8}%
\psfrag{x10}[t][t]{0.9}%
\psfrag{x11}[t][t]{1}%
\psfrag{x12}[t][t]{0}%
\psfrag{x13}[t][t]{10}%
\psfrag{x14}[t][t]{20}%
\psfrag{x15}[t][t]{30}%
\psfrag{x16}[t][t]{0}%
\psfrag{x17}[t][t]{10}%
\psfrag{x18}[t][t]{20}%
\psfrag{x19}[t][t]{30}%
%
\psfrag{v01}[r][r]{0}%
\psfrag{v02}[r][r]{0.1}%
\psfrag{v03}[r][r]{0.2}%
\psfrag{v04}[r][r]{0.3}%
\psfrag{v05}[r][r]{0.4}%
\psfrag{v06}[r][r]{0.5}%
\psfrag{v07}[r][r]{0.6}%
\psfrag{v08}[r][r]{0.7}%
\psfrag{v09}[r][r]{0.8}%
\psfrag{v10}[r][r]{0.9}%
\psfrag{v11}[r][r]{1}%
\psfrag{v12}[r][r]{0}%
\psfrag{v13}[r][r]{5}%
\psfrag{v14}[r][r]{10}%
\psfrag{v15}[r][r]{15}%
\psfrag{v16}[r][r]{0}%
\psfrag{v17}[r][r]{5}%
\psfrag{v18}[r][r]{10}%
\psfrag{v19}[r][r]{15}%
%
\resizebox{14.5cm}{!}{\includegraphics{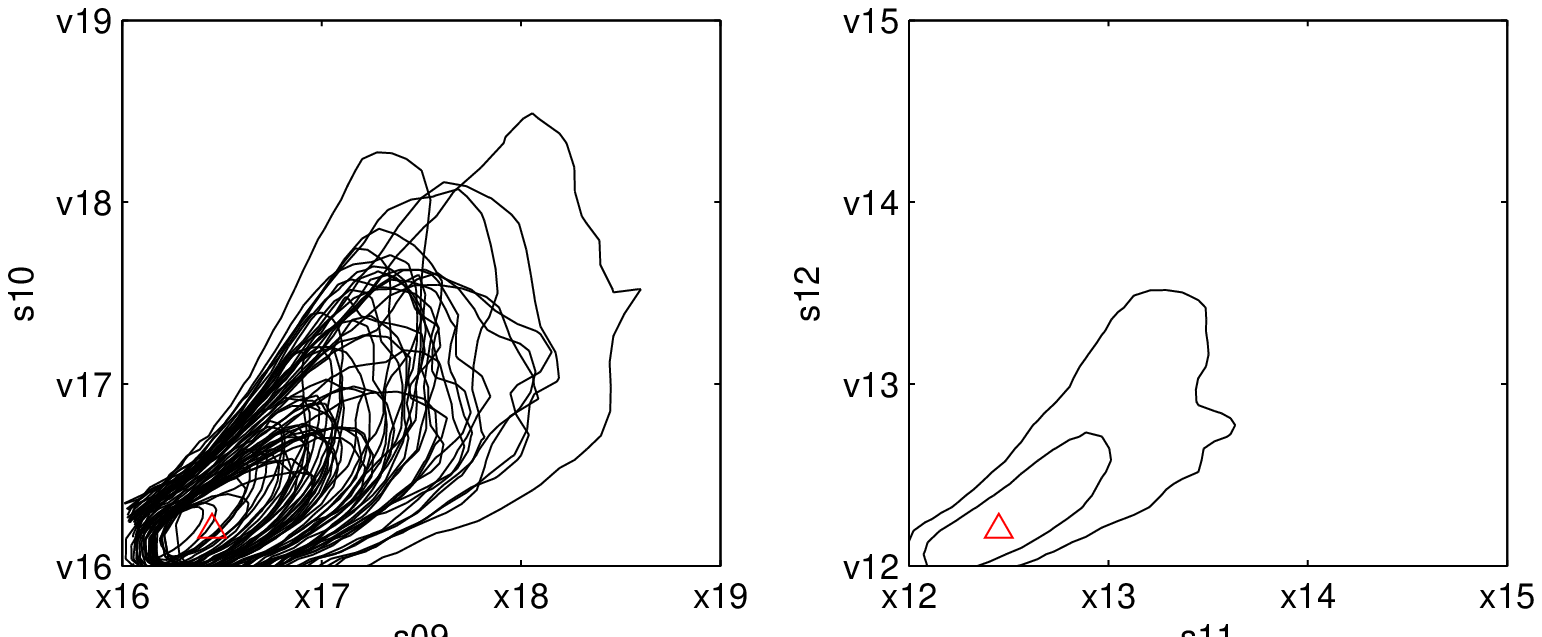}}%
\end{psfrags}%
%
}
    \centerline{
%
%
\begin{psfrags}%
\psfragscanon%
%
\psfrag{s09}[t][t]{\color[rgb]{0,0,0}\setlength{\tabcolsep}{0pt}\begin{tabular}{c}$\theta_s$ / arcmin\end{tabular}}%
\psfrag{s10}[b][b]{}%
\psfrag{s11}[t][t]{}%
\psfrag{s12}[b][b]{}%
%
\psfrag{x01}[t][t]{0}%
\psfrag{x02}[t][t]{0.1}%
\psfrag{x03}[t][t]{0.2}%
\psfrag{x04}[t][t]{0.3}%
\psfrag{x05}[t][t]{0.4}%
\psfrag{x06}[t][t]{0.5}%
\psfrag{x07}[t][t]{0.6}%
\psfrag{x08}[t][t]{0.7}%
\psfrag{x09}[t][t]{0.8}%
\psfrag{x10}[t][t]{0.9}%
\psfrag{x11}[t][t]{1}%
\psfrag{x12}[t][t]{0}%
\psfrag{x13}[t][t]{10}%
\psfrag{x14}[t][t]{20}%
\psfrag{x15}[t][t]{0}%
\psfrag{x16}[t][t]{10}%
\psfrag{x17}[t][t]{20}%
%
\psfrag{v01}[r][r]{0}%
\psfrag{v02}[r][r]{0.1}%
\psfrag{v03}[r][r]{0.2}%
\psfrag{v04}[r][r]{0.3}%
\psfrag{v05}[r][r]{0.4}%
\psfrag{v06}[r][r]{0.5}%
\psfrag{v07}[r][r]{0.6}%
\psfrag{v08}[r][r]{0.7}%
\psfrag{v09}[r][r]{0.8}%
\psfrag{v10}[r][r]{0.9}%
\psfrag{v11}[r][r]{1}%
\psfrag{v12}[r][r]{0}%
\psfrag{v13}[r][r]{5}%
\psfrag{v14}[r][r]{10}%
\psfrag{v15}[r][r]{15}%
\psfrag{v16}[r][r]{0}%
\psfrag{v17}[r][r]{5}%
\psfrag{v18}[r][r]{10}%
\psfrag{v19}[r][r]{15}%
%
\resizebox{14.5cm}{!}{\includegraphics{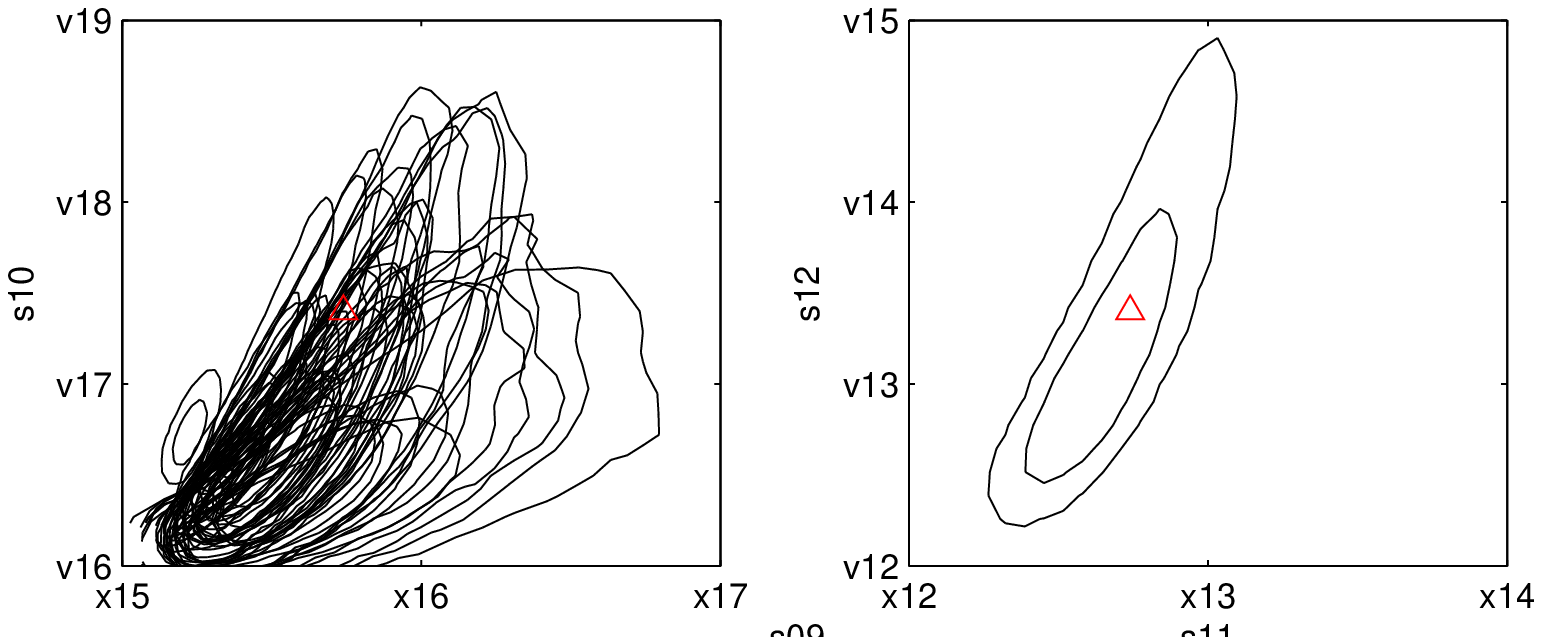}}%
\end{psfrags}%
%
}
    \centerline{(a) \hskip 0.3 \textwidth (b)}
    \caption{The posterior distribution for \Ytot \:and $\theta_s$ for simulated clusters with realistic CMB and noise levels (see text for details), and (a) differing GNFW shape parameter values ($\gamma, \alpha, \beta$) based on the REXCESS sample \citep{bohringer07, arnaud10}, and (b) simulated with the `universal' values.  In all cases the model used for recovering the parameters has the shape parameter values fixed to the `universal' values, and the joint two-dimensional prior on \Ytot \:and $\theta_s$ is used.  Results for three different angular sizes are shown (from top to bottom, $\theta_s = 1.8, 4.5$ and 7.4); the input parameter values are marked with red triangles.  The contours are at the 68\% and 95\% confidence boundaries.}\label{Fi:rexcess_sims}
\end{figure*}

As a first step to understanding how variation in the shape parameters affects constraints derived from AMI data, we generated a set of simulations with realistic thermal, CMB and source confusion noise levels.  We chose three representative values of $\theta_{s}$ based on the follow-up sample, and assigned realistic \Ytot \:values to each based on clusters in the sample with a similar angular size and that were well-detected by AMI, giving ($\theta_{s}$, \Ytot) = (1.8, 0.0009), (4.5, 0.001) and (7.4, 0.007).  For each ($\theta_{s}$, \Ytot), we generated simulations with $\alpha$ and $\gamma$ set to the 31 individual fitted values from the REXCESS sample \citep{bohringer07, arnaud10}, and with $\beta$ drawn from a uniform distribution between 4.5 and 6.5.  Fig.~\ref{Fi:rexcess_sims}(a) shows the result of analysing these simulations with the standard AMI analysis pipeline, assuming the `universal' profile parameters, whereas Fig.~\ref{Fi:rexcess_sims}(b) shows the results when the simulation is both generated and analysed with the `universal' profile.  In the former case, for the two smaller clusters, the true value is within the 68\% confidence limit 29 times out of 31, but it is clear that the size and degeneracy direction of the contours varies wildly for different sets of ($\gamma, \alpha, \beta$); on the whole, the mean and MAP values of $\theta_s$ and \Ytot \:are biased upward slightly.  For the largest cluster, the true value is within the 68\% confidence limit only 2 times out of 31, and within the 95\% confidence limit only 14 out of 31 times.  Again, the size and degeneracy directions of the contours vary wildly; note that the very tight contours which are significantly discrepant from the rest correspond to the profile in the REXCESS sample that is most discrepant from the `universal' profile, with shape parameters $\gamma = 0.065, \alpha = 0.33$.  On the whole, the mean and MAP values of $\theta_s$ and \Ytot \:are biased downward significantly for this cluster.  

\begin{figure}
     \centerline{
%
%
\begin{psfrags}%
\psfragscanon%
\Large
%
\psfrag{s02}[t][t]{\color[rgb]{0,0,0}\setlength{\tabcolsep}{0pt}\begin{tabular}{c}$\theta_s$ / arcmin\end{tabular}}%
\psfrag{s03}[b][b]{\color[rgb]{0,0,0}\setlength{\tabcolsep}{0pt}\begin{tabular}{c}$\log(P_{\rm e})$\end{tabular}}%
\psfrag{s05}[t][t]{\color[rgb]{0,0,0}\setlength{\tabcolsep}{0pt}\begin{tabular}{c}$uv$-distance / $\lambda$\end{tabular}}%
\psfrag{s06}[t][t]{\color[rgb]{0,0,0}\setlength{\tabcolsep}{0pt}\begin{tabular}{c}Visibility amplitude / mJy\end{tabular}}%
%
\psfrag{x01}[t][t]{0}%
\psfrag{x02}[t][t]{0.1}%
\psfrag{x03}[t][t]{0.2}%
\psfrag{x04}[t][t]{0.3}%
\psfrag{x05}[t][t]{0.4}%
\psfrag{x06}[t][t]{0.5}%
\psfrag{x07}[t][t]{0.6}%
\psfrag{x08}[t][t]{0.7}%
\psfrag{x09}[t][t]{0.8}%
\psfrag{x10}[t][t]{0.9}%
\psfrag{x11}[t][t]{1}%
\psfrag{x12}[t][t]{0}%
\psfrag{x13}[t][t]{400}%
\psfrag{x14}[t][t]{800}%
\psfrag{x15}[t][t]{1200}%
\psfrag{x16}[t][t]{0}%
\psfrag{x17}[t][t]{5}%
\psfrag{x18}[t][t]{10}%
\psfrag{x19}[t][t]{15}%
\psfrag{x20}[t][t]{20}%
%
\psfrag{v01}[r][r]{0}%
\psfrag{v02}[r][r]{0.1}%
\psfrag{v03}[r][r]{0.2}%
\psfrag{v04}[r][r]{0.3}%
\psfrag{v05}[r][r]{0.4}%
\psfrag{v06}[r][r]{0.5}%
\psfrag{v07}[r][r]{0.6}%
\psfrag{v08}[r][r]{0.7}%
\psfrag{v09}[r][r]{0.8}%
\psfrag{v10}[r][r]{0.9}%
\psfrag{v11}[r][r]{1}%
\psfrag{v12}[r][r]{0}%
\psfrag{v13}[r][r]{1}%
\psfrag{v14}[r][r]{2}%
\psfrag{v15}[r][r]{3}%
%
\resizebox{\linewidth}{!}{\includegraphics{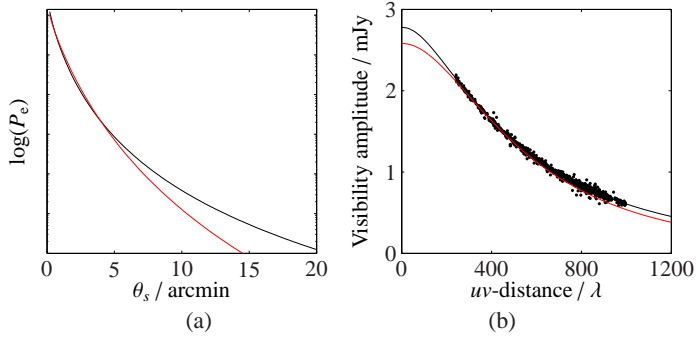}}%
\end{psfrags}%
%
}
     \medskip
     \centerline{(a) \hskip 0.4\linewidth (b)}
     \medskip
     \caption{A profile generated with $\beta = 5.4905, \theta_{s} = 1.8$ (the `universal' profile, black lines) can be mimicked for $\beta = 8.9$ using $\theta_{s} = 4.1$ and adjusting \Ytot \:downward (red lines).  The two profiles are almost identical over the AMI-SA range of baselines, while \emph{Planck} would measure the zero-spacing flux which differs by $\approx$\,7\% between the two models.  (a) shows the pressure profiles in radial coordinates (note that the $y$-axis scale is log), and (b) shows the profiles in $uv$-space for channel 5, with the simulated AMI data shown as dots.  Note that this simulation has been generated with an unrealistically small amount of thermal noise.}\label{Fi:degenerate_profiles_beta}
\end{figure}

To assess the potential for constraining $\alpha$ and $\beta$ using AMI data, we next analysed the simulations, allowing the shape parameters to vary one at a time and using wide, uniform priors on all parameters.  We found that, due to the lack of information on \Ytot \:in the data, there are very strong degeneracies between $\theta_{s}$ and $\alpha$ and $\beta$, even when data with very small amounts of noise are analysed.  For example, Fig.~\ref{Fi:degenerate_profiles_beta} shows that a profile generated with the `universal' value of $\beta$ and a small angular size can be mimicked almost identically across a given range of angular scales using a much larger $\beta$ and $\theta_{s}$ value.

\begin{figure}
     \centerline{
%
%
\begin{psfrags}%
\psfragscanon%
%
\psfrag{s01}[t][t]{\color[rgb]{0,0,0}\setlength{\tabcolsep}{0pt}\begin{tabular}{c}$\alpha$\end{tabular}}%
\psfrag{s02}[b][b]{\color[rgb]{0,0,0}\setlength{\tabcolsep}{0pt}\begin{tabular}{c}$\theta_s$\end{tabular}}%
\psfrag{s03}[t][t]{\color[rgb]{0,0,0}\setlength{\tabcolsep}{0pt}\begin{tabular}{c}$\alpha$\end{tabular}}%
\psfrag{s04}[t][t]{\color[rgb]{0,0,0}\setlength{\tabcolsep}{0pt}\begin{tabular}{c}$Y_{\rm tot} \times 10^3$\end{tabular}}%
\psfrag{s05}[t][t]{\color[rgb]{0,0,0}\setlength{\tabcolsep}{0pt}\begin{tabular}{c}$\theta_s$\end{tabular}}%
\psfrag{s06}[t][t]{\color[rgb]{0,0,0}\setlength{\tabcolsep}{0pt}\begin{tabular}{c}$Y_{\rm tot} \times 10^3$\end{tabular}}%
%
\psfrag{x01}[t][t]{5}%
\psfrag{x02}[t][t]{10}%
\psfrag{x03}[t][t]{15}%
\psfrag{x04}[t][t]{1}%
\psfrag{x05}[t][t]{2}%
\psfrag{x06}[t][t]{3}%
\psfrag{x07}[t][t]{1}%
\psfrag{x08}[t][t]{2}%
\psfrag{x09}[t][t]{3}%
%
\psfrag{v01}[r][r]{1}%
\psfrag{v02}[r][r]{3}%
\psfrag{v03}[r][r]{5}%
\psfrag{v04}[r][r]{1}%
\psfrag{v05}[r][r]{3}%
\psfrag{v06}[r][r]{5}%
\psfrag{v07}[r][r]{5}%
\psfrag{v08}[r][r]{10}%
\psfrag{v09}[r][r]{15}%
%
\resizebox{\linewidth}{!}{\includegraphics{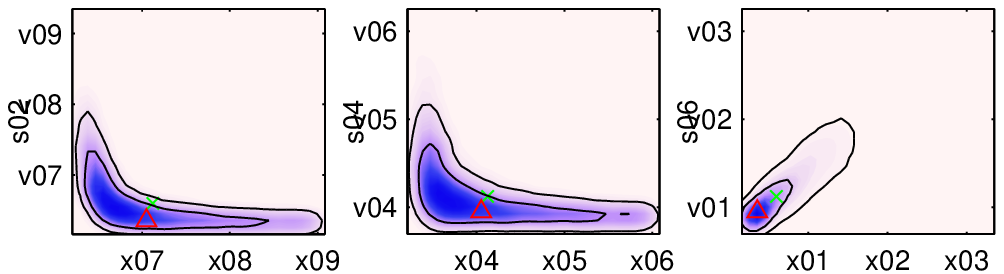}}%
\end{psfrags}%
%
}
     \bigskip
     \centerline{
%
%
\begin{psfrags}%
\psfragscanon%
%
\psfrag{s01}[t][t]{\color[rgb]{0,0,0}\setlength{\tabcolsep}{0pt}\begin{tabular}{c}$\alpha$\end{tabular}}%
\psfrag{s02}[b][b]{\color[rgb]{0,0,0}\setlength{\tabcolsep}{0pt}\begin{tabular}{c}$\theta_s$\end{tabular}}%
\psfrag{s03}[t][t]{\color[rgb]{0,0,0}\setlength{\tabcolsep}{0pt}\begin{tabular}{c}$\alpha$\end{tabular}}%
\psfrag{s04}[t][t]{\color[rgb]{0,0,0}\setlength{\tabcolsep}{0pt}\begin{tabular}{c}$Y_{\rm tot} \times 10^3$\end{tabular}}%
\psfrag{s05}[t][t]{\color[rgb]{0,0,0}\setlength{\tabcolsep}{0pt}\begin{tabular}{c}$\theta_s$\end{tabular}}%
\psfrag{s06}[t][t]{\color[rgb]{0,0,0}\setlength{\tabcolsep}{0pt}\begin{tabular}{c}$Y_{\rm tot} \times 10^3$\end{tabular}}%
%
\psfrag{x01}[t][t]{5}%
\psfrag{x02}[t][t]{10}%
\psfrag{x03}[t][t]{15}%
\psfrag{x04}[t][t]{20}%
\psfrag{x05}[t][t]{1}%
\psfrag{x06}[t][t]{2}%
\psfrag{x07}[t][t]{3}%
\psfrag{x08}[t][t]{1}%
\psfrag{x09}[t][t]{2}%
\psfrag{x10}[t][t]{3}%
%
\psfrag{v01}[r][r]{0}%
\psfrag{v02}[r][r]{4}%
\psfrag{v03}[r][r]{8}%
\psfrag{v04}[r][r]{12}%
\psfrag{v05}[r][r]{0}%
\psfrag{v06}[r][r]{4}%
\psfrag{v07}[r][r]{8}%
\psfrag{v08}[r][r]{12}%
\psfrag{v09}[r][r]{5}%
\psfrag{v10}[r][r]{10}%
\psfrag{v11}[r][r]{15}%
\psfrag{v12}[r][r]{20}%
%
\resizebox{\linewidth}{!}{\includegraphics{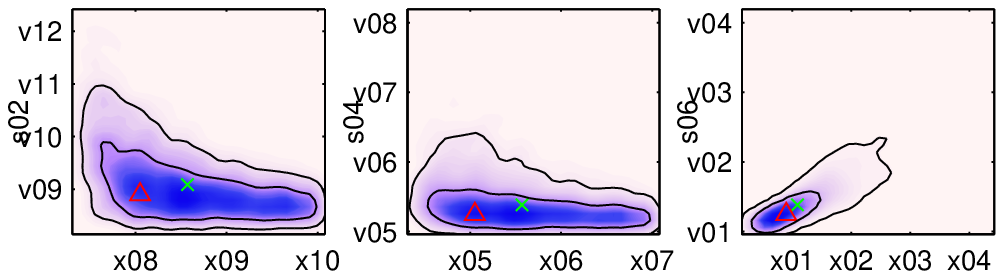}}%
\end{psfrags}%
%
}
     \bigskip
     \centerline{
%
%
\begin{psfrags}%
\psfragscanon%
%
\psfrag{s01}[t][t]{\color[rgb]{0,0,0}\setlength{\tabcolsep}{0pt}\begin{tabular}{c}$\alpha$\end{tabular}}%
\psfrag{s02}[t][t]{\color[rgb]{0,0,0}\setlength{\tabcolsep}{0pt}\begin{tabular}{c}$\theta_s$\end{tabular}}%
\psfrag{s03}[t][t]{\color[rgb]{0,0,0}\setlength{\tabcolsep}{0pt}\begin{tabular}{c}$\alpha$\end{tabular}}%
\psfrag{s04}[t][t]{\color[rgb]{0,0,0}\setlength{\tabcolsep}{0pt}\begin{tabular}{c}$Y_{\rm tot} \times 10^3$\end{tabular}}%
\psfrag{s05}[t][t]{\color[rgb]{0,0,0}\setlength{\tabcolsep}{0pt}\begin{tabular}{c}$\theta_s$\end{tabular}}%
\psfrag{s06}[t][t]{\color[rgb]{0,0,0}\setlength{\tabcolsep}{0pt}\begin{tabular}{c}$Y_{\rm tot} \times 10^3$\end{tabular}}%
%
\psfrag{x01}[t][t]{5}%
\psfrag{x02}[t][t]{10}%
\psfrag{x03}[t][t]{1}%
\psfrag{x04}[t][t]{2}%
\psfrag{x05}[t][t]{3}%
\psfrag{x06}[t][t]{1}%
\psfrag{x07}[t][t]{2}%
\psfrag{x08}[t][t]{3}%
%
\psfrag{v01}[r][r]{2}%
\psfrag{v02}[r][r]{6}%
\psfrag{v03}[r][r]{10}%
\psfrag{v04}[r][r]{2}%
\psfrag{v05}[r][r]{6}%
\psfrag{v06}[r][r]{10}%
\psfrag{v07}[r][r]{5}%
\psfrag{v08}[r][r]{10}%
%
\resizebox{\linewidth}{!}{\includegraphics{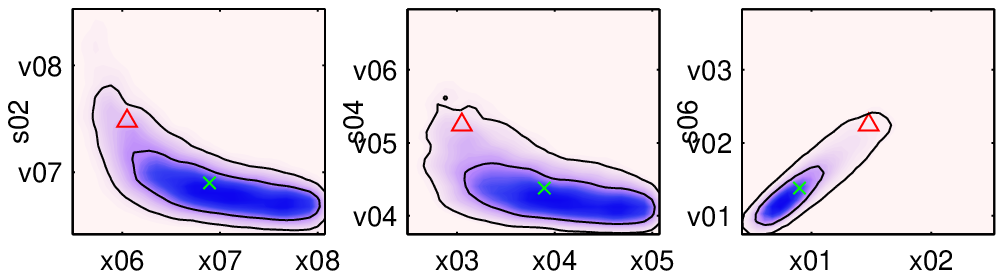}}%
\end{psfrags}%
%
}
     \medskip
     \caption{The posterior distributions for \Ytot, $\theta_{s}$ and $\alpha$ for simulated low-noise data, for clusters with $\theta_{s} = 1.8$ (top), 4.5 (centre) and 7.4 (bottom) arcmin and `universal' ($\gamma, \alpha, \beta$), with the two-dimensional prior on \Ytot \:and $\theta_{s}$ and a uniform prior on $\alpha$ between 0.1 and 3.0 ($\beta$ fixed to the correct, input value).  The input values are indicated by red triangles, and the posterior means with green crosses.  $\theta_{s}$ is in arcmin and {\Ytot} is in arcmin$^2$.}\label{Fi:lownoise_alpha2}
\end{figure}

\begin{figure}
     \centerline{
%
%
\begin{psfrags}%
\psfragscanon%
\Large
%
\psfrag{s01}[t][t]{\color[rgb]{0,0,0}\setlength{\tabcolsep}{0pt}\begin{tabular}{c}$\theta_s$\end{tabular}}%
\psfrag{s03}[b][b]{\color[rgb]{0,0,0}\setlength{\tabcolsep}{0pt}\begin{tabular}{c}Probability density\end{tabular}}%
\psfrag{s05}[t][t]{\color[rgb]{0,0,0}\setlength{\tabcolsep}{0pt}\begin{tabular}{c}$Y_{\rm tot} \times 10^3$\end{tabular}}%
\psfrag{s09}[t][t]{\color[rgb]{0,0,0}\setlength{\tabcolsep}{0pt}\begin{tabular}{c}$\alpha$\end{tabular}}%
%
\psfrag{x01}[t][t]{0}%
\psfrag{x02}[t][t]{0.1}%
\psfrag{x03}[t][t]{0.2}%
\psfrag{x04}[t][t]{0.3}%
\psfrag{x05}[t][t]{0.4}%
\psfrag{x06}[t][t]{0.5}%
\psfrag{x07}[t][t]{0.6}%
\psfrag{x08}[t][t]{0.7}%
\psfrag{x09}[t][t]{0.8}%
\psfrag{x10}[t][t]{0.9}%
\psfrag{x11}[t][t]{1}%
\psfrag{x12}[t][t]{1}%
\psfrag{x13}[t][t]{2}%
\psfrag{x14}[t][t]{3}%
\psfrag{x15}[t][t]{0}%
\psfrag{x16}[t][t]{4}%
\psfrag{x17}[t][t]{8}%
\psfrag{x18}[t][t]{12}%
\psfrag{x19}[t][t]{5}%
\psfrag{x20}[t][t]{10}%
\psfrag{x21}[t][t]{15}%
\psfrag{x22}[t][t]{20}%
%
\psfrag{v01}[r][r]{0}%
\psfrag{v02}[r][r]{0.1}%
\psfrag{v03}[r][r]{0.2}%
\psfrag{v04}[r][r]{0.3}%
\psfrag{v05}[r][r]{0.4}%
\psfrag{v06}[r][r]{0.5}%
\psfrag{v07}[r][r]{0.6}%
\psfrag{v08}[r][r]{0.7}%
\psfrag{v09}[r][r]{0.8}%
\psfrag{v10}[r][r]{0.9}%
\psfrag{v11}[r][r]{1}%
\psfrag{v12}[r][r]{0}%
\psfrag{v13}[r][r]{0.2}%
\psfrag{v14}[r][r]{0.4}%
\psfrag{v15}[r][r]{0.6}%
\psfrag{v16}[r][r]{0.8}%
\psfrag{v17}[r][r]{1}%
\psfrag{v18}[r][r]{0}%
\psfrag{v19}[r][r]{0.2}%
\psfrag{v20}[r][r]{0.4}%
\psfrag{v21}[r][r]{0.6}%
\psfrag{v22}[r][r]{0.8}%
\psfrag{v23}[r][r]{1}%
\psfrag{v24}[r][r]{0}%
\psfrag{v25}[r][r]{0.2}%
\psfrag{v26}[r][r]{0.4}%
\psfrag{v27}[r][r]{0.6}%
\psfrag{v28}[r][r]{0.8}%
\psfrag{v29}[r][r]{1}%
%
\resizebox{\linewidth}{!}{\includegraphics{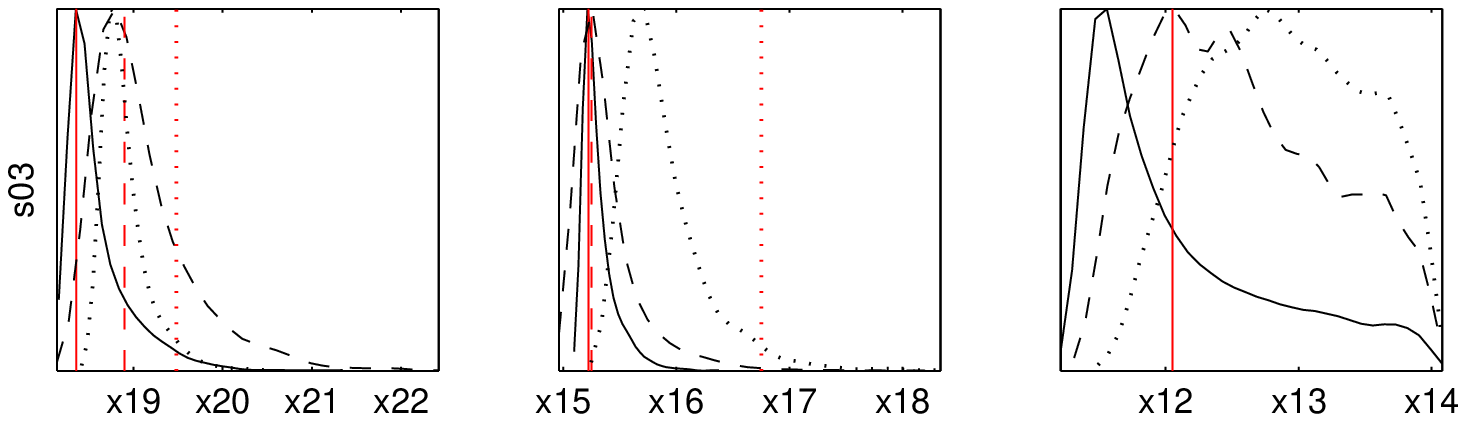}}%
\end{psfrags}%
%
}
     \medskip
     \caption{The one-dimensional marginal constraints on \Ytot, $\theta_{s}$ and $\alpha$ for simulated low-noise data, for clusters with $\theta_{s} = 1.8$ (solid lines), 4.5 (dashed lines) and 7.4 (dotted lines) arcmin and `universal' ($\gamma, \alpha, \beta$), with the two-dimensional prior on \Ytot \:and $\theta_{s}$ and a uniform prior on $\alpha$ between 0.1 and 3.0 ($\beta$ fixed to the correct, input value).  Input values are shown as red lines.  $\theta_{s}$ is in arcmin and {\Ytot} is in arcmin$^2$.}\label{Fi:lownoise_alpha_1d_2}
\end{figure}

\begin{figure}
     \centerline{
%
%
\begin{psfrags}%
\psfragscanon%
%
\psfrag{s01}[t][t]{\color[rgb]{0,0,0}\setlength{\tabcolsep}{0pt}\begin{tabular}{c}$\beta$\end{tabular}}%
\psfrag{s02}[t][t]{\color[rgb]{0,0,0}\setlength{\tabcolsep}{0pt}\begin{tabular}{c}$\theta_s$\end{tabular}}%
\psfrag{s03}[t][t]{\color[rgb]{0,0,0}\setlength{\tabcolsep}{0pt}\begin{tabular}{c}$\beta$\end{tabular}}%
\psfrag{s04}[t][t]{\color[rgb]{0,0,0}\setlength{\tabcolsep}{0pt}\begin{tabular}{c}$Y_{\rm tot} \times 10^3$\end{tabular}}%
\psfrag{s05}[t][t]{\color[rgb]{0,0,0}\setlength{\tabcolsep}{0pt}\begin{tabular}{c}$\theta_s$\end{tabular}}%
\psfrag{s06}[t][t]{\color[rgb]{0,0,0}\setlength{\tabcolsep}{0pt}\begin{tabular}{c}$Y_{\rm tot} \times 10^3$\end{tabular}}%
%
\psfrag{x01}[t][t]{0}%
\psfrag{x02}[t][t]{5}%
\psfrag{x03}[t][t]{10}%
\psfrag{x04}[t][t]{4}%
\psfrag{x05}[t][t]{6}%
\psfrag{x06}[t][t]{8}%
\psfrag{x07}[t][t]{4}%
\psfrag{x08}[t][t]{6}%
\psfrag{x09}[t][t]{8}%
%
\psfrag{v01}[r][r]{0}%
\psfrag{v02}[r][r]{1}%
\psfrag{v03}[r][r]{2}%
\psfrag{v04}[r][r]{3}%
\psfrag{v05}[r][r]{0}%
\psfrag{v06}[r][r]{1}%
\psfrag{v07}[r][r]{2}%
\psfrag{v08}[r][r]{3}%
\psfrag{v09}[r][r]{0}%
\psfrag{v10}[r][r]{5}%
\psfrag{v11}[r][r]{10}%
%
\resizebox{\linewidth}{!}{\includegraphics{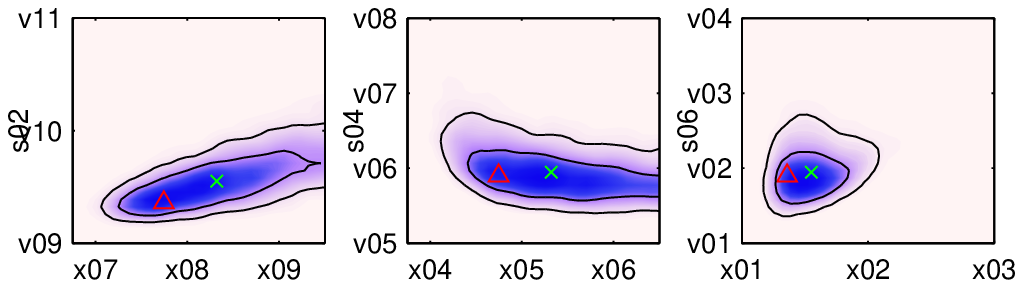}}%
\end{psfrags}%
%
}
     \bigskip
     \centerline{
%
%
\begin{psfrags}%
\psfragscanon%
%
\psfrag{s01}[t][t]{\color[rgb]{0,0,0}\setlength{\tabcolsep}{0pt}\begin{tabular}{c}$\beta$\end{tabular}}%
\psfrag{s02}[t][t]{\color[rgb]{0,0,0}\setlength{\tabcolsep}{0pt}\begin{tabular}{c}$\theta_s$\end{tabular}}%
\psfrag{s03}[t][t]{\color[rgb]{0,0,0}\setlength{\tabcolsep}{0pt}\begin{tabular}{c}$\beta$\end{tabular}}%
\psfrag{s04}[t][t]{\color[rgb]{0,0,0}\setlength{\tabcolsep}{0pt}\begin{tabular}{c}$Y_{\rm tot} \times 10^3$\end{tabular}}%
\psfrag{s05}[t][t]{\color[rgb]{0,0,0}\setlength{\tabcolsep}{0pt}\begin{tabular}{c}$\theta_s$\end{tabular}}%
\psfrag{s06}[t][t]{\color[rgb]{0,0,0}\setlength{\tabcolsep}{0pt}\begin{tabular}{c}$Y_{\rm tot} \times 10^3$\end{tabular}}%
%
\psfrag{x01}[t][t]{5}%
\psfrag{x02}[t][t]{10}%
\psfrag{x03}[t][t]{15}%
\psfrag{x04}[t][t]{4}%
\psfrag{x05}[t][t]{6}%
\psfrag{x06}[t][t]{8}%
\psfrag{x07}[t][t]{4}%
\psfrag{x08}[t][t]{6}%
\psfrag{x09}[t][t]{8}%
%
\psfrag{v01}[r][r]{0}%
\psfrag{v02}[r][r]{4}%
\psfrag{v03}[r][r]{8}%
\psfrag{v04}[r][r]{5}%
\psfrag{v05}[r][r]{10}%
\psfrag{v06}[r][r]{15}%
\psfrag{v07}[r][r]{0}%
\psfrag{v08}[r][r]{4}%
\psfrag{v09}[r][r]{8}%
%
\resizebox{\linewidth}{!}{\includegraphics{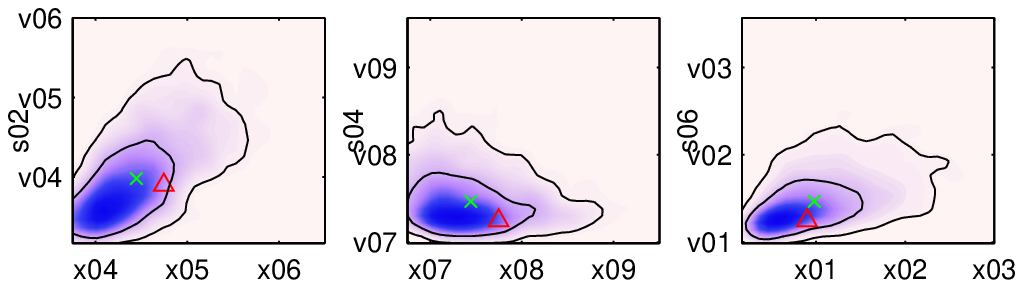}}%
\end{psfrags}%
%
}
     \bigskip
     \centerline{
%
%
\begin{psfrags}%
\psfragscanon%
%
\psfrag{s01}[t][t]{\color[rgb]{0,0,0}\setlength{\tabcolsep}{0pt}\begin{tabular}{c}$\beta$\end{tabular}}%
\psfrag{s02}[b][b]{\color[rgb]{0,0,0}\setlength{\tabcolsep}{0pt}\begin{tabular}{c}$\theta_s$\end{tabular}}%
\psfrag{s03}[t][t]{\color[rgb]{0,0,0}\setlength{\tabcolsep}{0pt}\begin{tabular}{c}$\beta$\end{tabular}}%
\psfrag{s04}[t][t]{\color[rgb]{0,0,0}\setlength{\tabcolsep}{0pt}\begin{tabular}{c}$Y_{\rm tot} \times 10^3$\end{tabular}}%
\psfrag{s05}[t][t]{\color[rgb]{0,0,0}\setlength{\tabcolsep}{0pt}\begin{tabular}{c}$\theta_s$\end{tabular}}%
\psfrag{s06}[t][t]{\color[rgb]{0,0,0}\setlength{\tabcolsep}{0pt}\begin{tabular}{c}$Y_{\rm tot} \times 10^3$\end{tabular}}%
%
\psfrag{x01}[t][t]{5}%
\psfrag{x02}[t][t]{10}%
\psfrag{x03}[t][t]{15}%
\psfrag{x04}[t][t]{4}%
\psfrag{x05}[t][t]{6}%
\psfrag{x06}[t][t]{8}%
\psfrag{x07}[t][t]{4}%
\psfrag{x08}[t][t]{6}%
\psfrag{x09}[t][t]{8}%
%
\psfrag{v01}[r][r]{2}%
\psfrag{v02}[r][r]{6}%
\psfrag{v03}[r][r]{10}%
\psfrag{v04}[r][r]{2}%
\psfrag{v05}[r][r]{6}%
\psfrag{v06}[r][r]{10}%
\psfrag{v07}[r][r]{5}%
\psfrag{v08}[r][r]{10}%
\psfrag{v09}[r][r]{15}%
%
\resizebox{\linewidth}{!}{\includegraphics{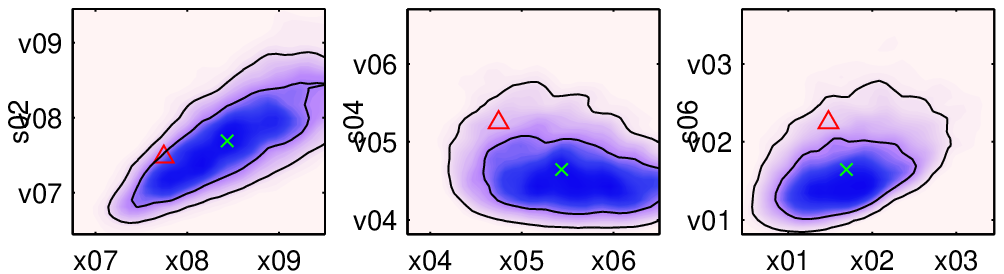}}%
\end{psfrags}%
%
}
     \medskip
     \caption{The posterior distributions for \Ytot, $\theta_{s}$ and $\beta$ for simulated low-noise data, for clusters with $\theta_{s} = 1.8$ (top), 4.5 (centre) and 7.4 (bottom) arcmin and `universal' ($\gamma, \alpha, \beta$), with the two-dimensional prior on \Ytot \:and $\theta_{s}$ and a uniform prior on $\beta$ between 3.5 and 9.0 ($\alpha$ fixed to the correct, input value).  The input values are indicated by red triangles, and the posterior means with green crosses.  $\theta_{s}$ is in arcmin and {\Ytot} is in arcmin$^2$.}\label{Fi:lownoise_beta2}
\end{figure}

\begin{figure}
     \centerline{
%
%
\begin{psfrags}%
\psfragscanon%
\Large
%
\psfrag{s01}[t][t]{\color[rgb]{0,0,0}\setlength{\tabcolsep}{0pt}\begin{tabular}{c}$\theta_s$\end{tabular}}%
\psfrag{s02}[b][b]{\color[rgb]{0,0,0}\setlength{\tabcolsep}{0pt}\begin{tabular}{c}Probability density\end{tabular}}%
\psfrag{s05}[t][t]{\color[rgb]{0,0,0}\setlength{\tabcolsep}{0pt}\begin{tabular}{c}$Y_{\rm tot} \times 10^3$\end{tabular}}%
\psfrag{s09}[t][t]{\color[rgb]{0,0,0}\setlength{\tabcolsep}{0pt}\begin{tabular}{c}$\beta$\end{tabular}}%
%
\psfrag{x01}[t][t]{0}%
\psfrag{x02}[t][t]{0.1}%
\psfrag{x03}[t][t]{0.2}%
\psfrag{x04}[t][t]{0.3}%
\psfrag{x05}[t][t]{0.4}%
\psfrag{x06}[t][t]{0.5}%
\psfrag{x07}[t][t]{0.6}%
\psfrag{x08}[t][t]{0.7}%
\psfrag{x09}[t][t]{0.8}%
\psfrag{x10}[t][t]{0.9}%
\psfrag{x11}[t][t]{1}%
\psfrag{x12}[t][t]{4}%
\psfrag{x13}[t][t]{6}%
\psfrag{x14}[t][t]{8}%
\psfrag{x15}[t][t]{0}%
\psfrag{x16}[t][t]{4}%
\psfrag{x17}[t][t]{8}%
\psfrag{x18}[t][t]{0}%
\psfrag{x19}[t][t]{5}%
\psfrag{x20}[t][t]{10}%
\psfrag{x21}[t][t]{15}%
%
\psfrag{v01}[r][r]{0}%
\psfrag{v02}[r][r]{0.1}%
\psfrag{v03}[r][r]{0.2}%
\psfrag{v04}[r][r]{0.3}%
\psfrag{v05}[r][r]{0.4}%
\psfrag{v06}[r][r]{0.5}%
\psfrag{v07}[r][r]{0.6}%
\psfrag{v08}[r][r]{0.7}%
\psfrag{v09}[r][r]{0.8}%
\psfrag{v10}[r][r]{0.9}%
\psfrag{v11}[r][r]{1}%
\psfrag{v12}[r][r]{0}%
\psfrag{v13}[r][r]{0.1}%
\psfrag{v14}[r][r]{0.2}%
\psfrag{v15}[r][r]{0.3}%
\psfrag{v16}[r][r]{0.4}%
\psfrag{v17}[r][r]{0.5}%
\psfrag{v18}[r][r]{0.6}%
\psfrag{v19}[r][r]{0.7}%
\psfrag{v20}[r][r]{0.8}%
\psfrag{v21}[r][r]{0.9}%
\psfrag{v22}[r][r]{1}%
\psfrag{v23}[r][r]{0}%
\psfrag{v24}[r][r]{0.1}%
\psfrag{v25}[r][r]{0.2}%
\psfrag{v26}[r][r]{0.3}%
\psfrag{v27}[r][r]{0.4}%
\psfrag{v28}[r][r]{0.5}%
\psfrag{v29}[r][r]{0.6}%
\psfrag{v30}[r][r]{0.7}%
\psfrag{v31}[r][r]{0.8}%
\psfrag{v32}[r][r]{0.9}%
\psfrag{v33}[r][r]{1}%
\psfrag{v34}[r][r]{0}%
\psfrag{v35}[r][r]{0.1}%
\psfrag{v36}[r][r]{0.2}%
\psfrag{v37}[r][r]{0.3}%
\psfrag{v38}[r][r]{0.4}%
\psfrag{v39}[r][r]{0.5}%
\psfrag{v40}[r][r]{0.6}%
\psfrag{v41}[r][r]{0.7}%
\psfrag{v42}[r][r]{0.8}%
\psfrag{v43}[r][r]{0.9}%
\psfrag{v44}[r][r]{1}%
%
\resizebox{\linewidth}{!}{\includegraphics{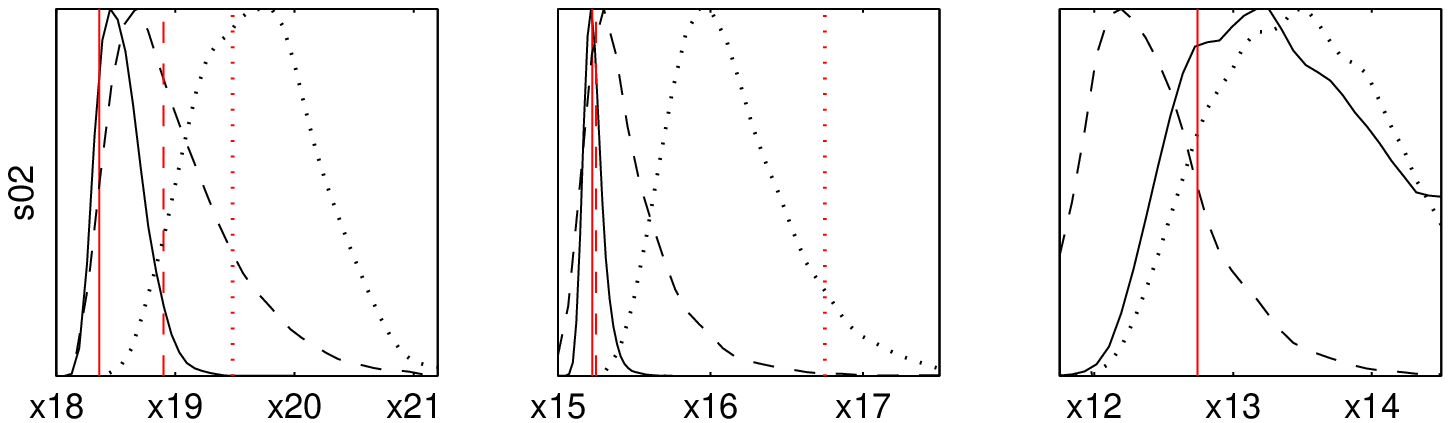}}%
\end{psfrags}%
%
}
     \medskip
     \caption{The one-dimensional marginal constraints on \Ytot, $\theta_{s}$ and $\beta$ for simulated low-noise data, for clusters with $\theta_{s} = 1.8$ (solid lines), 4.5 (dashed lines) and 7.4 (dotted lines) arcmin and `universal' ($\gamma, \alpha, \beta$), with the two-dimensional prior on \Ytot \:and $\theta_{s}$ and a uniform prior on $\beta$ between 3.5 and 9.0 ($\alpha$ fixed to the correct, input value).  Input values are shown as red lines.  $\theta_{s}$ is in arcmin and {\Ytot} is in arcmin$^2$.}\label{Fi:lownoise_beta_1d_2}
\end{figure}

In practice, these strong degeneracies were found to lead to spurious constraints in $\alpha$ and $\beta$ in the one-dimensional marginalised posterior distribution.  This is simply due to the shape of the three-dimensional posterior; more \Ytot-$\theta_{s}$ space becomes available for lower values of $\alpha$ and $\beta$.  Applying the two-dimensional prior on \Ytot \:and $\theta_{s}$ to ensure that physically motivated parts of the \Ytot-$\theta_{s}$ space are selected reduces, but does not eliminate, the problem.  This is illustrated in Fig.~\ref{Fi:lownoise_alpha2} and Fig.~\ref{Fi:lownoise_alpha_1d_2} where the two- and one-dimensional posterior distributions are shown for $\theta_{s}$, \Ytot \:and $\alpha$, with the standard two-dimensional prior on $\theta_{s}$ and \Ytot \:and a uniform prior between 0.1 and 3.0 on $\alpha$ (with $\beta$ fixed to the correct, input value of 5.4905).  When there is little information on $\alpha$ in the data (particularly for the smallest cluster), the shape of the two-dimensional posteriors produces an apparent (and incorrect) constraint on $\alpha$ in the one-dimensional posteriors.  Similar effects occur in the constraints on $\beta$, shown in Fig.~\ref{Fi:lownoise_beta2} and \ref{Fi:lownoise_beta_1d_2} (in which $\alpha$ is fixed to the correct, input value of 1.0510).

\begin{figure*}
    \centerline{
%
%
\begin{psfrags}%
\psfragscanon%
%
\psfrag{s07}[t][t]{\color[rgb]{0,0,0}\setlength{\tabcolsep}{0pt}\begin{tabular}{c}$\theta_s$\end{tabular}}%
\psfrag{s08}[b][b]{\color[rgb]{0,0,0}\setlength{\tabcolsep}{0pt}\begin{tabular}{c}$Y_{tot} \times 10^3$\end{tabular}}%
\psfrag{s09}[t][t]{\color[rgb]{0,0,0}\setlength{\tabcolsep}{0pt}\begin{tabular}{c}$\alpha - \alpha_{\rm true}$\end{tabular}}%
\psfrag{s10}[b][b]{\color[rgb]{0,0,0}\setlength{\tabcolsep}{0pt}\begin{tabular}{c}Probability density\end{tabular}}%
\psfrag{s11}[t][t]{\color[rgb]{0,0,0}\setlength{\tabcolsep}{0pt}\begin{tabular}{c}$\beta - \beta_{\rm true}$\end{tabular}}%
\psfrag{s12}[b][b]{\color[rgb]{0,0,0}\setlength{\tabcolsep}{0pt}\begin{tabular}{c}Probability density\end{tabular}}%
%
\psfrag{x01}[t][t]{0}%
\psfrag{x02}[t][t]{0.1}%
\psfrag{x03}[t][t]{0.2}%
\psfrag{x04}[t][t]{0.3}%
\psfrag{x05}[t][t]{0.4}%
\psfrag{x06}[t][t]{0.5}%
\psfrag{x07}[t][t]{0.6}%
\psfrag{x08}[t][t]{0.7}%
\psfrag{x09}[t][t]{0.8}%
\psfrag{x10}[t][t]{0.9}%
\psfrag{x11}[t][t]{1}%
\psfrag{x12}[t][t]{-2}%
\psfrag{x13}[t][t]{-1}%
\psfrag{x14}[t][t]{0}%
\psfrag{x15}[t][t]{1}%
\psfrag{x16}[t][t]{-2}%
\psfrag{x17}[t][t]{-1}%
\psfrag{x18}[t][t]{0}%
\psfrag{x19}[t][t]{1}%
\psfrag{x20}[t][t]{0}%
\psfrag{x21}[t][t]{5}%
\psfrag{x22}[t][t]{10}%
\psfrag{x23}[t][t]{15}%
%
\psfrag{v01}[r][r]{0}%
\psfrag{v02}[r][r]{0.1}%
\psfrag{v03}[r][r]{0.2}%
\psfrag{v04}[r][r]{0.3}%
\psfrag{v05}[r][r]{0.4}%
\psfrag{v06}[r][r]{0.5}%
\psfrag{v07}[r][r]{0.6}%
\psfrag{v08}[r][r]{0.7}%
\psfrag{v09}[r][r]{0.8}%
\psfrag{v10}[r][r]{0.9}%
\psfrag{v11}[r][r]{1}%
\psfrag{v12}[r][r]{0}%
\psfrag{v13}[r][r]{0.1}%
\psfrag{v14}[r][r]{0.2}%
\psfrag{v15}[r][r]{0.3}%
\psfrag{v16}[r][r]{0.4}%
\psfrag{v17}[r][r]{0.5}%
\psfrag{v18}[r][r]{0.6}%
\psfrag{v19}[r][r]{0.7}%
\psfrag{v20}[r][r]{0.8}%
\psfrag{v21}[r][r]{0.9}%
\psfrag{v22}[r][r]{1}%
\psfrag{v23}[r][r]{0}%
\psfrag{v24}[r][r]{0.1}%
\psfrag{v25}[r][r]{0.2}%
\psfrag{v26}[r][r]{0.3}%
\psfrag{v27}[r][r]{0.4}%
\psfrag{v28}[r][r]{0.5}%
\psfrag{v29}[r][r]{0.6}%
\psfrag{v30}[r][r]{0.7}%
\psfrag{v31}[r][r]{0.8}%
\psfrag{v32}[r][r]{0.9}%
\psfrag{v33}[r][r]{1}%
\psfrag{v34}[r][r]{0}%
\psfrag{v35}[r][r]{3}%
\psfrag{v36}[r][r]{6}%
%
\resizebox{14.5cm}{!}{\includegraphics{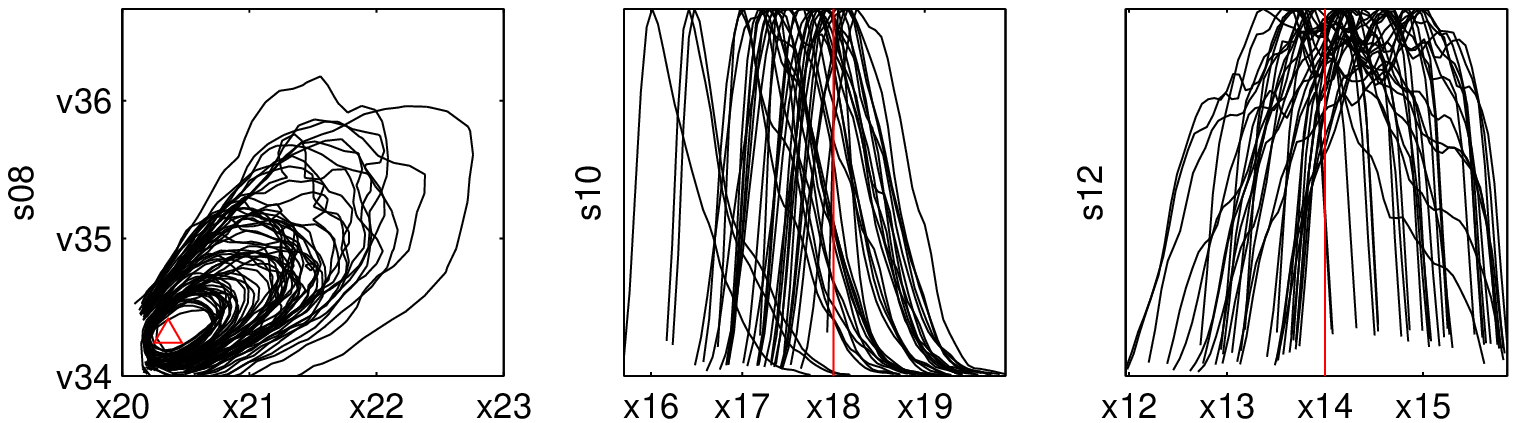}}%
\end{psfrags}%
%
}
    \centerline{
%
%
\begin{psfrags}%
\psfragscanon%
%
\psfrag{s11}[t][t]{}%
\psfrag{s12}[b][b]{\color[rgb]{0,0,0}\setlength{\tabcolsep}{0pt}\begin{tabular}{c}$Y_{tot} \times 10^3$ / arcmin$^2$\end{tabular}}%
\psfrag{s13}[t][t]{}%
\psfrag{s14}[b][b]{\color[rgb]{0,0,0}\setlength{\tabcolsep}{0pt}\begin{tabular}{c}Probability density\end{tabular}}%
\psfrag{s15}[t][t]{}%
\psfrag{s16}[b][b]{\color[rgb]{0,0,0}\setlength{\tabcolsep}{0pt}\begin{tabular}{c}Probability density\end{tabular}}%
%
\psfrag{x01}[t][t]{0}%
\psfrag{x02}[t][t]{0.1}%
\psfrag{x03}[t][t]{0.2}%
\psfrag{x04}[t][t]{0.3}%
\psfrag{x05}[t][t]{0.4}%
\psfrag{x06}[t][t]{0.5}%
\psfrag{x07}[t][t]{0.6}%
\psfrag{x08}[t][t]{0.7}%
\psfrag{x09}[t][t]{0.8}%
\psfrag{x10}[t][t]{0.9}%
\psfrag{x11}[t][t]{1}%
\psfrag{x12}[t][t]{-2}%
\psfrag{x13}[t][t]{-1}%
\psfrag{x14}[t][t]{0}%
\psfrag{x15}[t][t]{1}%
\psfrag{x16}[t][t]{-2}%
\psfrag{x17}[t][t]{-1}%
\psfrag{x18}[t][t]{0}%
\psfrag{x19}[t][t]{1}%
\psfrag{x20}[t][t]{2}%
\psfrag{x21}[t][t]{0}%
\psfrag{x22}[t][t]{10}%
\psfrag{x23}[t][t]{20}%
\psfrag{x24}[t][t]{30}%
%
\psfrag{v01}[r][r]{0}%
\psfrag{v02}[r][r]{0.1}%
\psfrag{v03}[r][r]{0.2}%
\psfrag{v04}[r][r]{0.3}%
\psfrag{v05}[r][r]{0.4}%
\psfrag{v06}[r][r]{0.5}%
\psfrag{v07}[r][r]{0.6}%
\psfrag{v08}[r][r]{0.7}%
\psfrag{v09}[r][r]{0.8}%
\psfrag{v10}[r][r]{0.9}%
\psfrag{v11}[r][r]{1}%
\psfrag{v12}[r][r]{0}%
\psfrag{v13}[r][r]{0.1}%
\psfrag{v14}[r][r]{0.2}%
\psfrag{v15}[r][r]{0.3}%
\psfrag{v16}[r][r]{0.4}%
\psfrag{v17}[r][r]{0.5}%
\psfrag{v18}[r][r]{0.6}%
\psfrag{v19}[r][r]{0.7}%
\psfrag{v20}[r][r]{0.8}%
\psfrag{v21}[r][r]{0.9}%
\psfrag{v22}[r][r]{1}%
\psfrag{v23}[r][r]{0}%
\psfrag{v24}[r][r]{0.1}%
\psfrag{v25}[r][r]{0.2}%
\psfrag{v26}[r][r]{0.3}%
\psfrag{v27}[r][r]{0.4}%
\psfrag{v28}[r][r]{0.5}%
\psfrag{v29}[r][r]{0.6}%
\psfrag{v30}[r][r]{0.7}%
\psfrag{v31}[r][r]{0.8}%
\psfrag{v32}[r][r]{0.9}%
\psfrag{v33}[r][r]{1}%
\psfrag{v34}[r][r]{0}%
\psfrag{v35}[r][r]{5}%
\psfrag{v36}[r][r]{10}%
\psfrag{v37}[r][r]{15}%
%
\resizebox{14.5cm}{!}{\includegraphics{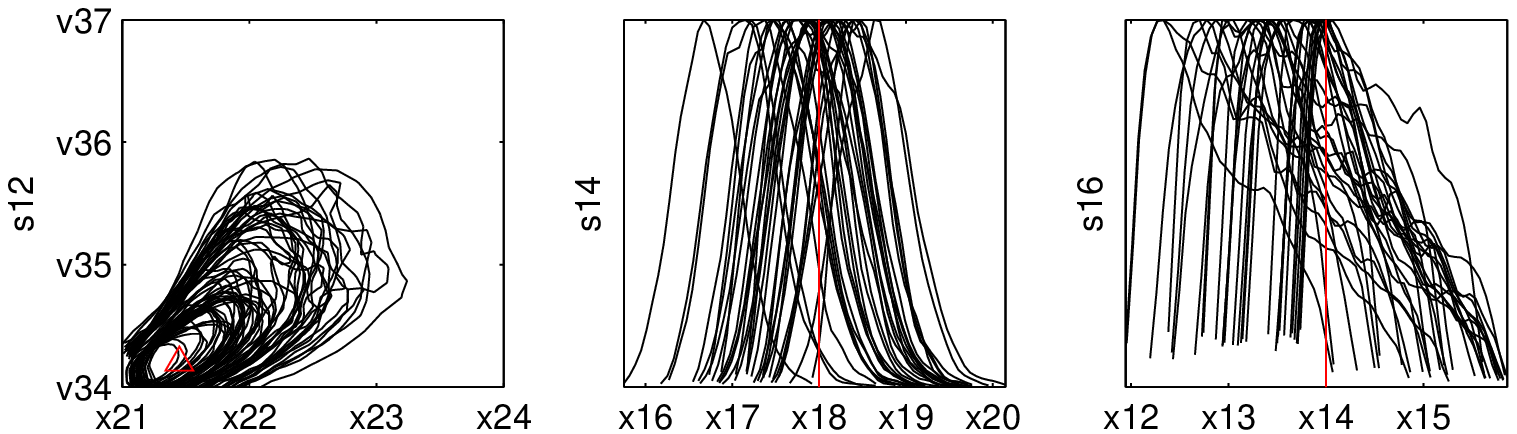}}%
\end{psfrags}%
%
}
    \centerline{
%
%
\begin{psfrags}%
\psfragscanon%
%
\psfrag{s11}[t][t]{\color[rgb]{0,0,0}\setlength{\tabcolsep}{0pt}\begin{tabular}{c}$\theta_s$ / arcmin\end{tabular}}%
\psfrag{s12}[b][b]{}%
\psfrag{s13}[t][t]{\color[rgb]{0,0,0}\setlength{\tabcolsep}{0pt}\begin{tabular}{c}$\alpha - \alpha_{\rm true}$\end{tabular}}%
\psfrag{s14}[b][b]{}%
\psfrag{s15}[t][t]{\color[rgb]{0,0,0}\setlength{\tabcolsep}{0pt}\begin{tabular}{c}$\beta - \beta_{\rm true}$\end{tabular}}%
\psfrag{s16}[b][b]{}%
%
\psfrag{x01}[t][t]{0}%
\psfrag{x02}[t][t]{0.1}%
\psfrag{x03}[t][t]{0.2}%
\psfrag{x04}[t][t]{0.3}%
\psfrag{x05}[t][t]{0.4}%
\psfrag{x06}[t][t]{0.5}%
\psfrag{x07}[t][t]{0.6}%
\psfrag{x08}[t][t]{0.7}%
\psfrag{x09}[t][t]{0.8}%
\psfrag{x10}[t][t]{0.9}%
\psfrag{x11}[t][t]{1}%
\psfrag{x12}[t][t]{-2}%
\psfrag{x13}[t][t]{-1}%
\psfrag{x14}[t][t]{0}%
\psfrag{x15}[t][t]{1}%
\psfrag{x16}[t][t]{-2}%
\psfrag{x17}[t][t]{-1}%
\psfrag{x18}[t][t]{0}%
\psfrag{x19}[t][t]{1}%
\psfrag{x20}[t][t]{0}%
\psfrag{x21}[t][t]{10}%
\psfrag{x22}[t][t]{20}%
%
\psfrag{v01}[r][r]{0}%
\psfrag{v02}[r][r]{0.1}%
\psfrag{v03}[r][r]{0.2}%
\psfrag{v04}[r][r]{0.3}%
\psfrag{v05}[r][r]{0.4}%
\psfrag{v06}[r][r]{0.5}%
\psfrag{v07}[r][r]{0.6}%
\psfrag{v08}[r][r]{0.7}%
\psfrag{v09}[r][r]{0.8}%
\psfrag{v10}[r][r]{0.9}%
\psfrag{v11}[r][r]{1}%
\psfrag{v12}[r][r]{0}%
\psfrag{v13}[r][r]{0.1}%
\psfrag{v14}[r][r]{0.2}%
\psfrag{v15}[r][r]{0.3}%
\psfrag{v16}[r][r]{0.4}%
\psfrag{v17}[r][r]{0.5}%
\psfrag{v18}[r][r]{0.6}%
\psfrag{v19}[r][r]{0.7}%
\psfrag{v20}[r][r]{0.8}%
\psfrag{v21}[r][r]{0.9}%
\psfrag{v22}[r][r]{1}%
\psfrag{v23}[r][r]{0}%
\psfrag{v24}[r][r]{0.1}%
\psfrag{v25}[r][r]{0.2}%
\psfrag{v26}[r][r]{0.3}%
\psfrag{v27}[r][r]{0.4}%
\psfrag{v28}[r][r]{0.5}%
\psfrag{v29}[r][r]{0.6}%
\psfrag{v30}[r][r]{0.7}%
\psfrag{v31}[r][r]{0.8}%
\psfrag{v32}[r][r]{0.9}%
\psfrag{v33}[r][r]{1}%
\psfrag{v34}[r][r]{0}%
\psfrag{v35}[r][r]{5}%
\psfrag{v36}[r][r]{10}%
\psfrag{v37}[r][r]{15}%
%
\resizebox{14.5cm}{!}{\includegraphics{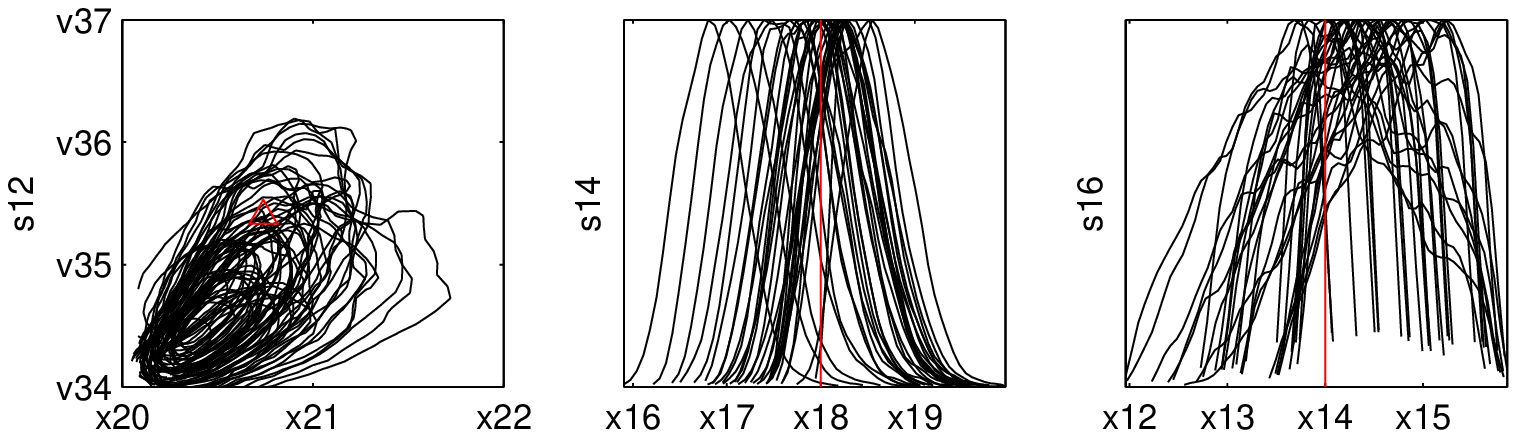}}%
\end{psfrags}%
%
}
    \bigskip
    \centerline{\hskip 0.05 \textwidth (a) \hskip 0.25 \textwidth (b) \hskip 0.25 \textwidth (c)}
    \caption{The posterior distributions for simulated clusters with realistic noise levels (see text for details), and varying GNFW shape parameter values based on the REXCESS sample \citep{bohringer07, arnaud10}.  (a) shows the two-dimensional $\theta_s$ and \Ytot \:posterior, and (b) and (c) show the one-dimensional posteriors for $\alpha$ and $\beta$, shifted to be centred on the appropriate true value.  In all cases $\gamma$ is fixed to the `universal' value, $\alpha$ has a truncated Gaussian prior based on the REXCESS sample, $\beta$ is varied uniformly between 4.5 and 6.5, and the joint two-dimensional prior on \Ytot \:and $\theta_s$ is used.  Results for three different angular sizes are shown (from top to bottom, $\theta_s = 1.8, 4.5$ and 7.4 arcmin); the input parameter values are marked with red triangles and lines.}\label{Fi:rexcess_sims_ab}
\end{figure*}

To attempt to control these biases, we reanalysed the simulations using a Gaussian prior based on the REXCESS sample on $\alpha$, namely $\mathcal{N}(1.0510, 0.47)$ truncated at 0.3, and a tighter uniform prior on $\beta$, $\mathcal{U}[4.5, 6.5]$.  Fig.~\ref{Fi:rexcess_sims_ab} shows the resulting posterior distributions, varying both $\alpha$ and $\beta$ (but with $\gamma$ fixed to the `universal' value).  For the two smaller angular-size clusters, this results in correct recovery of $\theta_{s}$ and \Ytot, and reduces the biassing considerably in $\alpha$ and $\beta$.  For the largest angular-size cluster, the input values of $\theta_{s}$ and \Ytot \:are \emph{not} recovered correctly, because there is not enough information available in the angular scales measured by the SA to constrain these parameters simultaneously, so the prior on \Ytot \:and $\theta_{s}$ biases the recovered posteriors downwards.

\begin{figure}
  \centerline{
%
%
\begin{psfrags}%
\psfragscanon%
\Large
%
\psfrag{s05}[t][t]{\color[rgb]{0,0,0}\setlength{\tabcolsep}{0pt}\begin{tabular}{c}$\gamma$\end{tabular}}%
\psfrag{s06}[b][b]{\color[rgb]{0,0,0}\setlength{\tabcolsep}{0pt}\begin{tabular}{c}$\Delta \theta_s$\end{tabular}}%
\psfrag{s07}[t][t]{\color[rgb]{0,0,0}\setlength{\tabcolsep}{0pt}\begin{tabular}{c}$\gamma$\end{tabular}}%
\psfrag{s08}[b][b]{\color[rgb]{0,0,0}\setlength{\tabcolsep}{0pt}\begin{tabular}{c}$\Delta Y_{\rm tot}$\end{tabular}}%
%
\psfrag{x01}[t][t]{0}%
\psfrag{x02}[t][t]{0.3}%
\psfrag{x03}[t][t]{0.6}%
\psfrag{x04}[t][t]{0.9}%
\psfrag{x05}[t][t]{0}%
\psfrag{x06}[t][t]{0.3}%
\psfrag{x07}[t][t]{0.6}%
\psfrag{x08}[t][t]{0.9}%
%
\psfrag{v01}[r][r]{-1}%
\psfrag{v02}[r][r]{0}%
\psfrag{v03}[r][r]{1}%
\psfrag{v04}[r][r]{-0.5}%
\psfrag{v05}[r][r]{0}%
\psfrag{v06}[r][r]{0.5}%
\psfrag{v07}[r][r]{1}%
%
\resizebox{\linewidth}{!}{\includegraphics{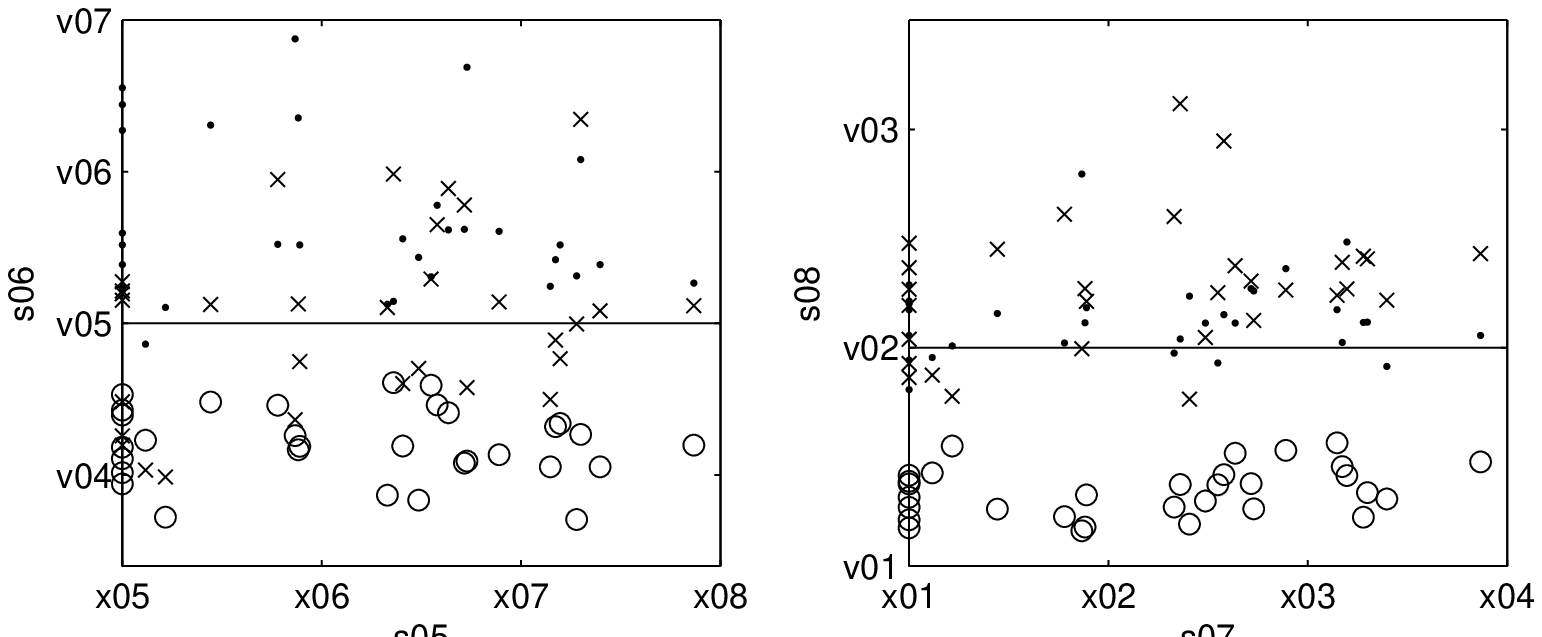}}%
\end{psfrags}%
%
}\medskip
  \centerline{(a) \hskip 0.4\linewidth (b)}
  \caption{The fractional difference [(MAP value - true value)/(true value)] in $\theta_s$ (a) and \Ytot \:(b) as a function of the input value of $\gamma$.  Clusters with $\theta_{s} = 1.8$ are plotted as dots, $\theta_s = 4.5$ as crosses and $\theta_s = 7.4$ as open circles.}\label{Fi:rexcess_sims_gamma}
\end{figure}

We check for any biases due to $\gamma$ being fixed (incorrectly) to the `universal' value by plotting the error in the recovered values of $\theta_{s}$ and \Ytot \:as a function of the true input $\gamma$ value.  There is some correlation between the fractional difference in $\theta_s$ and $\gamma$, especially for the two smaller clusters, but mostly any correlation is beneath the level of the noise (Fig.~\ref{Fi:rexcess_sims_gamma}).

We also add point sources of varying flux densities and at varying distances from the phase centre to test for any issues in decorrelating point source flux from cluster flux when varying the shape parameters; the parameter estimation is unaffected.

\subsubsection{Adding \emph{Planck} information}

\begin{figure*}
    \centerline{\input{figs/rexcess_sim_CAJ0441py.tex}}
    \centerline{
%
%
\begin{psfrags}%
\psfragscanon%
%
\psfrag{s11}[t][t]{}%
\psfrag{s12}[b][b]{\color[rgb]{0,0,0}\setlength{\tabcolsep}{0pt}\begin{tabular}{c}$Y_{tot} \times 10^3$ / arcmin$^2$\end{tabular}}%
\psfrag{s13}[t][t]{}%
\psfrag{s14}[b][b]{\color[rgb]{0,0,0}\setlength{\tabcolsep}{0pt}\begin{tabular}{c}Probability density\end{tabular}}%
\psfrag{s15}[t][t]{}%
\psfrag{s16}[b][b]{\color[rgb]{0,0,0}\setlength{\tabcolsep}{0pt}\begin{tabular}{c}Probability density\end{tabular}}%
%
\psfrag{x01}[t][t]{0}%
\psfrag{x02}[t][t]{0.1}%
\psfrag{x03}[t][t]{0.2}%
\psfrag{x04}[t][t]{0.3}%
\psfrag{x05}[t][t]{0.4}%
\psfrag{x06}[t][t]{0.5}%
\psfrag{x07}[t][t]{0.6}%
\psfrag{x08}[t][t]{0.7}%
\psfrag{x09}[t][t]{0.8}%
\psfrag{x10}[t][t]{0.9}%
\psfrag{x11}[t][t]{1}%
\psfrag{x12}[t][t]{-2}%
\psfrag{x13}[t][t]{-1}%
\psfrag{x14}[t][t]{0}%
\psfrag{x15}[t][t]{1}%
\psfrag{x16}[t][t]{-2}%
\psfrag{x17}[t][t]{-1}%
\psfrag{x18}[t][t]{0}%
\psfrag{x19}[t][t]{1}%
\psfrag{x20}[t][t]{2}%
\psfrag{x21}[t][t]{0}%
\psfrag{x22}[t][t]{10}%
\psfrag{x23}[t][t]{20}%
%
\psfrag{v01}[r][r]{0}%
\psfrag{v02}[r][r]{0.1}%
\psfrag{v03}[r][r]{0.2}%
\psfrag{v04}[r][r]{0.3}%
\psfrag{v05}[r][r]{0.4}%
\psfrag{v06}[r][r]{0.5}%
\psfrag{v07}[r][r]{0.6}%
\psfrag{v08}[r][r]{0.7}%
\psfrag{v09}[r][r]{0.8}%
\psfrag{v10}[r][r]{0.9}%
\psfrag{v11}[r][r]{1}%
\psfrag{v12}[r][r]{0}%
\psfrag{v13}[r][r]{0.1}%
\psfrag{v14}[r][r]{0.2}%
\psfrag{v15}[r][r]{0.3}%
\psfrag{v16}[r][r]{0.4}%
\psfrag{v17}[r][r]{0.5}%
\psfrag{v18}[r][r]{0.6}%
\psfrag{v19}[r][r]{0.7}%
\psfrag{v20}[r][r]{0.8}%
\psfrag{v21}[r][r]{0.9}%
\psfrag{v22}[r][r]{1}%
\psfrag{v23}[r][r]{0}%
\psfrag{v24}[r][r]{0.1}%
\psfrag{v25}[r][r]{0.2}%
\psfrag{v26}[r][r]{0.3}%
\psfrag{v27}[r][r]{0.4}%
\psfrag{v28}[r][r]{0.5}%
\psfrag{v29}[r][r]{0.6}%
\psfrag{v30}[r][r]{0.7}%
\psfrag{v31}[r][r]{0.8}%
\psfrag{v32}[r][r]{0.9}%
\psfrag{v33}[r][r]{1}%
\psfrag{v34}[r][r]{0}%
\psfrag{v35}[r][r]{2}%
\psfrag{v36}[r][r]{4}%
%
\resizebox{14.5cm}{!}{\includegraphics{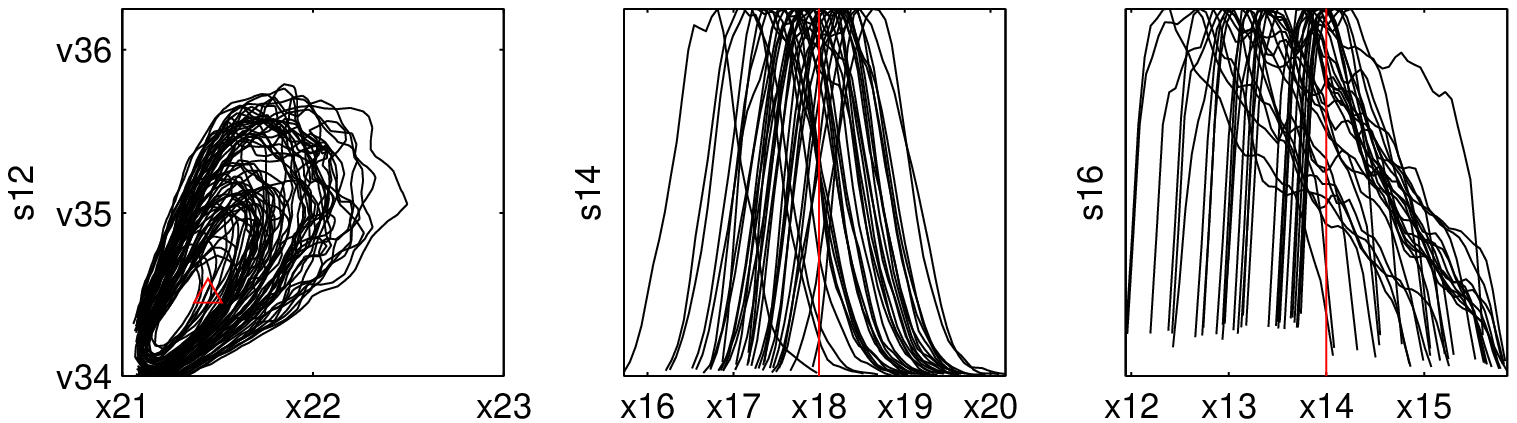}}%
\end{psfrags}%
%
}
    \centerline{
%
%
\begin{psfrags}%
\psfragscanon%
%
\psfrag{s11}[t][t]{\color[rgb]{0,0,0}\setlength{\tabcolsep}{0pt}\begin{tabular}{c}$\theta_s$ / arcmin\end{tabular}}%
\psfrag{s12}[b][b]{}%
\psfrag{s13}[t][t]{\color[rgb]{0,0,0}\setlength{\tabcolsep}{0pt}\begin{tabular}{c}$\alpha - \alpha_{\rm true}$\end{tabular}}%
\psfrag{s14}[b][b]{}%
\psfrag{s15}[t][t]{\color[rgb]{0,0,0}\setlength{\tabcolsep}{0pt}\begin{tabular}{c}$\beta - \beta_{\rm true}$\end{tabular}}%
\psfrag{s16}[b][b]{}%
%
\psfrag{x01}[t][t]{0}%
\psfrag{x02}[t][t]{0.1}%
\psfrag{x03}[t][t]{0.2}%
\psfrag{x04}[t][t]{0.3}%
\psfrag{x05}[t][t]{0.4}%
\psfrag{x06}[t][t]{0.5}%
\psfrag{x07}[t][t]{0.6}%
\psfrag{x08}[t][t]{0.7}%
\psfrag{x09}[t][t]{0.8}%
\psfrag{x10}[t][t]{0.9}%
\psfrag{x11}[t][t]{1}%
\psfrag{x12}[t][t]{-2}%
\psfrag{x13}[t][t]{-1}%
\psfrag{x14}[t][t]{0}%
\psfrag{x15}[t][t]{1}%
\psfrag{x16}[t][t]{-2}%
\psfrag{x17}[t][t]{-1}%
\psfrag{x18}[t][t]{0}%
\psfrag{x19}[t][t]{1}%
\psfrag{x20}[t][t]{0}%
\psfrag{x21}[t][t]{10}%
\psfrag{x22}[t][t]{20}%
\psfrag{x23}[t][t]{30}%
%
\psfrag{v01}[r][r]{0}%
\psfrag{v02}[r][r]{0.1}%
\psfrag{v03}[r][r]{0.2}%
\psfrag{v04}[r][r]{0.3}%
\psfrag{v05}[r][r]{0.4}%
\psfrag{v06}[r][r]{0.5}%
\psfrag{v07}[r][r]{0.6}%
\psfrag{v08}[r][r]{0.7}%
\psfrag{v09}[r][r]{0.8}%
\psfrag{v10}[r][r]{0.9}%
\psfrag{v11}[r][r]{1}%
\psfrag{v12}[r][r]{0}%
\psfrag{v13}[r][r]{0.1}%
\psfrag{v14}[r][r]{0.2}%
\psfrag{v15}[r][r]{0.3}%
\psfrag{v16}[r][r]{0.4}%
\psfrag{v17}[r][r]{0.5}%
\psfrag{v18}[r][r]{0.6}%
\psfrag{v19}[r][r]{0.7}%
\psfrag{v20}[r][r]{0.8}%
\psfrag{v21}[r][r]{0.9}%
\psfrag{v22}[r][r]{1}%
\psfrag{v23}[r][r]{0}%
\psfrag{v24}[r][r]{0.1}%
\psfrag{v25}[r][r]{0.2}%
\psfrag{v26}[r][r]{0.3}%
\psfrag{v27}[r][r]{0.4}%
\psfrag{v28}[r][r]{0.5}%
\psfrag{v29}[r][r]{0.6}%
\psfrag{v30}[r][r]{0.7}%
\psfrag{v31}[r][r]{0.8}%
\psfrag{v32}[r][r]{0.9}%
\psfrag{v33}[r][r]{1}%
\psfrag{v34}[r][r]{5}%
\psfrag{v35}[r][r]{7}%
\psfrag{v36}[r][r]{9}%
%
\resizebox{14.5cm}{!}{\includegraphics{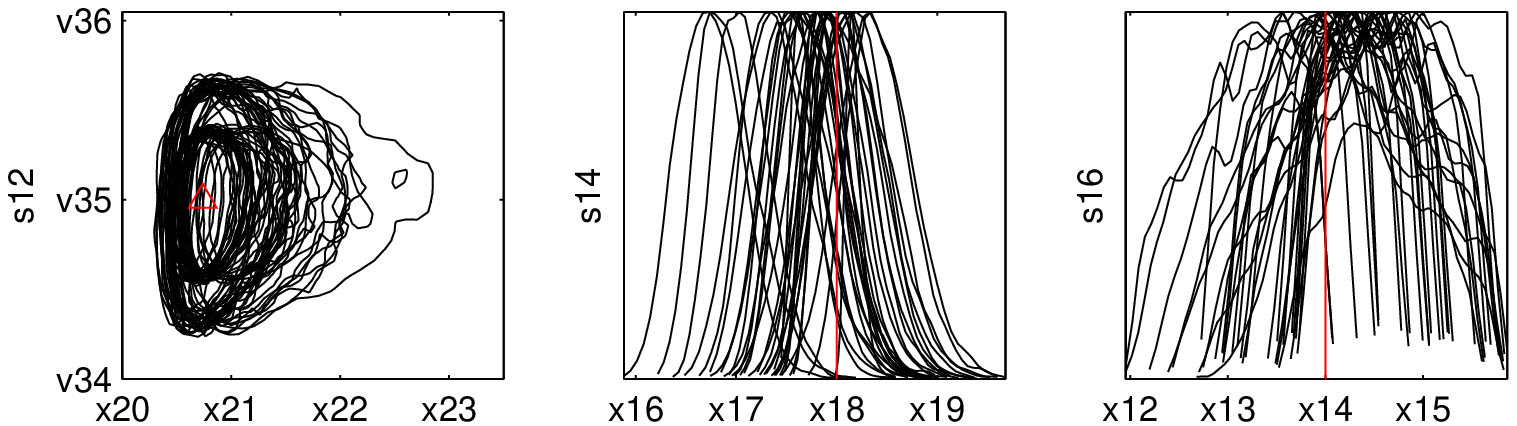}}%
\end{psfrags}%
%
}
    \bigskip
    \centerline{\hskip 0.05 \textwidth (a) \hskip 0.25 \textwidth (b) \hskip 0.25 \textwidth (c)}
    \caption{The posterior distributions for simulated clusters with realistic noise levels (see text for details), and varying GNFW shape parameter values based on the REXCESS sample \citep{bohringer07, arnaud10}.  (a) shows the two-dimensional $\theta_s$ and \Ytot \:posterior, and (b) and (c) show the one-dimensional posteriors for $\alpha$ and $\beta$, shifted to be centred on the appropriate true value.  In all cases $\gamma$ is fixed to the `universal' value, $\alpha$ has a truncated Gaussian prior based on the REXCESS sample, $\beta$ is varied uniformly between 4.5 and 6.5, a \emph{Planck}-like Gaussian prior is used on \Ytot \:and $\theta_s$ has the conditional prior drawn from the two-dimensional prior.  Results for three different angular sizes are shown (from top to bottom, $\theta_s = 1.8, 4.5$ and 7.4); the input parameter values are marked with red triangles and lines.}\label{Fi:rexcess_sims_py}
\end{figure*}

Although the immediate issue is to check whether we can achieve consistency between AMI and \emph{Planck} results, it is also interesting to consider whether we can take advantage of the complementary nature of the two instruments to derive better constraints on the behaviour of the pressure profile over a range of radii.  To this end, for each of our three simulated cluster sizes we derived a \emph{Planck}-like prior on \Ytot \:by marginalising over the $\theta_{s}$ dimension of the two-dimensional constraint produced by \emph{Planck} for a cluster with similar angular size, and approximating as a Gaussian.  We use this marginalised constraint as a prior rather than the full two-dimensional constraint since \emph{Planck} \Ytot \:estimation is more robust to changes in the profile shape parameters \citep{Harrison2015}.  We then use our standard two-dimensional prior on $\theta_{s}$ conditioned on values drawn from the \emph{Planck}-like \Ytot \:prior; priors on $\alpha$ and $\beta$ are as in the previous section.  Fig.~\ref{Fi:rexcess_sims_py} shows the resulting posterior distributions.  For all three clusters, the constraints on $\theta_{s}$ and \Ytot \:are much tighter, and for the large angular-size cluster, the true values of $\theta_{s}$ and \Ytot \:are now recovered correctly.  However, the constraints on the shape parameters are not very different.

This is a fairly crude way of including \emph{Planck} information in the analysis and does not make the best use of the information available in the \emph{Planck} data on the cluster shape.  A full joint analysis of AMI and \emph{Planck} data would fill in the gap in $uv$-coverage between the zero-spacing flux and the shortest AMI-SA baselines, and there would be some overlap with the shortest baselines since the resolution of \emph{Planck} is $\approx$\,5\,arcmin; this should produce better constraints on the profile shape parameters.  This will be addressed in a future paper.

\subsection{Summary of simulation results}

We have shown with the simulated bank of clusters based on the REXCESS sample, that when a cluster has an angular size $\theta_{s} \gtrapprox 5$\,arcmin, the true input values of $\theta_{s}$ and \Ytot \:can only be recovered correctly using AMI data when the model for the pressure profile used for parameter extraction is a good match to the actual pressure profile of the cluster.  This is not surprising since, as we have mentioned, an interferometer does not measure zero-spacing flux directly and so the \Ytot \:value `measured' by AMI is actually an extrapolation based on the assumed profile.  This is also consistent with what we observe in the real sample; clusters with high \emph{Planck} SNR (and therefore large $\theta_{s}$) are consistently measured to be smaller and fainter by AMI.

When attempting to vary the GNFW shape parameters, we must be careful to avoid over-interpretation of apparent constraints on parameters which are actually just caused by the shape of the two-dimensional degeneracies.  Reducing the range of $\beta$ and imposing a prior based on the REXCESS sample on $\alpha$ reduces these problems significantly.  However, it is clear from Fig.~\ref{Fi:rexcess_sims_ab} that in some cases these spurious constraints still do occur, particularly in $\alpha$ for small angular-size clusters, and $\beta$ for medium angular-size clusters.  Surprisingly, $\beta$ is often recovered correctly for large angular-size clusters -- this is due to the intersection of the physically motivated prior on $\theta_{s}$ and \Ytot \:and the degeneracy direction between $\theta_{s}$ and $\beta$.

It is also clear from Fig.~\ref{Fi:rexcess_sims_ab} that varying the shape parameters does not aid in recovering the correct $\theta_{s}$ and \Ytot \:values for large angular-size clusters; joint analysis of \emph{Planck} and AMI data is required to achieve this.  As a first approximation, using a \emph{Planck}-derived prior on \Ytot \:can help, but does not improve the constraints on $\alpha$ and $\beta$.

It is also interesting to note that our parameter constraints are not very reliant on noise level.  Our initial tests were made on simulated data with unrealistically small noise levels of 100\,$\mu$Jy per visibility; when we moved to simulations with more realistic noise levels (of $\approx$\,120\,$\mu$Jy\,beam$^{-1}$ across the channel-averaged map), the constraints changed very little.  As long as one has a good detection, it seems that the limiting factor on our parameter constraints is very much the range of angular scales present in the data with respect to the size of the cluster, rather than the detection significance.

\subsection{Analysis of real data}\label{S:real_data_vary_ab}

For all the clear detections in the sample, we re-extract the cluster parameters allowing $\alpha$ and $\beta$ to vary as described in Section~\ref{S:simulation}.  The constraints on \Ytot \:and $\theta_{s}$ are on the whole broader but the positions of the maxima are unchanged.  The full two-dimensional constraints for the whole sample are available online at \url{http://www.astro.phy.cam.ac.uk/surveys/ami-planck/}; here we present a few examples.

\subsubsection{Abell 1413 (PSZ1~G226.19+76.78)}

\begin{figure}
  \fbox{
%
%
\begin{psfrags}%
\psfragscanon%
%
\psfrag{s09}[t][t]{\color[rgb]{0,0,0}\setlength{\tabcolsep}{0pt}\begin{tabular}{c}$\beta$\end{tabular}}%
\psfrag{s11}[t][t]{\color[rgb]{0,0,0}\setlength{\tabcolsep}{0pt}\begin{tabular}{c}$Y_{\rm tot}$\\$\times 10^3$\end{tabular}}%
\psfrag{s13}[t][t]{\color[rgb]{0,0,0}\setlength{\tabcolsep}{0pt}\begin{tabular}{c}$\alpha$\end{tabular}}%
\psfrag{s15}[t][t]{\color[rgb]{0,0,0}\setlength{\tabcolsep}{0pt}\begin{tabular}{c}$\theta_s$\end{tabular}}%
\psfrag{s16}[t][t]{\color[rgb]{0,0,0}\setlength{\tabcolsep}{0pt}\begin{tabular}{c}$\beta$\end{tabular}}%
\psfrag{s19}[t][t]{\color[rgb]{0,0,0}\setlength{\tabcolsep}{0pt}\begin{tabular}{c}$Y_{\rm tot} \times 10^3$\end{tabular}}%
\psfrag{s21}[t][t]{\color[rgb]{0,0,0}\setlength{\tabcolsep}{0pt}\begin{tabular}{c}$\alpha$\end{tabular}}%
\psfrag{s28}[][]{\color[rgb]{0,0,0}\setlength{\tabcolsep}{0pt}\begin{tabular}{c} \end{tabular}}%
\psfrag{s29}[][]{\color[rgb]{0,0,0}\setlength{\tabcolsep}{0pt}\begin{tabular}{c} \end{tabular}}%
%
\psfrag{x01}[t][t]{0}%
\psfrag{x02}[t][t]{0.1}%
\psfrag{x03}[t][t]{0.2}%
\psfrag{x04}[t][t]{0.3}%
\psfrag{x05}[t][t]{0.4}%
\psfrag{x06}[t][t]{0.5}%
\psfrag{x07}[t][t]{0.6}%
\psfrag{x08}[t][t]{0.7}%
\psfrag{x09}[t][t]{0.8}%
\psfrag{x10}[t][t]{0.9}%
\psfrag{x11}[t][t]{1}%
\psfrag{x12}[t][t]{0}%
\psfrag{x13}[t][t]{0.5}%
\psfrag{x14}[t][t]{1}%
\psfrag{x15}[t][t]{1}%
\psfrag{x16}[t][t]{2}%
\psfrag{x17}[t][t]{2}%
\psfrag{x18}[t][t]{4}%
\psfrag{x19}[t][t]{0}%
\psfrag{x20}[t][t]{0.5}%
\psfrag{x21}[t][t]{1}%
\psfrag{x22}[t][t]{5}%
\psfrag{x23}[t][t]{10}%
\psfrag{x24}[t][t]{0}%
\psfrag{x25}[t][t]{0.5}%
\psfrag{x26}[t][t]{1}%
\psfrag{x27}[t][t]{0}%
\psfrag{x28}[t][t]{0.5}%
\psfrag{x29}[t][t]{1}%
\psfrag{x30}[t][t]{4}%
\psfrag{x31}[t][t]{5}%
\psfrag{x32}[t][t]{6}%
\psfrag{x33}[t][t]{0}%
\psfrag{x34}[t][t]{0.5}%
\psfrag{x35}[t][t]{1}%
\psfrag{x36}[t][t]{0}%
\psfrag{x37}[t][t]{0.5}%
\psfrag{x38}[t][t]{1}%
\psfrag{x39}[t][t]{0}%
\psfrag{x40}[t][t]{0.5}%
\psfrag{x41}[t][t]{1}%
%
\psfrag{v01}[r][r]{0}%
\psfrag{v02}[r][r]{0.1}%
\psfrag{v03}[r][r]{0.2}%
\psfrag{v04}[r][r]{0.3}%
\psfrag{v05}[r][r]{0.4}%
\psfrag{v06}[r][r]{0.5}%
\psfrag{v07}[r][r]{0.6}%
\psfrag{v08}[r][r]{0.7}%
\psfrag{v09}[r][r]{0.8}%
\psfrag{v10}[r][r]{0.9}%
\psfrag{v11}[r][r]{1}%
\psfrag{v12}[r][r]{0}%
\psfrag{v13}[r][r]{0.5}%
\psfrag{v14}[r][r]{1}%
\psfrag{v15}[r][r]{0}%
\psfrag{v16}[r][r]{0.5}%
\psfrag{v17}[r][r]{1}%
\psfrag{v18}[r][r]{0}%
\psfrag{v19}[r][r]{0.5}%
\psfrag{v20}[r][r]{1}%
\psfrag{v21}[r][r]{0}%
\psfrag{v22}[r][r]{0.5}%
\psfrag{v23}[r][r]{1}%
\psfrag{v24}[r][r]{4}%
\psfrag{v25}[r][r]{5}%
\psfrag{v26}[r][r]{6}%
\psfrag{v27}[r][r]{1}%
\psfrag{v28}[r][r]{2}%
\psfrag{v29}[r][r]{2}%
\psfrag{v30}[r][r]{4}%
\psfrag{v31}[r][r]{0}%
\psfrag{v32}[r][r]{0.5}%
\psfrag{v33}[r][r]{1}%
\psfrag{v34}[r][r]{0}%
\psfrag{v35}[r][r]{0.5}%
\psfrag{v36}[r][r]{1}%
\psfrag{v37}[r][r]{0}%
\psfrag{v38}[r][r]{5}%
\psfrag{v39}[r][r]{0}%
\psfrag{v40}[r][r]{0.5}%
\psfrag{v41}[r][r]{1}%
%
\resizebox{\linewidth}{!}{\includegraphics{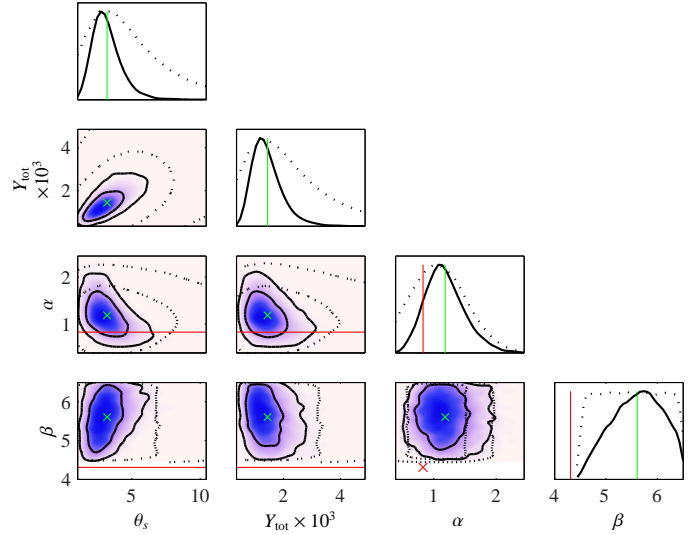}}%
\end{psfrags}%
%
}
  \caption{AMI posterior distributions for A1413, allowing $\alpha$ and $\beta$ to vary.  Posterior means are indicated with green lines and crosses, and the \emph{Planck} $+$ \emph{XMM-Newton} estimates of $\alpha$ and $\beta$ from \citet{planck2012-V} are shown with red lines and crosses.  The priors on the parameters in the AMI analysis are shown as black dashed lines.  $\theta_s$ is in arcmin and {\Ytot} is in arcmin$^2$.}\label{Fi:A1413_post}
\end{figure}

Abell 1413 is well-detected by AMI, with an evidence ratio of $\Delta \ln(\mathcal{Z}) = 26$, and \emph{Planck}, with a PwS SNR of 9.8 and detections by all three algorithms.  It is at redshift $z = 0.143$ (e.g.\ \citealt{1987ApJS...63..543S}) so could be expected to have a large angular size; the $\theta_{500}$ value inferred from the X-ray luminosity is $\approx$\,7.9\,arcmin (\citealt{bohringer00}, \citealt{2011A&A...534A.109P}), corresponding to $\theta_{s} \approx$\,6.7\,arcmin for $c_{500} = 1.177$.  The AMI constraints on $\theta_{s}$ and \Ytot \:could therefore be expected to be biased downward if the profile differs from the `universal' profile.  Indeed, the \emph{Planck} constraints indicate much higher values of both (see Fig.~\ref{Fi:all_post} under the \emph{Planck} name of PSZ1~G226.19+76.78).  From the simulation results we can therefore expect to produce some constraints on $\alpha$ and $\beta$ from the AMI data, although not to recover the correct values of $\theta_s$ and \Ytot; the posterior distributions for the real data are shown in Fig.~\ref{Fi:A1413_post}.  

\begin{figure}
  \centerline{\includegraphics[bb=306 282 558 521, clip=, width=0.8\linewidth]{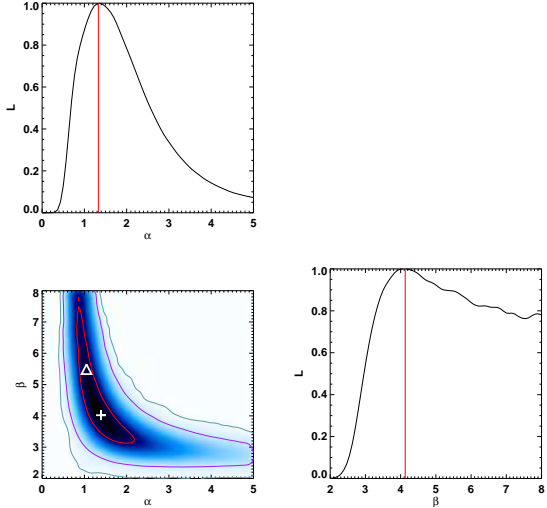}}
  \caption{Marginalised posterior likelihood distribution for $\alpha$ and $\beta$ based on stacked \emph{Planck} and \emph{XMM-Newton} data for a sample of high-SNR \emph{Planck} clusters (from \citealt{planck2012-V}).  The white cross marks the position of the best-fit value, and the white triangle marks the `universal' values.}\label{Fi:planck_like}
\end{figure}

In \citet{planck2012-V}, \emph{Planck} and \emph{XMM-Newton} data were used to produce fitted values for $\alpha$ and $\beta$ for a sample of high-SNR \emph{Planck} clusters.  The sample includes seven of the clear detections in our SZ sample (however we note that for three of these, \citet{planck2012-V} report non-physical values for ($\gamma, \alpha, \beta$) producing negative values of \Ytot \:because of the $\Gamma$ functions in Equation~\ref{Eq:Ytot}).  Their reported values for Abell 1413 are $\alpha = 0.83$ and $\beta = 4.31$ ($\gamma$ fixed at 0.31), which are plotted for comparison in Fig.~\ref{Fi:A1413_post}.  The AMI analysis produces a somewhat higher (but consistent) value for $\alpha$; although the \emph{Planck} $\beta$ estimate is outside our prior range for $\beta$, our analysis shows no tendency to push toward the lower limit, toward the \emph{Planck} value.  However, assuming the shape of the \emph{Planck} $\alpha$-$\beta$ degeneracy for the individual clusters is similar to that for their stacked profile (reproduced in Fig.~\ref{Fi:planck_like}), the AMI and \emph{Planck} constraints on $\beta$ could be consistent.

\subsubsection{RXC~J2228.6+2036 (PSZ1~G083.30-31.01)}

\begin{figure}
  \fbox{
%
%
\begin{psfrags}%
\psfragscanon%
%
\psfrag{s02}[t][t]{\color[rgb]{0,0,0}\setlength{\tabcolsep}{0pt}\begin{tabular}{c}$\beta$\end{tabular}}%
\psfrag{s03}[t][t]{\color[rgb]{0,0,0}\setlength{\tabcolsep}{0pt}\begin{tabular}{c}$Y_{\rm tot}$\\$\times 10^3$\end{tabular}}%
\psfrag{s04}[t][t]{\color[rgb]{0,0,0}\setlength{\tabcolsep}{0pt}\begin{tabular}{c}$\alpha$\end{tabular}}%
\psfrag{s05}[t][t]{\color[rgb]{0,0,0}\setlength{\tabcolsep}{0pt}\begin{tabular}{c}$\theta_s$\end{tabular}}%
\psfrag{s06}[t][t]{\color[rgb]{0,0,0}\setlength{\tabcolsep}{0pt}\begin{tabular}{c}$\beta$\end{tabular}}%
\psfrag{s07}[t][t]{\color[rgb]{0,0,0}\setlength{\tabcolsep}{0pt}\begin{tabular}{c}$Y_{\rm tot} \times 10^3$\end{tabular}}%
\psfrag{s08}[t][t]{\color[rgb]{0,0,0}\setlength{\tabcolsep}{0pt}\begin{tabular}{c}$\alpha$\end{tabular}}%
%
\psfrag{x01}[t][t]{0}%
\psfrag{x02}[t][t]{0.1}%
\psfrag{x03}[t][t]{0.2}%
\psfrag{x04}[t][t]{0.3}%
\psfrag{x05}[t][t]{0.4}%
\psfrag{x06}[t][t]{0.5}%
\psfrag{x07}[t][t]{0.6}%
\psfrag{x08}[t][t]{0.7}%
\psfrag{x09}[t][t]{0.8}%
\psfrag{x10}[t][t]{0.9}%
\psfrag{x11}[t][t]{1}%
\psfrag{x12}[t][t]{1}%
\psfrag{x13}[t][t]{2}%
\psfrag{x14}[t][t]{1}%
\psfrag{x15}[t][t]{2}%
\psfrag{x16}[t][t]{3}%
\psfrag{x17}[t][t]{1}%
\psfrag{x18}[t][t]{2}%
\psfrag{x19}[t][t]{3}%
\psfrag{x20}[t][t]{2}%
\psfrag{x21}[t][t]{4}%
\psfrag{x22}[t][t]{6}%
\psfrag{x23}[t][t]{2}%
\psfrag{x24}[t][t]{2}%
\psfrag{x25}[t][t]{4}%
\psfrag{x26}[t][t]{6}%
\psfrag{x27}[t][t]{4.5}%
\psfrag{x28}[t][t]{5.5}%
\psfrag{x29}[t][t]{6.5}%
\psfrag{x30}[t][t]{0.5}%
\psfrag{x31}[t][t]{1}%
\psfrag{x32}[t][t]{1.5}%
\psfrag{x33}[t][t]{2}%
\psfrag{x34}[t][t]{1}%
\psfrag{x35}[t][t]{2}%
\psfrag{x36}[t][t]{3}%
\psfrag{x37}[t][t]{2}%
\psfrag{x38}[t][t]{4}%
\psfrag{x39}[t][t]{6}%
%
\psfrag{v01}[r][r]{0}%
\psfrag{v02}[r][r]{0.1}%
\psfrag{v03}[r][r]{0.2}%
\psfrag{v04}[r][r]{0.3}%
\psfrag{v05}[r][r]{0.4}%
\psfrag{v06}[r][r]{0.5}%
\psfrag{v07}[r][r]{0.6}%
\psfrag{v08}[r][r]{0.7}%
\psfrag{v09}[r][r]{0.8}%
\psfrag{v10}[r][r]{0.9}%
\psfrag{v11}[r][r]{1}%
\psfrag{v12}[r][r]{4.5}%
\psfrag{v13}[r][r]{5}%
\psfrag{v14}[r][r]{5.5}%
\psfrag{v15}[r][r]{6}%
\psfrag{v16}[r][r]{6.5}%
\psfrag{v17}[r][r]{4.5}%
\psfrag{v18}[r][r]{5}%
\psfrag{v19}[r][r]{5.5}%
\psfrag{v20}[r][r]{6}%
\psfrag{v21}[r][r]{6.5}%
\psfrag{v22}[r][r]{0.5}%
\psfrag{v23}[r][r]{1}%
\psfrag{v24}[r][r]{1.5}%
\psfrag{v25}[r][r]{2}%
\psfrag{v26}[r][r]{4.5}%
\psfrag{v27}[r][r]{5.5}%
\psfrag{v28}[r][r]{6.5}%
\psfrag{v29}[r][r]{1}%
\psfrag{v30}[r][r]{2}%
\psfrag{v31}[r][r]{1}%
\psfrag{v32}[r][r]{2}%
\psfrag{v33}[r][r]{3}%
\psfrag{v34}[r][r]{0}%
\psfrag{v35}[r][r]{0.5}%
\psfrag{v36}[r][r]{1}%
\psfrag{v37}[r][r]{0}%
\psfrag{v38}[r][r]{0.5}%
\psfrag{v39}[r][r]{1}%
\psfrag{v40}[r][r]{0}%
\psfrag{v41}[r][r]{0.5}%
\psfrag{v42}[r][r]{1}%
\psfrag{v43}[r][r]{0}%
\psfrag{v44}[r][r]{0.5}%
\psfrag{v45}[r][r]{1}%
%
\resizebox{0.95\linewidth}{!}{\includegraphics{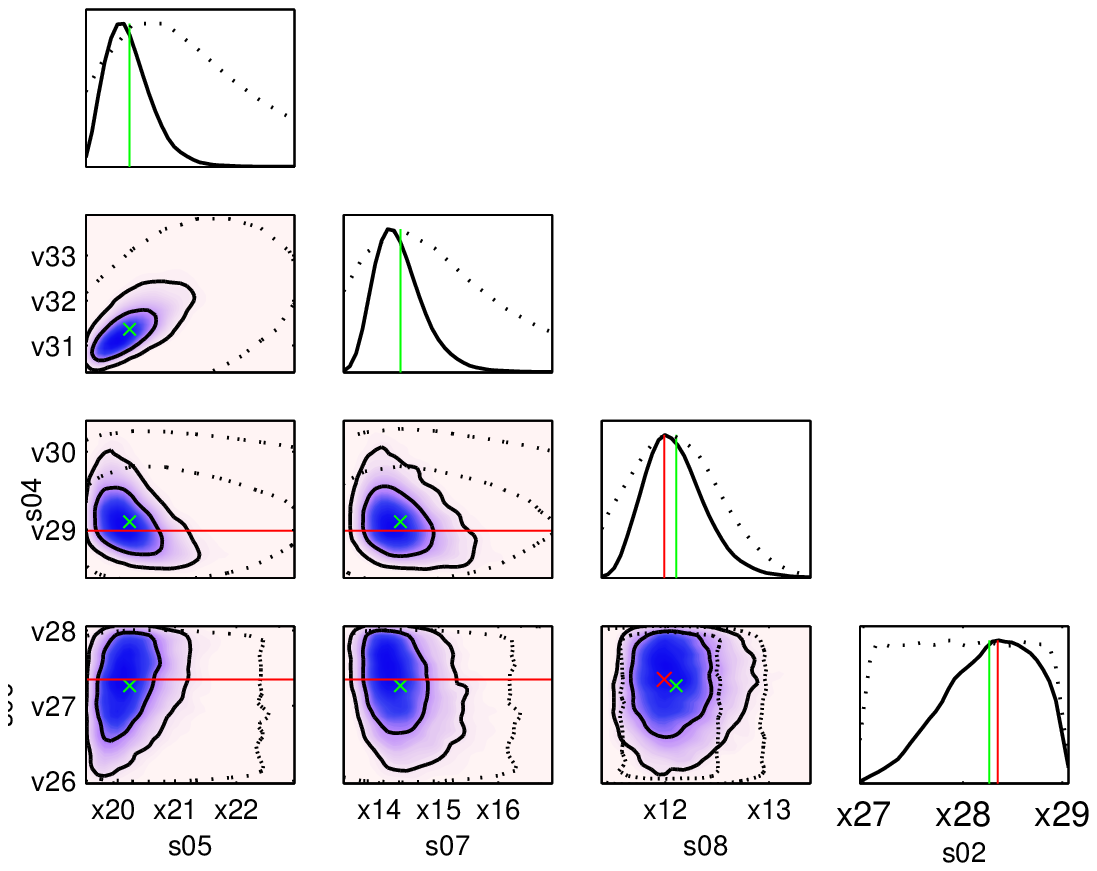}}%
\end{psfrags}%
%
}\hspace{-0.42\linewidth}
  \fbox{\raisebox{\height}{\includegraphics[width=0.4\linewidth]{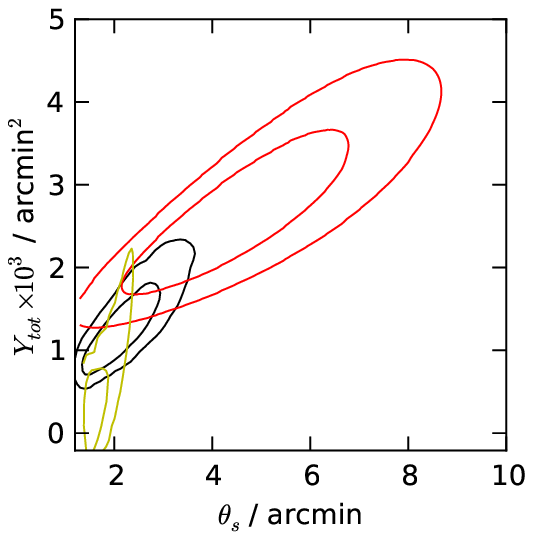}}}
  \caption{AMI posterior distributions for RXC~J2228.6+2036, allowing $\alpha$ and $\beta$ to vary.  Posterior means are indicated with green lines and crosses, and the \emph{Planck} values for $\alpha$ and $\beta$ from \citet{planck2012-V} are shown with red lines and crosses.  The priors on the parameters in the AMI analysis are shown as black dashed lines.  $\theta_s$ is in arcmin and {\Ytot} is in arcmin$^2$.  Also shown in the upper right hand corner are the posteriors produced by AMI (black) and \emph{Planck} (red) using the `universal' profile, and a prediction produced by the physical model described in \citet{2012MNRAS.423.1534O} based on a redshift of $z=0.412$ (yellow).}\label{Fi:CAJ2228_post}
\end{figure}

Similarly to Abell 1413, RXC~J2228.6+2036 is well-detected by AMI ($\Delta \ln(\mathcal{Z}) = 28$) and \emph{Planck} (SNR = 7.3, detected by all algorithms).  It is at higher redshift, $z = 0.412$ \citep{bohringer00} so the large value for $\theta_{s}$ of $\approx$\,4.5\,arcmin preferred by \emph{Planck} is slightly surprising.  Fig.~\ref{Fi:CAJ2228_post} shows the posteriors on $\theta_{s}$ and \Ytot \:produced by AMI and \emph{Planck}, as well as the region of the space predicted by the physical model described in \citet{2012MNRAS.423.1534O} assuming the `universal' pressure profile for the gas, and the Tinker mass function \citep{2008ApJ...688..709T}.  The AMI posterior is much more consistent with the prediction than the \emph{Planck} posterior; also our simulations have shown that if the correct value were $\theta_{s} \approx 4.5$, we should recover it even if the profile deviates from the `universal' profile.  In addition, $\theta_{500}$ determined from the X-ray luminosity is 3.9\,arcmin (\citealt{bohringer00}, \citealt{2011A&A...534A.109P}), corresponding to $\theta_{s} = 3.3$\,arcmin for $c_{500} = 1.177$ is consistent with the AMI mean value of 2.3\,arcmin.  We therefore conclude that in this case the \emph{Planck} $\theta_{s}$ estimate is likely to be an over-estimate.

Fig.~\ref{Fi:CAJ2228_post} also shows the posteriors on $\alpha$ and $\beta$ resulting from the AMI analysis.  Assuming the AMI value of $\theta_{s}$ is correct, we should be able to produce some constraint on $\beta$; indeed, there is a weak preference for higher values of $\beta$, while the posterior distribution for $\alpha$ mostly recovers the prior.  The fitted $\alpha$ and $\beta$ values from \citet{planck2012-V} are also shown and in this case are very consistent with the AMI constraints.

\subsection{PSZ1~G134.31-06.57}

\begin{figure}
  \fbox{
%
%
\begin{psfrags}%
\psfragscanon%
%
\psfrag{s02}[t][t]{\color[rgb]{0,0,0}\setlength{\tabcolsep}{0pt}\begin{tabular}{c}$\beta$\end{tabular}}%
\psfrag{s03}[t][t]{\color[rgb]{0,0,0}\setlength{\tabcolsep}{0pt}\begin{tabular}{c}$Y_{\rm tot}$\\$\times 10^3$\end{tabular}}%
\psfrag{s04}[t][t]{\color[rgb]{0,0,0}\setlength{\tabcolsep}{0pt}\begin{tabular}{c}$\alpha$\end{tabular}}%
\psfrag{s05}[t][t]{\color[rgb]{0,0,0}\setlength{\tabcolsep}{0pt}\begin{tabular}{c}$\theta_s$\end{tabular}}%
\psfrag{s06}[t][t]{\color[rgb]{0,0,0}\setlength{\tabcolsep}{0pt}\begin{tabular}{c}$\beta$\end{tabular}}%
\psfrag{s07}[t][t]{\color[rgb]{0,0,0}\setlength{\tabcolsep}{0pt}\begin{tabular}{c}$Y_{\rm tot} \times 10^3$\end{tabular}}%
\psfrag{s08}[t][t]{\color[rgb]{0,0,0}\setlength{\tabcolsep}{0pt}\begin{tabular}{c}$\alpha$\end{tabular}}%
%
\psfrag{x01}[t][t]{0}%
\psfrag{x02}[t][t]{0.1}%
\psfrag{x03}[t][t]{0.2}%
\psfrag{x04}[t][t]{0.3}%
\psfrag{x05}[t][t]{0.4}%
\psfrag{x06}[t][t]{0.5}%
\psfrag{x07}[t][t]{0.6}%
\psfrag{x08}[t][t]{0.7}%
\psfrag{x09}[t][t]{0.8}%
\psfrag{x10}[t][t]{0.9}%
\psfrag{x11}[t][t]{1}%
\psfrag{x12}[t][t]{1}%
\psfrag{x13}[t][t]{2}%
\psfrag{x14}[t][t]{2}%
\psfrag{x15}[t][t]{6}%
\psfrag{x16}[t][t]{10}%
\psfrag{x17}[t][t]{2}%
\psfrag{x18}[t][t]{4}%
\psfrag{x19}[t][t]{6}%
\psfrag{x20}[t][t]{8}%
\psfrag{x21}[t][t]{10}%
\psfrag{x22}[t][t]{4}%
\psfrag{x23}[t][t]{8}%
\psfrag{x24}[t][t]{12}%
\psfrag{x25}[t][t]{4}%
\psfrag{x26}[t][t]{6}%
\psfrag{x27}[t][t]{8}%
\psfrag{x28}[t][t]{10}%
\psfrag{x29}[t][t]{12}%
\psfrag{x30}[t][t]{4}%
\psfrag{x31}[t][t]{6}%
\psfrag{x32}[t][t]{8}%
\psfrag{x33}[t][t]{10}%
\psfrag{x34}[t][t]{12}%
\psfrag{x35}[t][t]{4.5}%
\psfrag{x36}[t][t]{5.5}%
\psfrag{x37}[t][t]{6.5}%
\psfrag{x38}[t][t]{1}%
\psfrag{x39}[t][t]{1.5}%
\psfrag{x40}[t][t]{2}%
\psfrag{x41}[t][t]{2.5}%
\psfrag{x42}[t][t]{2}%
\psfrag{x43}[t][t]{4}%
\psfrag{x44}[t][t]{6}%
\psfrag{x45}[t][t]{8}%
\psfrag{x46}[t][t]{10}%
\psfrag{x47}[t][t]{4}%
\psfrag{x48}[t][t]{6}%
\psfrag{x49}[t][t]{8}%
\psfrag{x50}[t][t]{10}%
\psfrag{x51}[t][t]{12}%
%
\psfrag{v01}[r][r]{0}%
\psfrag{v02}[r][r]{0.1}%
\psfrag{v03}[r][r]{0.2}%
\psfrag{v04}[r][r]{0.3}%
\psfrag{v05}[r][r]{0.4}%
\psfrag{v06}[r][r]{0.5}%
\psfrag{v07}[r][r]{0.6}%
\psfrag{v08}[r][r]{0.7}%
\psfrag{v09}[r][r]{0.8}%
\psfrag{v10}[r][r]{0.9}%
\psfrag{v11}[r][r]{1}%
\psfrag{v12}[r][r]{4.5}%
\psfrag{v13}[r][r]{5}%
\psfrag{v14}[r][r]{5.5}%
\psfrag{v15}[r][r]{6}%
\psfrag{v16}[r][r]{6.5}%
\psfrag{v17}[r][r]{4.5}%
\psfrag{v18}[r][r]{5}%
\psfrag{v19}[r][r]{5.5}%
\psfrag{v20}[r][r]{6}%
\psfrag{v21}[r][r]{6.5}%
\psfrag{v22}[r][r]{1}%
\psfrag{v23}[r][r]{1.5}%
\psfrag{v24}[r][r]{2}%
\psfrag{v25}[r][r]{2.5}%
\psfrag{v26}[r][r]{4.5}%
\psfrag{v27}[r][r]{5.5}%
\psfrag{v28}[r][r]{6.5}%
\psfrag{v29}[r][r]{1}%
\psfrag{v30}[r][r]{2}%
\psfrag{v31}[r][r]{2}%
\psfrag{v32}[r][r]{6}%
\psfrag{v33}[r][r]{10}%
\psfrag{v34}[r][r]{0}%
\psfrag{v35}[r][r]{0.5}%
\psfrag{v36}[r][r]{1}%
\psfrag{v37}[r][r]{0}%
\psfrag{v38}[r][r]{0.5}%
\psfrag{v39}[r][r]{1}%
\psfrag{v40}[r][r]{0}%
\psfrag{v41}[r][r]{0.5}%
\psfrag{v42}[r][r]{1}%
\psfrag{v43}[r][r]{0}%
\psfrag{v44}[r][r]{0.5}%
\psfrag{v45}[r][r]{1}%
%
\resizebox{\linewidth}{!}{\includegraphics{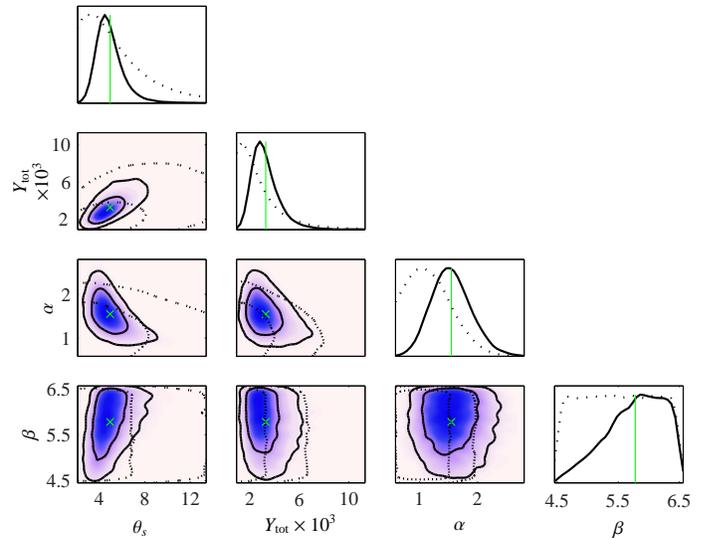}}%
\end{psfrags}%
%
}
  \caption{AMI posterior distributions for PSZ1~G134.31-06.57, allowing $\alpha$ and $\beta$ to vary.  Posterior means are indicated with green lines and crosses.  The priors on the parameters in the AMI analysis are shown as black dashed lines.  $\theta_s$ is in arcmin and {\Ytot} is in arcmin$^2$.}\label{Fi:CAJ0210_post}
\end{figure}

PSZ1~G134.31-06.57 is a new \emph{Planck} cluster at unknown redshift, with \emph{Planck} SNR = 5.4 and AMI $\Delta \ln(\mathcal{Z}) = 31$.  The \emph{Planck} and AMI constraints for this cluster overlap, and the different degeneracy directions result in a considerably tighter joint constraint, giving $\theta_{s}$\,$\approx$\,4.5\,arcmin (see Fig.~\ref{Fi:all_post}).  At this angular size, AMI data should produce constraints on $\alpha$.  Fig.~\ref{Fi:CAJ0210_post} shows the parameter constraints -- $\alpha$ moves away from the prior to a higher value of $\approx$\,1.5, while $\beta$ also shows a weak constraint to values higher than the `universal' value.

\subsubsection{Properties of $\alpha$ and $\beta$ in the sample}

\begin{figure}
  \centerline{
%
%
\begin{psfrags}%
\psfragscanon%
%
\psfrag{s06}[t][t]{\color[rgb]{0,0,0}\setlength{\tabcolsep}{0pt}\begin{tabular}{c}Mean $\alpha$\end{tabular}}%
\psfrag{s07}[b][b]{\color[rgb]{0,0,0}\setlength{\tabcolsep}{0pt}\begin{tabular}{c}Number of clusters\end{tabular}}%
\psfrag{s08}[t][t]{\color[rgb]{0,0,0}\setlength{\tabcolsep}{0pt}\begin{tabular}{c}Mean $\beta$\end{tabular}}%
%
\psfrag{x01}[t][t]{4.5}%
\psfrag{x02}[t][t]{5.5}%
\psfrag{x03}[t][t]{6.5}%
\psfrag{x04}[t][t]{0.5}%
\psfrag{x05}[t][t]{1}%
\psfrag{x06}[t][t]{1.5}%
%
\psfrag{v01}[r][r]{0}%
\psfrag{v02}[r][r]{5}%
\psfrag{v03}[r][r]{10}%
\psfrag{v04}[r][r]{15}%
\psfrag{v05}[r][r]{20}%
\psfrag{v06}[r][r]{0}%
\psfrag{v07}[r][r]{5}%
\psfrag{v08}[r][r]{10}%
\psfrag{v09}[r][r]{15}%
\psfrag{v10}[r][r]{20}%
%
\resizebox{8.9cm}{!}{\includegraphics{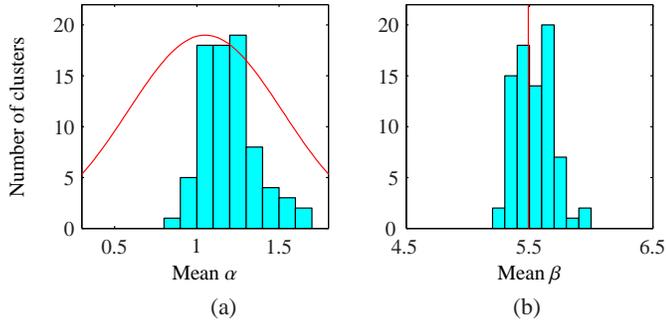}}%
\end{psfrags}%
%
}\bigskip
  \centerline{\hskip 0.1\linewidth (a) \hskip 0.4\linewidth (b)}
  \caption{The distribution of mean values of $\alpha$ and $\beta$ obtained for all the clear detections in the SZ sample.  For comparison, the REXCESS-based prior on $\alpha$ (scaled arbitrarily) is also plotted in red, and the `universal' value of $\beta$ predicted from numerical simulations is indicated with a red line.}\label{Fi:real_alpha_beta}
\end{figure}

Fig.~\ref{Fi:real_alpha_beta} shows a histogram of the recovered mean $\alpha$ and $\beta$ values for all the clear detections in the sample.

\citet{planck2012-V} and \citet{2013ApJ...768..177S} both derive average pressure profiles for smaller samples of clusters using SZ data from \emph{Planck} and BOLOCAM respectively.  In both analyses, the radial profiles derived from the SZ maps are scaled by X-ray-determined $r_{500}$ values and then stacked; a GNFW model is fitted to the stacked profiles ($+$ X-ray points for the inner part of the profile in \citealt{planck2012-V}).  Their final best fit parameters are given by ($c_{500}, \gamma, \alpha, \beta$) = (1.81, 0.31, 1.33, 4.1) and (1.18, 0.67, 0.86, 3.67) respectively.  In contrast, the AMI analysis does not rely on X-ray estimates of $r_{500}$, being based purely on the AMI SZ data.  The AMI preferred values for $\beta$ are on the whole centred around the `universal' value predicted by simulations, and do not show a trend towards the lower values derived from the \emph{Planck} and BOLOCAM analyses.  The AMI mean $\alpha$ estimates show a slight trend toward higher values, in agreement with the \emph{Planck} value and in disagreement with the BOLOCAM value.  However, since there are large (and different) degeneracies between the GNFW model parameters in the three analyses it is difficult to judge whether the analyses truly disagree (see Fig.~\ref{Fi:planck_like}, where it is clear that higher $\beta$ values are not ruled out by the \emph{Planck} likelihood).

The AMI data for these clusters therefore indicate that although individual clusters do not necessarily conform to the `universal' profile and it is important to take this into account when analysing AMI data, on the whole the average profile for the sample remains close to the `universal' profile shape even though these clusters are selected via a very different selection function compared to the REXCESS sample.

\section{Conclusions}
\label{sec:conclusions}

We have followed up the 195 clusters from the \emph{Planck} union catalogue that are visible to AMI, lie at $z>0.100$ and have \emph{Planck} SNR $\ge 5$.  Of these, we reject 72 due to difficult radio source environment, leaving a total SZ sample of 123.  We find that:

\begin{enumerate}
\item{We detect 99 of the clusters, including 79 very good detections.}
\item{We re-confirm 14 of 16 new clusters already confirmed by other observations, and validate 14 of 25 new clusters which were not confirmed at the time the \emph{Planck} catalogue was published.}
\item{We do not detect 24 of the clusters, which may be too extended for AMI to detect, be significantly offset from the phase centre, have a gas pressure profile deviating significantly from the `universal' profile, or be spurious detections by \emph{Planck}.  75\% of the AMI non-detections are detected by $<3$ \emph{Planck} algorithms, as opposed to 18\% of the AMI detections; none of the AMI non-detections have quality flag values of 1.  These correlations indicate that an AMI non-detection is a good indicator for a spurious \emph{Planck} detection.}
\item{Comparing the AMI positional estimates to those produced by PwS and the MMF algorithms shows that PwS positional estimates are generally more accurate, a more reliable function of SNR, and have a positional error estimate consistent with the true uncertainty in the positions; in contrast, the MMF3 positional errors are over-estimated by a factor of $\approx$\,3.}
\item{The trend seen in AP2013 where \emph{Planck} consistently characterises clusters to be of larger angular size and brighter is continued in the larger sample, particularly for high \emph{Planck} SNR clusters; our simulation results suggest that this may be caused by deviation from the `universal' profile used for parameter recovery.}
\item{We can generalise the model used for parameter extraction from AMI data to consider variation in $\alpha$ and $\beta$, however the priors on the shape parameters must be considered carefully since degeneracies with $\theta_{s}$ and \Ytot \:can produce spurious one-dimensional constraints on the shape parameters.}
\item{AMI data alone cannot reliably constrain $\theta_{s}$ and \Ytot \:for clusters of angular size $\theta_{s} \gtrapprox 5$\,arcmin when there is uncertainty in the pressure profile of the cluster; it can however be used to constrain $\alpha$ and $\beta$.}
\item{AMI data can be used to constrain $\theta_{s}$, \Ytot \:and $\beta$ ($\alpha$) simultaneously for clusters of angular size $\approx$\,3\,arcmin ($\approx$\,5\,arcmin), given a careful choice of priors on $\alpha$ and $\beta$.}
\item{While deviation from the `universal' profile has been shown to be important for analysing AMI data on a cluster-by-cluster basis, overall the $\beta$ values obtained by re-analysing all of the clear detections from the \emph{Planck} sample with varying $\alpha$ and $\beta$ do not show support for deviation from the `universal' $\beta$ value derived from numerical simulations.}
\end{enumerate}

\begin{acknowledgements}

We are grateful to a substitute referee who responded rapidly, with useful comments, to our paper.  The AMI telescope is supported by Cambridge University.  YCP and CR acknowledge support from CCT/Cavendish Laboratory and STFC studentships, respectively.  YCP and MO acknowledge support from Research Fellowships from Trinity College and Sidney Sussex College, Cambridge, respectively.  This work was partially undertaken on the COSMOS Shared Memory system at DAMTP, University of Cambridge operated on behalf of the STFC DiRAC HPC Facility. This equipment is funded by BIS National E-infrastructure capital grant ST/J005673/1 and STFC grants ST/H008586/1, ST/K00333X/1.  CM acknowledges her KICC Fellowship grant funding for the procurement of the cluster used for computational work.  In addition, we would like to thank the IOA computing support team for maintaining the cluster.

\end{acknowledgements}

\bibliographystyle{aa}
\bibliography{ms}

\clearpage

\appendix
\onecolumn
\section{Results table}\label{sec:results_table}

\footnotesize
\tabulinesep=_1mm
\extrarowsep=1mm
\LTcapwidth=\textwidth
\begin{longtabu} to \textwidth{X[1,l]X[0.5,r]X[0.3,c]X[0.3,c]X[0.5,c]X[1,l]X[0.3,c]X[0.5,c]X[1,l]}

\caption[Summary of results for all clusters between $20^{\circ} \le \delta < 87^{\circ}$  with \emph{Planck} SNR $>5$.]{Summary of results for all clusters between $20^{\circ} \le \delta < 87^{\circ}$ with \emph{Planck} SNR $>5$.  The rejection reason (LZ = low redshift, R = rejected by automated point-source criteria, SE = rejected manually for difficult source environment) or detection category (Y = clear detection, M = moderate detection, N = non-detection, NN = clear non-detection) is given in each case.  Also given is the \emph{Planck} SNR and the pipelines detecting the cluster (e.g.\ 110 indicates that the cluster was detected by MMF3, MMF1 but not \textsc{PwS}).  Redshifts are taken from the \emph{Planck} 2013 SZ catalogue.  Some aliases for previously-known clusters are given from: \citet{1937ApJ....86..217Z} (and references therein), \citet{1958ApJS....3..211A}, \citet{1961cgcg.book.....Z}, \citet{1967MmRAS..71...49G}, \citet{1977ApJ...211..309A}, \citet{1981SvA....25..647F}, \citet{1984ApJ...281..570P}, \citet{appenzeller98}, \citet{1999A&A...349..389V}, \citet{2001ApJ...553..668E}, \citet{2002ApJ...580..774E}, \citet{2009ApJS..183..197W}, \citet{2011A&A...534A.109P} (and references therein), \citet{2012MNRAS.423.1024M}, \citet{2012ApJS..199...34W}, \citet{planck2011-5.1a}.  Reference numbers refer to previously published AMI analyses, (1) \citet{2006MNRAS.369L...1A}, (2) \citet{2011MNRAS.414L..75A}, (3) \citet{2011MNRAS.418.2754Z}, (4) \citet{2012MNRAS.419.2921A}, (5) \citet{2012MNRAS.425..162L}, (6) AP2013, (7) \citet{2013MNRAS.433.2920S}.  $\Delta \ln(\mathcal{Z})$ is the Bayesian evidence difference.  For non-detections, predicted signal-to-noise ratios in the naturally-weighted ($\sigma_{\rm NW}$) and $uv$-tapered ($\sigma_{\rm tap}$) maps are also given based on the \emph{Planck} mean posterior parameter values.} \\

\hline Cluster name & \emph{Planck} SNR & \emph{Planck} det. & Category & $\Delta \ln(\mathcal{Z})$ & Aliases & Previous AMI & Redshift & Notes \\ \hline
\endfirsthead

\multicolumn{9}{c}
{{\tablename\ \thetable{} -- continued from previous page}} \\
\hline Cluster name & \emph{Planck} SNR & \emph{Planck} det. & Category & $\Delta \ln(\mathcal{Z})$ & Aliases & Previous AMI & Redshift & Notes \\ \hline
\endhead

\hline \multicolumn{9}{r}{{Continued on next page}} \\ \hline
\endfoot

\hline
\endlastfoot

PSZ1~G075.71+13.51 & 25.96 & 111 & LZ &  & RXC~J1921.1+4357, A2319 &  & 0.0557 & \\ 
PSZ1~G110.99+31.74 & 22.70 & 111 & LZ &  & RXC~J1703.8+7838, A2256 &  & 0.0581 & \\ 
PSZ1~G044.24+48.66 & 19.56 & 111 & LZ &  & RXC~J1558.3+2713, A2142 &  & 0.0894 & \\ 
PSZ1~G072.61+41.47 & 19.42 & 111 & R &  & RXC~J1640.3+4642, A2219 &  & 0.228 & \\ 
PSZ1~G093.93+34.92 & 18.07 & 111 & LZ &  & RXC~J1712.7+6403, A2255 &  & 0.0809 & \\ 
PSZ1~G097.72+38.13 & 17.21 & 111 & Y & 33.77 & RXC~J1635.8+6612, A2218 & 2,5,6 & 0.1709 & \\ 
PSZ1~G186.37+37.26 & 15.51 & 111 & R &  & RXC~J0842.9+3621, A697 &  & 0.282 & \\ 
PSZ1~G057.84+87.98 & 15.25 & 111 & LZ &  & RXC~J1259.7+2756, Coma, A1656 &  & 0.0231 & \\ 
PSZ1~G086.47+15.31 & 14.97 & 111 & Y & 14.58 & RXC~J1938.3+5409 &  & 0.26 & \\ 
PSZ1~G033.84+77.17 & 14.20 & 111 & LZ &  & RXC~J1348.8+2635, A1795 &  & 0.0622 & \\ 
PSZ1~G170.22+09.74 & 14.12 & 111 & R &  & 1RXS~J060313.4+421231 &  &  & \\ 
PSZ1~G149.21+54.17 & 13.60 & 111 & R &  & RXC~J1058.4+5647, A1132 &  & 0.1369 & \\ 
PSZ1~G092.67+73.44 & 13.41 & 111 & R &  & RXC~J1335.3+4059, A1763 &  & 0.2279 & \\ 
PSZ1~G072.78-18.70 & 13.09 & 111 & M & 1.78 & ZwCl~2120+2256, ZwCl~8503 &  & 0.143 & \\ 
PSZ1~G149.75+34.68 & 12.97 & 111 & Y & 46.38 & RXC~J0830.9+6551, A665 &  & 0.1818 & \\ 
PSZ1~G191.00+06.65 & 12.44 & 111 & LZ &  & RXC~J0635.0+2231 &  & 0.068 & \\ 
PSZ1~G058.29+18.57 & 11.78 & 111 & LZ &  & RXC~J1825.3+3026, CIZA~J1825.3+3026 &  & 0.065 & \\ 
PSZ1~G067.19+67.44 & 11.76 & 111 & Y & 28.83 & RXC~J1426.0+3749, A1914 & 1,2,4,6 & 0.1712 & \\ 
PSZ1~G107.14+65.29 & 11.20 & 111 & R &  & RXC~J1332.7+5032, A1758 & 5 & 0.2799 & \\ 
PSZ1~G055.58+31.87 & 10.83 & 111 & R &  & RXC~J1722.4+3208, A2261 &  & 0.224 & \\ 
PSZ1~G062.94+43.69 & 10.78 & 111 & LZ &  & RXC~J1628.6+3932, A2199 &  & 0.0299 & \\ 
PSZ1~G042.85+56.63 & 10.67 & 111 & LZ &  & RXC~J1522.4+2742, A2065 &  & 0.0723 & \\ 
PSZ1~G094.00+27.41 & 10.56 & 111 & R &  & H1821+643 &  & 0.3315 & \\ 
PSZ1~G180.25+21.03 & 10.54 & 111 & R &  & RXC~J0717.5+3745, MCS~J0717.5+3745 & 2 & 0.546 & \\ 
PSZ1~G053.52+59.52 & 10.46 & 111 & Y & 31.24 & RXC~J1510.1+3330, A2034 & 6 & 0.113 & \\ 
PSZ1~G125.34-08.65 & 10.22 & 111 & Y & 12.26 & RXC~J0107.7+5408, ZwCl~0104+5350 &  & 0.1066 & \\ 
PSZ1~G124.20-36.47 & 10.13 & 111 & R &  & RXC~J0055.9+2622, A115 & 4 & 0.1971 & \\ 
PSZ1~G112.48+57.02 & 9.97 & 111 & LZ &  & RXC~J1336.1+5912, A1767 &  & 0.0701 & \\ 
PSZ1~G049.22+30.84 & 9.90 & 111 & M & 1.61 & RXC~J1720.1+2637 & 5 & 0.1644 & \\ 
PSZ1~G226.19+76.78 & 9.79 & 111 & Y & 25.52 & RXC~J1155.3+2324, A1413 & 5,6 & 0.1427 & \\ 
PSZ1~G067.36+10.74 & 9.61 & 111 & Y & 10.47 & RXC~J1916.1+3525 &  & 0.209 & \\ 
PSZ1~G056.79+36.30 & 9.58 & 111 & LZ &  & RXC~J1702.7+3403, A2244 &  & 0.0953 & \\ 
PSZ1~G084.47+12.63 & 9.54 & 111 & Y & 4.75 & RXC~J1948.3+5113 &  & 0.185 & \\ 
PSZ1~G166.11+43.40 & 9.53 & 111 & Y & 27.21 & RXC~J0917.8+5143, A773 & 2,5,6 & 0.2172 & \\ 
PSZ1~G139.17+56.37 & 9.48 & 111 & R &  & RXC~J1142.5+5832, A1351, MCS~J1142.4+5831 &  & 0.322 & \\ 
PSZ1~G167.64+17.63 & 9.43 & 111 & Y & 4.74 & RXC~J0638.1+4747, ZwCl~0634+4750, ZwCl~1133 &  & 0.174 & \\ 
PSZ1~G057.63+34.92 & 9.03 & 111 & LZ &  & RXC~J1709.8+3426, A2249 &  & 0.0802 & \\ 
PSZ1~G113.84+44.33 & 8.98 & 111 & Y & 3.10 & RXC~J1414.2+7115, A1895 &  & 0.225 & \\ 
PSZ1~G046.90+56.48 & 8.96 & 111 & M & 0.88 & RXC~J1524.1+2955, A2069 &  & 0.1145 & \\ 
PSZ1~G077.89-26.62 & 8.74 & 111 & Y & 35.33 & RXC~J2200.8+2058, A2409 & 5,6 & 0.147 & \\ 
PSZ1~G139.61+24.20 & 8.66 & 111 & Y & 27.08 &  & 6 & 0.2671 & \\ 
PSZ1~G118.58+28.57 & 8.57 & 111 & Y & 4.83 & RXC~J1723.7+8553, A2294 &  & 0.178 & \\ 
PSZ1~G071.21+28.86 & 8.46 & 011 & Y & 12.60 & RXC~J1752.0+4440, MCS~J1752.0+4440 &  & 0.366 & \\ 
PSZ1~G125.72+53.87 & 8.45 & 111 & R &  & RXC~J1236.9+6311, A1576, MCS~J1236.9+6311 &  & 0.3019 & \\ 
PSZ1~G098.12+30.30 & 8.45 & 111 & LZ &  & RXC~J1754.6+6803, ZwCl~1754+680 &  & 0.077 & \\ 
PSZ1~G165.06+54.13 & 8.44 & 111 & Y & 16.86 & RXC~J1023.6+4907, A990 & 5,6 & 0.144 & \\ 
PSZ1~G180.56+76.66 & 8.43 & 111 & Y & 7.13 & RXC~J1157.3+3336, A1423 & 5 & 0.2138 & \\ 
PSZ1~G048.08+57.17 & 8.36 & 101 & LZ &  & RXC~J1521.2+3038, A2061 &  & 0.0777 & \\ 
PSZ1~G157.32-26.77 & 8.35 & 111 & Y & 25.87 & RXC~J0308.9+2645, MCS~J0308.9+2645 & 2 & 0.356 & \\ 
PSZ1~G163.69+53.52 & 8.26 & 111 & Y & 5.10 & RXC~J1022.5+5006, A980 &  & 0.158 & \\ 
PSZ1~G157.44+30.34 & 8.19 & 011 & Y & 32.51 & [ATZ98]~B100, RXC~J0748.7+5941 & 6 &  & \\ 
PSZ1~G143.28+65.22 & 8.19 & 111 & Y & 5.85 & RXC~J1159.2+4947, A1430 &  & 0.211 & \\ 
PSZ1~G046.09+27.16 & 8.19 & 111 & R &  & RXC~J1731.6+2251, MCS~J1731.6+2252 &  & 0.389 & \\ 
PSZ1~G229.70+77.97 & 8.18 & 111 & R &  & RXC~J1201.3+2306, A1443 &  & 0.269 & \\ 
PSZ1~G132.49-17.29 & 8.09 & 111 & Y & 33.24 & RXC~J0142.9+4438 &  & 0.341 & \\ 
PSZ1~G114.78-33.72 & 7.92 & 111 & LZ &  & RXC~J0020.6+2840, A21 &  & 0.094 & \\ 
PSZ1~G088.83-12.99 & 7.70 & 111 & R &  & ClG~2153.8+3746 &  & 0.292 & \\ 
PSZ1~G150.56+58.32 & 7.61 & 111 & Y & 8.63 & RXC~J1115.2+5320, XMMXCS~J1115.2+5319, MCS~J1115.2+5320 & 7 & 0.47 & \\ 
PSZ1~G114.29+64.91 & 7.48 & 111 & Y & 6.48 & RXC~J1315.1+5149, A1703 &  & 0.2836 & \\ 
PSZ1~G182.55+55.83 & 7.46 & 111 & R &  & RXC~J1017.0+3902, A963 &  & 0.206 & \\ 
PSZ1~G134.73+48.89 & 7.41 & 111 & SE &  & RXC~J1133.2+6622, A1302 &  & 0.116 & 63 mJy source at 17 arcmin causes artifacts in the SA map\\ 
PSZ1~G080.38+14.65 & 7.41 & 111 & LZ &  & RXC~J1926.1+4832 &  & 0.098 & \\ 
PSZ1~G114.99+70.36 & 7.40 & 111 & R &  & RXC~J1306.9+4633, A1682 &  & 0.2259 & \\ 
PSZ1~G091.82+26.11 & 7.26 & 111 & SE &  &  &  & 0.24 & \\ 
PSZ1~G083.30-31.01 & 7.26 & 111 & Y & 28.09 & RXC~J2228.6+2036 &  & 0.412 & \\ 
PSZ1~G161.39+26.24 & 7.24 & 111 & LZ &  & RXC~J0721.3+5547, A576 &  & 0.0381 & \\ 
PSZ1~G060.12+11.42 & 7.22 & 111 & Y & 16.07 &  &  &  & \\ 
PSZ1~G207.87+81.31 & 7.19 & 111 & Y & 19.99 & RXC~J1212.3+2733, A1489 &  & 0.353 & \\ 
PSZ1~G085.98+26.69 & 7.13 & 111 & M & 2.89 & RXC~J1819.9+5710, A2302 &  & 0.179 & Positional error increased to 5 arcmin to encompass visible decrement in map\\ 
PSZ1~G228.21+75.20 & 7.12 & 111 & Y & 112.81 & RXC~J1149.5+2224, MCS~J1149.5+2223 & 6 & 0.545 & \\ 
PSZ1~G099.48+55.62 & 7.06 & 111 & N & -0.01 & RXC~J1428.4+5652, A1925 &  & 0.1051 & Predicted $\sigma_{\rm NW} =  4.4; \sigma_{\rm tap} =  6.9$\\ 
PSZ1~G071.63+29.78 & 7.01 & 111 & Y & 3.01 & RXC~J1747.2+4512, ZwCl~8284, ZwCl~1745+4513 &  & 0.1565 & \\ 
PSZ1~G115.70+17.51 & 7.00 & 111 & M & 0.76 &  &  &  & \\ 
PSZ1~G133.56+69.05 & 6.97 & 111 & Y & 5.05 & RXC~J1229.0+4737, A1550 &  & 0.254 & \\ 
PSZ1~G359.99+78.04 & 6.96 & 111 & R &  & RXC~J1334.1+2013, A1759 &  & 0.171 & \\ 
PSZ1~G140.67+29.44 & 6.94 & 111 & Y & 5.93 & RXC~J0741.7+7414, ZwCl~1370, ZwCl~0735+7421 &  & 0.2149 & \\ 
PSZ1~G318.61+83.80 & 6.93 & 001 & SE &  &  &  &  & 33mJy source (extended to LA) at 10 arcmin leaves significant residuals in the SA map\\ 
PSZ1~G067.52+34.75 & 6.92 & 111 & R &  & RXC~J1717.3+4226, ZwCl~8193, ZwCl~1715+4229 &  & 0.1754 & \\ 
PSZ1~G113.26-29.69 & 6.91 & 111 & R &  & RXC~J0011.7+3225, A7 &  & 0.1073 & \\ 
PSZ1~G098.85-07.27 & 6.89 & 011 & SE &  &  &  &  & \\ 
PSZ1~G096.89+24.17 & 6.89 & 111 & Y & 3.54 & ZwCl~1856+6616, PLCKESZ~G096.87+24.21 &  & 0.3 & \\ 
PSZ1~G138.60-10.85 & 6.86 & 111 & Y & 6.15 &  &  &  & \\ 
PSZ1~G153.41+36.58 & 6.85 & 010 & N & -2.70 &  &  &  & Predicted $\sigma_{\rm NW} =  3.0; \sigma_{\rm tap} =  3.7$\\ 
PSZ1~G146.37-15.57 & 6.83 & 111 & LZ &  & RXC~J0254.4+4134, AWM7 &  & 0.0172 & \\ 
PSZ1~G148.20+23.49 & 6.77 & 111 & Y & 3.19 &  &  &  & \\ 
PSZ1~G121.09+57.02 & 6.72 & 111 & Y & 10.37 &  & 3,6 & 0.3436 & \\ 
PSZ1~G118.46+39.31 & 6.67 & 111 & Y & 4.73 & RXC~J1354.6+7715 &  & 0.3967 & \\ 
PSZ1~G094.69+26.34 & 6.66 & 111 & N & -0.26 & RXC~J1832.5+6449 &  & 0.1623 & Predicted $\sigma_{\rm NW} =  4.1; \sigma_{\rm tap} =  5.3$\\ 
PSZ1~G084.41-12.43 & 6.59 & 011 & Y & 15.82 &  &  &  & \\ 
PSZ1~G102.97-04.77 & 6.56 & 011 & Y & 4.27 &  &  &  & \\ 
PSZ1~G162.30-26.92 & 6.56 & 100 & R &  &  &  &  & \\ 
PSZ1~G109.14-28.02 & 6.56 & 111 & SE &  & WHL~J358.303+33.2696 &  & 0.4709 & \\ 
PSZ1~G127.55+20.84 & 6.55 & 011 & R &  &  &  &  & \\ 
PSZ1~G100.18-29.68 & 6.54 & 111 & R &  &  &  & 0.485 & \\ 
PSZ1~G049.35+44.36 & 6.53 & 111 & LZ &  & RXC~J1620.5+2953, A2175 &  & 0.0972 & \\ 
PSZ1~G063.80+11.42 & 6.53 & 111 & Y & 3.78 &  &  &  & \\ 
PSZ1~G098.96+24.87 & 6.52 & 111 & LZ &  & RXC~J1853.9+6822 &  & 0.0928 & \\ 
PSZ1~G108.18-11.53 & 6.49 & 111 & Y & 16.62 &  &  &  & \\ 
PSZ1~G066.41+27.03 & 6.48 & 111 & Y & 16.80 & WHL~J269.219+40.1353 &  & 0.5699 & \\ 
PSZ1~G100.16+41.66 & 6.43 & 111 & R &  & RXC~J1556.1+6621, A2146 & 5 & 0.2339 & \\ 
PSZ1~G068.23+15.20 & 6.42 & 011 & LZ &  & RXC~J1857.6+3800 &  & 0.0567 & \\ 
PSZ1~G166.61+42.12 & 6.38 & 111 & Y & 3.79 & RXC~J0909.3+5133, A746 &  & 0.23225 & \\ 
PSZ1~G099.84+58.45 & 6.35 & 111 & Y & 29.56 & WHL~J213.697+54.7844 &  & 0.6305 & \\ 
PSZ1~G054.99+53.42 & 6.31 & 111 & Y & 16.98 & RXC~J1539.7+3424, A2111 & 4,5 & 0.229 & \\ 
PSZ1~G136.94+59.46 & 6.31 & 111 & LZ &  & RXC~J1200.3+5613, A1436 &  & 0.065 & \\ 
PSZ1~G057.91+27.62 & 6.30 & 111 & LZ &  & RXC~J1744.2+3259, ZwCl~8276 &  & 0.0757 & \\ 
PSZ1~G105.25-17.96 & 6.29 & 111 & R &  & RXC~J2320.2+4146 &  & 0.14 & \\ 
PSZ1~G195.60+44.03 & 6.27 & 111 & R &  & RXC~J0920.4+3030, A781 & 5 & 0.2952 & \\ 
PSZ1~G068.32+81.81 & 6.27 & 111 & SE &  & RXC~J1322.8+3138 &  & 0.3083 & Extended source to south-east\\ 
PSZ1~G118.88+52.40 & 6.25 & 111 & Y & 21.80 & RXC~J1314.4+6434, A1704 & 5 & 0.22 & \\ 
PSZ1~G186.98+38.66 & 6.22 & 111 & Y & 4.68 & RXC~J0850.2+3603, ZwCl~1953 &  & 0.378 & \\ 
PSZ1~G083.62+85.08 & 6.17 & 111 & R &  & RXC~J1305.9+3054, A1677 &  & 0.1832 & \\ 
PSZ1~G143.67+42.63 & 6.16 & 111 & R &  & RXC~J1003.1+6709, A910 &  & 0.206 & \\ 
PSZ1~G192.19+56.12 & 6.14 & 111 & M & 0.22 & RXC~J1016.3+3338, A961 &  & 0.124 & \\ 
PSZ1~G135.03+36.03 & 6.12 & 111 & Y & 6.11 & RXC~J0947.2+7623, MCS~J0947.2+7623 &  & 0.345 & \\ 
PSZ1~G074.75-24.59 & 6.10 & 111 & N & -2.58 & ZwCl~2143+2014 &  & 0.25 & Predicted $\sigma_{\rm NW} =  7.7; \sigma_{\rm tap} =  9.2$\\ 
PSZ1~G152.68+25.43 & 6.10 & 111 & LZ &  & RXC~J0704.4+6318, A566 &  & 0.098 & \\ 
PSZ1~G223.97+69.31 & 6.09 & 111 & M & 1.34 & RXC~J1123.9+2129, A1246 &  & 0.1904 & \\ 
PSZ1~G184.70+28.92 & 6.06 & 101 & Y & 20.64 & RXC~J0800.9+3602, A611 & 2,4,5 & 0.288 & \\ 
PSZ1~G040.63+77.13 & 6.05 & 111 & LZ &  & RXC~J1349.3+2806, A1800 &  & 0.0748 & \\ 
PSZ1~G131.02+29.98 & 6.02 & 111 & M & 2.98 & RXC~J0825.7+8218, A625 &  & 0.2 & \\ 
PSZ1~G171.01+39.44 & 6.01 & 111 & Y & 27.90 &  &  & 0.5131 & \\ 
PSZ1~G050.41+31.18 & 5.98 & 111 & Y & 10.30 & RXC~J1720.1+2740, A2259 & 4 & 0.164 & \\ 
PSZ1~G153.56+36.23 & 5.96 & 110 & M & 0.64 &  &  &  & \\ 
PSZ1~G205.85+73.77 & 5.96 & 111 & Y & 17.84 & WHL~J174.518+27.9773 &  & 0.4474 & \\ 
PSZ1~G031.94+78.71 & 5.95 & 111 & LZ &  & RXC~J1341.8+2622 &  & 0.0724 & \\ 
PSZ1~G187.53+21.92 & 5.88 & 111 & Y & 12.52 & RXC~J0732.3+3137, A586 & 5 & 0.171 & \\ 
PSZ1~G201.50+30.63 & 5.87 & 111 & Y & 15.32 & ZwCl~0824+2244 &  & 0.287 & \\ 
PSZ1~G096.87+52.48 & 5.85 & 111 & M & 1.25 & RXC~J1452.9+5802, A1995 &  & 0.3179 & \\ 
PSZ1~G078.67+20.06 & 5.84 & 011 & R &  &  &  & 0.45 & \\ 
PSZ1~G040.06+74.94 & 5.84 & 111 & LZ &  & RXC~J1359.2+2758, A1831 &  & 0.0612 & \\ 
PSZ1~G142.38+22.82 & 5.81 & 110 & Y & 7.22 &  &  &  & \\ 
PSZ1~G142.17+37.28 & 5.79 & 100 & NN & -5.05 &  &  &  & Predicted $\sigma_{\rm NW} =  6.5; \sigma_{\rm tap} =  8.6$\\ 
PSZ1~G186.81+07.31 & 5.79 & 001 & R &  & WHL~J97.3409+26.5054 &  & 0.2577 & \\ 
PSZ1~G105.91-38.39 & 5.77 & 111 & Y & 13.03 &  &  &  & Positional uncertainty increased to 5 arcmin to encompass large decrement visible in map\\ 
PSZ1~G099.31+20.89 & 5.75 & 111 & Y & 7.41 & RXC~J1935.3+6734 &  & 0.1706 & \\ 
PSZ1~G137.56+53.88 & 5.73 & 001 & NN & -4.14 &  &  &  & Predicted $\sigma_{\rm NW} = 17.3; \sigma_{\rm tap} = 17.6$\\ 
PSZ1~G189.27+59.24 & 5.73 & 111 & R &  & RXC~J1031.7+3502, A1033 &  & 0.1259 & \\ 
PSZ1~G095.37+14.42 & 5.72 & 011 & R &  &  &  & 0.1188 & \\ 
PSZ1~G183.27+34.97 & 5.69 & 111 & Y & 9.64 & WHL~J127.437+38.4651 &  & 0.3919 & \\ 
PSZ1~G069.92-18.89 & 5.68 & 111 & R &  &  &  & 0.3076 & \\ 
PSZ1~G156.88+13.48 & 5.67 & 111 & Y & 7.57 &  &  &  & \\ 
PSZ1~G109.99+52.87 & 5.64 & 111 & Y & 17.90 & RXC~J1359.8+6231, ZwCl~6429, ZwCl~1358+6245 &  & 0.3259 & \\ 
PSZ1~G179.13+60.14 & 5.61 & 111 & R &  & RXC~J1040.7+3956, A1068 &  & 0.1372 & \\ 
PSZ1~G107.32-31.51 & 5.60 & 111 & N & -2.03 & RXC~J2350.5+2929 &  & 0.1498 & Predicted $\sigma_{\rm NW} =  7.0; \sigma_{\rm tap} =  9.8$\\ 
PSZ1~G084.62-15.86 & 5.59 & 111 & M & 1.47 &  &  &  & \\ 
PSZ1~G145.19+32.14 & 5.58 & 001 & R &  & RXC~J0811.1+7002, A621 & 5 & 0.223 & \\ 
PSZ1~G127.36-10.69 & 5.58 & 100 & R &  &  &  &  & \\ 
PSZ1~G097.93+19.46 & 5.54 & 111 & M & 1.30 & 4C~65.28 &  & 0.25 & \\ 
PSZ1~G136.62-25.05 & 5.52 & 111 & LZ &  & RXC~J0152.7+3609, A262 &  & 0.0163 & \\ 
PSZ1~G094.54+51.01 & 5.52 & 011 & Y & 24.04 & WHL~J227.050+57.9005 &  & 0.5392 & \\ 
PSZ1~G123.55-10.34 & 5.51 & 111 & SE &  &  &  & 0.1 & Lots of extended emission across the centre of the map\\ 
PSZ1~G100.82+24.61 & 5.50 & 011 & LZ &  & RXC~J1900.4+6958, A2315 &  & 0.0877 & \\ 
PSZ1~G103.58+24.78 & 5.48 & 011 & SE &  &  &  & 0.33 & 30mJy source at 11 arcmin leaves substantial residuals at map centre\\ 
PSZ1~G092.46-35.25 & 5.47 & 100 & SE &  &  &  &  & Large amounts of extended emission present on the map after point source subtraction\\ 
PSZ1~G151.19+48.29 & 5.45 & 111 & R &  & RXC~J1017.5+5934, A959 &  & 0.353 & \\ 
PSZ1~G109.88+27.94 & 5.44 & 111 & Y & 3.41 &  &  & 0.4 & \\ 
PSZ1~G134.31-06.57 & 5.44 & 011 & Y & 30.68 &  &  &  & \\ 
PSZ1~G172.64+65.29 & 5.43 & 111 & LZ &  & RXC~J1111.6+4050 &  & 0.0794 & \\ 
PSZ1~G101.52-29.96 & 5.43 & 111 & R &  &  &  & 0.227 & \\ 
PSZ1~G168.34+69.73 & 5.42 & 011 & SE &  & A1319 &  & 0.288 & Many radio sources close together and unresolved on the SA map, plus some extended emission, make source subtraction too difficult\\ 
PSZ1~G134.59+53.41 & 5.42 & 011 & N & -2.05 & WHL~J177.705+62.3301 &  & 0.3452 & Predicted $\sigma_{\rm NW} = 19.3; \sigma_{\rm tap} = 20.8$\\ 
PSZ1~G135.03+54.38 & 5.40 & 001 & SE &  & WHL~J178.058+61.3331 &  & 0.3169 & Lots of extended emission across the centre of the map\\ 
PSZ1~G106.49-10.43 & 5.40 & 110 & R &  &  &  &  & \\ 
PSZ1~G188.41+07.04 & 5.39 & 001 & LZ &  & RXC~J0631.3+2500, ZwCl~0628+2502 &  & 0.081 & \\ 
PSZ1~G108.13-09.21 & 5.39 & 110 & Y & 29.88 &  &  &  & \\ 
PSZ1~G090.82+44.13 & 5.37 & 110 & N & -0.97 & ZwCl~1602+5917 &  & 0.2544 & Predicted $\sigma_{\rm NW} =  2.4; \sigma_{\rm tap} =  2.2$\\ 
PSZ1~G127.02+26.21 & 5.37 & 111 & M & 1.98 &  &  &  & \\ 
PSZ1~G164.63+46.37 & 5.36 & 111 & M & 0.93 & ZwCl~0934+5216, PLCKESZ~G164.61+46.38 &  & 0.3605 & \\ 
PSZ1~G085.71+10.67 & 5.35 & 001 & R &  &  &  &  & \\ 
PSZ1~G050.46+67.54 & 5.35 & 111 & N & -2.30 & RXC~J1432.4+3137, A1930 &  & 0.1313 & Predicted $\sigma_{\rm NW} = 11.1; \sigma_{\rm tap} = 15.1$\\ 
PSZ1~G137.51-10.01 & 5.33 & 010 & R &  &  &  &  & \\ 
PSZ1~G098.64+23.20 & 5.33 & 011 & Y & 5.57 & RXC~J1910.4+6741 &  & 0.2471 & \\ 
PSZ1~G060.50+26.94 & 5.33 & 110 & R &  & RXC~J1750.2+3504 &  & 0.1712 & \\ 
PSZ1~G169.80+26.10 & 5.32 & 010 & N & -1.46 &  &  &  & Predicted $\sigma_{\rm NW} = 14.4; \sigma_{\rm tap} = 16.1$\\ 
PSZ1~G135.12+57.90 & 5.29 & 010 & SE &  & RXC~J1201.9+5802, A1446 &  & 0.1031 & Only observed on SA, 64 mJy source on pointing centre\\ 
PSZ1~G157.67+77.99 & 5.28 & 111 & R &  & WHL~J184.380+36.6865 &  & 0.3732 & \\ 
PSZ1~G101.36+32.39 & 5.27 & 011 & N & -1.60 & RXC~J1727.4+7035 &  & 0.3059 & Predicted $\sigma_{\rm NW} =  4.6; \sigma_{\rm tap} =  5.8$\\ 
PSZ1~G121.75+51.81 & 5.26 & 111 & Y & 119.16 & ZwCl~1256+6537 &  & 0.23765 & Lots of unsubtracted extended emission on the maps; the cluster is clearly detected, but parameter estimation may be unreliable\\ 
PSZ1~G130.26-26.53 & 5.25 & 010 & SE &  & ZwCl~0120+3538 &  & 0.2159 & \\ 
PSZ1~G084.85+20.63 & 5.25 & 111 & Y & 8.75 &  &  & 0.29 & \\ 
PSZ1~G149.38-36.86 & 5.25 & 111 & Y & 11.63 & A344 &  & 0.1696 & \\ 
PSZ1~G138.11+42.03 & 5.24 & 011 & R &  &  &  & 0.4961 & \\ 
PSZ1~G198.50+46.01 & 5.24 & 111 & M & 1.01 & ZwCl~0928+2904 &  & 0.222 & \\ 
PSZ1~G091.81-26.97 & 5.23 & 011 & R &  & RXC~J2245.4+2808 &  & 0.3551 & \\ 
PSZ1~G031.91+67.94 & 5.23 & 100 & N & -0.29 &  &  &  & Predicted $\sigma_{\rm NW} =  5.8; \sigma_{\rm tap} =  6.5$\\ 
PSZ1~G213.37+80.60 & 5.23 & 111 & Y & 22.50 & WHL~J182.349+26.6796 &  & 0.5586 & \\ 
PSZ1~G100.03+23.73 & 5.22 & 001 & Y & 8.84 & RXC~J1908.3+6903, A2317 &  & 0.2103 & \\ 
PSZ1~G135.92+76.21 & 5.22 & 010 & N & -2.91 &  &  &  & Predicted $\sigma_{\rm NW} =  2.6; \sigma_{\rm tap} =  4.0$\\ 
PSZ1~G071.44+59.57 & 5.21 & 111 & R &  & RXC~J1501.3+4220, ZwCl~7215, ZwCl~1459+4240 &  & 0.2917 & \\ 
PSZ1~G164.26+08.91 & 5.21 & 111 & Y & 14.38 & WHL~J85.8665+46.9358 &  & 0.2505 & \\ 
PSZ1~G084.84+35.04 & 5.21 & 111 & N & -0.66 & RXC~J1718.1+5639, ZwCl~8197 &  & 0.1138 & Predicted $\sigma_{\rm NW} =  5.0; \sigma_{\rm tap} =  5.3$\\ 
PSZ1~G119.37+46.84 & 5.21 & 111 & SE &  & RXC~J1320.0+7003, A1722, MCS~J1319.9+7003 &  & 0.3275 & Extended structure to the west not detected in LA map\\ 
PSZ1~G076.44+23.53 & 5.21 & 111 & SE &  &  &  & 0.1685 & \\ 
PSZ1~G077.71+26.72 & 5.20 & 011 & LZ &  & RXC~J1811.0+4954, ZwCl~8338 &  & 0.0501 & \\ 
PSZ1~G183.26+12.25 & 5.20 & 011 & N & -1.52 &  &  &  & Predicted $\sigma_{\rm NW} = 17.1; \sigma_{\rm tap} = 18.2$\\ 
PSZ1~G085.85+35.45 & 5.20 & 011 & LZ &  & RXC~J1715.3+5724 &  & 0.0276 & \\ 
PSZ1~G114.98+19.10 & 5.19 & 010 & N & -0.75 &  &  &  & Predicted $\sigma_{\rm NW} = 14.6; \sigma_{\rm tap} = 17.0$\\ 
PSZ1~G059.51+33.06 & 5.18 & 011 & SE &  & RXC~J1720.2+3536, MCS~J1720.2+3536 &  & 0.387 & 280 mJy source at 13 arcmin produces artifacts on SA map\\ 
PSZ1~G172.93+21.31 & 5.18 & 011 & Y & 4.86 &  &  & 0.3309 & \\ 
PSZ1~G091.93+35.48 & 5.18 & 100 & N & -2.80 &  &  &  & Predicted $\sigma_{\rm NW} = 14.1; \sigma_{\rm tap} = 12.4$\\ 
PSZ1~G075.29+26.66 & 5.17 & 100 & N & -2.85 &  &  &  & Predicted $\sigma_{\rm NW} = 17.3; \sigma_{\rm tap} = 17.1$\\ 
PSZ1~G175.89+24.24 & 5.16 & 010 & N & -0.65 & ZwCl~0723+4239 &  & 0.19175 & Predicted $\sigma_{\rm NW} =  2.7; \sigma_{\rm tap} =  2.5$\\ 
PSZ1~G144.86+25.09 & 5.15 & 111 & Y & 44.36 & RXC~J0647.8+7014, MCS~J0647.6+7015 &  & 0.584 & \\ 
PSZ1~G123.72+34.65 & 5.14 & 100 & R &  & RXC~J1231.3+8225 &  & 0.2053 & \\ 
PSZ1~G197.13+33.46 & 5.13 & 110 & R &  & WHL~J128.694+26.9757 &  & 0.4561 & \\ 
PSZ1~G122.98-35.52 & 5.11 & 001 & Y & 11.74 & RXC~J0051.6+2720 &  & 0.3615 & \\ 
PSZ1~G053.50+09.56 & 5.11 & 101 & NN & -4.20 &  &  &  & Predicted $\sigma_{\rm NW} = 15.9; \sigma_{\rm tap} = 19.3$\\ 
PSZ1~G045.07+67.80 & 5.11 & 100 & N & -2.05 & A1929 &  & 0.2191 & Predicted $\sigma_{\rm NW} = 13.8; \sigma_{\rm tap} = 13.4$\\ 
PSZ1~G116.79-09.82 & 5.11 & 011 & R &  & ZwCl~0008+5215 &  & 0.104 & \\ 
PSZ1~G189.29+07.44 & 5.10 & 001 & R &  &  &  &  & \\ 
PSZ1~G103.16-14.95 & 5.08 & 110 & SE &  &  &  &  & \\ 
PSZ1~G157.84+21.23 & 5.08 & 111 & M & 2.15 &  &  &  & \\ 
PSZ1~G048.09+27.18 & 5.07 & 111 & M & 1.04 &  &  & 0.73608 & \\ 
PSZ1~G087.47+37.65 & 5.07 & 010 & R &  &  &  & 0.1132 & \\ 
PSZ1~G111.74+70.35 & 5.07 & 111 & M & 1.06 & RXC~J1313.1+4616, A1697 &  & 0.183 & \\ 
PSZ1~G066.20+12.87 & 5.06 & 001 & N & -0.95 &  &  & 0.23 & Predicted $\sigma_{\rm NW} =  9.3; \sigma_{\rm tap} = 11.2$\\ 
PSZ1~G045.85+57.71 & 5.06 & 111 & Y & 10.15 &  &  & 0.611 & \\ 
PSZ1~G079.33+28.33 & 5.06 & 011 & SE &  & ZwCl~1801+5136 &  & 0.2036 & Too many radio sources near the cluster centre to be sure of a non-detection\\ 
PSZ1~G097.52-14.92 & 5.06 & 010 & Y & 35.39 &  &  &  & Bright, extended radio galaxy at about 10 arcmin removed from the SA data manually using CLEAN components leaving significant residuals in the source-subtracted map; cluster is clearly detected but parameter estimation is suspect\\ 
PSZ1~G118.06+31.10 & 5.05 & 011 & SE &  &  &  &  & Extended emission near the cluster centre\\ 
PSZ1~G056.13+28.06 & 5.05 & 011 & Y & 4.13 & WHL~J265.066+31.6026 &  & 0.426 & \\ 
PSZ1~G083.35+76.41 & 5.03 & 011 & R &  &  &  &  & \\ 
PSZ1~G073.64+36.49 & 5.03 & 001 & N & -0.05 &  &  & 0.56 & Predicted $\sigma_{\rm NW} = 21.1; \sigma_{\rm tap} = 22.2$\\ 
PSZ1~G129.81+16.85 & 5.03 & 100 & Y & 3.91 &  &  & 0.41159 & \\ 
PSZ1~G134.64-11.77 & 5.02 & 111 & Y & 31.69 &  &  &  & 66mJy source at 10 arcmin leaves residuals in the source-subtracted map; cluster is clearly detected but parameter estimation is suspect\\ 
PSZ1~G178.10+18.58 & 5.01 & 101 & SE &  &  &  &  & \\ 
PSZ1~G165.41+66.17 & 5.00 & 111 & M & 1.96 & WHL~J170.907+43.0578 &  & 0.1957 & \\ 
PSZ1~G099.48+37.72 & 5.00 & 101 & M & 0.79 & RXC~J1634.6+6738, A2216 &  & 0.1668 & \\ 

\label{T:res}
\end{longtabu}

\clearpage

\section{\Ytot-$\theta_{s}$ posterior comparison}
\begin{figure*}[hb]
\threefigures{posts/CAJ1635+6612bm1_posteriors.ps}{posts/CAJ1938+5409bm1_posteriors.ps}{posts/no_comb/CAJ2122+2311b_posteriors.ps}
\threefigures{posts/CAJ0830+6551b_posteriors.ps}{posts/CAJ1425+3750bm1_posteriors.ps}{posts/CAJ1510+3329bm1_posteriors.ps}
\threefigures{posts/no_comb/CAJ0107+5407bm1_posteriors.ps}{posts/no_comb/CAJ1720+2637b_posteriors.ps}{posts/no_comb/CAJ1155+2324bm1_posteriors.ps}
\caption{$Y_{\rm tot}$-$\theta_{s}$ posterior distributions for AMI and \emph{Planck}, in descending \emph{Planck} SNR order (note that this is the `compatibility' SNR for \textsc{PwS}), for all AMI detections ($\Delta \ln(\mathcal{Z}) \ge 0$).  Contours mark the 68\% and 95\% confidence limits of the posterior distributions.  Where available, X-ray values for $\theta_s$ converted from $\theta_{500}$ values from \citealt{2011A&A...534A.109P} using the `universal' $c_{500} = 1.177$ are shown with black dotted lines.}\label{Fi:all_post}
\end{figure*}
\begin{figure*}[hb]
\ContinuedFloat
\threefigures{posts/CAJ1916+3524bm1_posteriors.ps}{posts/CAJ1948+5114b_posteriors.ps}{posts/CAJ0917+5143bm1_posteriors.ps}
\threefigures{posts/CAJ0638+4748bm1_posteriors.ps}{posts/CAJ1414+7116bm1a_posteriors.ps}{posts/no_comb/CAJ1524+2954b_posteriors.ps}
\threefigures{posts/CAJ2200+2058bm1_posteriors.ps}{posts/no_comb/CAJ0622+7442b_posteriors.ps}{posts/CAJ1724+8553b_posteriors.ps}
\threefigures{posts/no_comb/CAJ1752+4440bm1_posteriors.ps}{posts/no_comb/CAJ1023+4907bm1_posteriors.ps}{posts/CAJ1157+3336bm1_posteriors.ps}
\caption{Continued.}
\end{figure*}
\begin{figure*}[hb]
\ContinuedFloat
\threefigures{posts/CAJ0308+2645b_posteriors.ps}{posts/CAJ1022+5006b_posteriors.ps}{posts/CAJ0748+5941bm1_posteriors.ps}
\threefigures{posts/CAJ1159+4946b_posteriors.ps}{posts/CAJ0142+4438b_posteriors.ps}{posts/CAJ1115+5320bm1_posteriors.ps}
\threefigures{posts/CAJ1315+5148cm1_posteriors.ps}{posts/CAJ2228+2037b_posteriors.ps}{posts/CAJ1858+2916b_posteriors.ps}
\threefigures{posts/CAJ1212+2732bm1_posteriors.ps}{posts/CAJ1819+5711cm1_posteriors.ps}{posts/CAJ1149+2223bm1_posteriors.ps}
\caption{Continued.}
\end{figure*}
\begin{figure*}[hb]
\ContinuedFloat
\threefigures{posts/CAJ1747+4512bm1_posteriors.ps}{posts/no_comb/CAJ2226+7818b_posteriors.ps}{posts/CAJ1229+4737b_posteriors.ps}
\threefigures{posts/CAJ0742+7414bm1_posteriors.ps}{posts/CAJ1856+6622b_posteriors.ps}{posts/no_comb/CAJ0227+4904bm1_posteriors.ps}
\threefigures{posts/CAJ0637+6654b_posteriors.ps}{posts/CAJ1259+6004bm1_posteriors.ps}{posts/CAJ1354+7714b_posteriors.ps}
\threefigures{posts/no_comb/CAJ2137+3531bm1_posteriors.ps}{posts/no_comb/CAJ2234+5243b_posteriors.ps}{posts/no_comb/CAJ1905+3233b_posteriors.ps}
\caption{Continued.}
\end{figure*}
\begin{figure*}[hb]
\ContinuedFloat
\threefigures{posts/CAJ2322+4845bm1_posteriors.ps}{posts/CAJ1756+4007bm1_posteriors.ps}{posts/CAJ0909+5133b_posteriors.ps}
\threefigures{posts/CAJ1414+5447c_posteriors.ps}{posts/CAJ1539+3426b_posteriors.ps}{posts/CAJ1314+6433b_posteriors.ps}
\threefigures{posts/CAJ0850+3604b_posteriors.ps}{posts/CAJ1016+3339b_posteriors.ps}{posts/CAJ0947+7622bm1_posteriors.ps}
\threefigures{posts/CAJ1123+2128bm1_posteriors.ps}{posts/CAJ0801+3605bm1_posteriors.ps}{posts/CAJ0826+8219bm1_posteriors.ps}
\caption{Continued.}
\end{figure*}
\begin{figure*}[hb]
\ContinuedFloat
\threefigures{posts/CAJ0851+4829b_posteriors.ps}{posts/CAJ1720+2741b_posteriors.ps}{posts/CAJ0839+6231b_posteriors.ps}
\threefigures{posts/CAJ1138+2755bm1_posteriors.ps}{posts/CAJ0732+3136b_posteriors.ps}{posts/CAJ0827+2229b_posteriors.ps}
\threefigures{posts/CAJ1453+5802bm1_posteriors.ps}{posts/CAJ0613+7152bm3_posteriors.ps}{posts/CAJ2353+2234c_posteriors.ps}
\threefigures{posts/no_comb/CAJ1935+6735b_posteriors.ps}{posts/CAJ0829+3826b_posteriors.ps}{posts/CAJ0545+5531bm1_posteriors.ps}
\caption{Continued.}
\end{figure*}
\begin{figure*}[hb]
\ContinuedFloat
\threefigures{posts/CAJ1359+6230bm1_posteriors.ps}{posts/CAJ2149+3310b_posteriors.ps}{posts/CAJ1944+6550b_posteriors.ps}
\threefigures{posts/CAJ1508+5753bm1_posteriors.ps}{posts/CAJ1823+7823bm1_posteriors.ps}{posts/CAJ0210+5432b_posteriors.ps}
\threefigures{posts/CAJ2317+5054b_posteriors.ps}{posts/CAJ0559+8613b_posteriors.ps}{posts/CAJ0938+5202b_posteriors.ps}
\threefigures{posts/CAJ1910+6742b_posteriors.ps}{posts/no_comb/CAJ1258+6519bm1_posteriors.ps}{posts/CAJ1900+5442b_posteriors.ps}
\caption{Continued.}
\end{figure*}
\begin{figure*}[hb]
\ContinuedFloat
\threefigures{posts/CAJ0221+2121b_posteriors.ps}{posts/CAJ0930+2848b_posteriors.ps}{posts/CAJ1209+2640b_posteriors.ps}
\threefigures{posts/CAJ1908+6904b_posteriors.ps}{posts/CAJ0543+4656b_posteriors.ps}{posts/CAJ0707+4420bm1_posteriors.ps}
\threefigures{posts/CAJ0647+7014bm1_posteriors.ps}{posts/CAJ0051+2720bm1_posteriors.ps}{posts/CAJ0640+5744bm1_posteriors.ps}
\threefigures{posts/CAJ1734+2432b_posteriors.ps}{posts/CAJ1313+4617b_posteriors.ps}{posts/CAJ1518+2926b_posteriors.ps}
\caption{Continued.}
\end{figure*}
\begin{figure*}[hb]
\ContinuedFloat
\threefigures{posts/CAJ2237+4115b_posteriors.ps}{posts/CAJ1740+3136bm1_posteriors.ps}{posts/CAJ0303+7755b_posteriors.ps}
\threefigures{posts/CAJ0202+4926b_posteriors.ps}{posts/CAJ1123+4302b_posteriors.ps}{posts/CAJ1634+6738b_posteriors.ps}
\caption{Continued.}
\end{figure*}

\begin{figure*}[hb]
\threefigures{posts/no_comb/CAJ1428+5651b_posteriors.ps}{posts/no_comb/CAJ0842+6234bm2_posteriors.ps}{posts/no_comb/CAJ1832+6449b_posteriors.ps}
\threefigures{posts/no_comb/CAJ2146+2029bm1_posteriors.ps}{posts/no_comb/CAJ0918+7051b_posteriors.ps}{posts/no_comb/CAJ1139+6109b_posteriors.ps}
\threefigures{posts/no_comb/CAJ2350+2929b_posteriors.ps}{posts/no_comb/CAJ1151+6220b_posteriors.ps}{posts/no_comb/CAJ1603+5908b_posteriors.ps}
\threefigures{posts/no_comb/CAJ1432+3135bm1_posteriors.ps}{posts/no_comb/CAJ0730+4820b_posteriors.ps}{posts/no_comb/CAJ1727+7036b_posteriors.ps}
\caption{$Y_{\rm tot}$-$\theta_{s}$ posterior distributions for AMI and \emph{Planck}, in descending \emph{Planck} SNR order (note that this is the `compatibility' SNR for \textsc{PwS}), for all AMI non-detections ($\Delta \ln(\mathcal{Z}) < 0$).  Contours mark the 68\% and 95\% confidence limits of the posterior distributions. Where available, X-ray values for $\theta_s$ converted from $\theta_{500}$ values from \citealt{2011A&A...534A.109P} using the `universal' $c_{500} = 1.177$ are shown with black dotted lines.}\label{Fi:all_post2}
\end{figure*}
\begin{figure*}[hb]
\ContinuedFloat
\threefigures{posts/no_comb/CAJ1429+2437b_posteriors.ps}{posts/no_comb/CAJ1235+4031bm1_posteriors.ps}{posts/no_comb/CAJ1718+5639b_posteriors.ps}
\threefigures{posts/no_comb/CAJ0642+3150b_posteriors.ps}{posts/no_comb/CAJ2155+7910bm2_posteriors.ps}{posts/no_comb/CAJ1710+6221b_posteriors.ps}
\threefigures{posts/no_comb/CAJ1808+4746b_posteriors.ps}{posts/no_comb/CAJ0727+4232bm2_posteriors.ps}{posts/no_comb/CAJ1853+2236bm1_posteriors.ps}
\threefigures{posts/no_comb/CAJ1431+2932b_posteriors.ps}{posts/no_comb/CAJ1904+3516b_posteriors.ps}{posts/no_comb/CAJ1709+4733b_posteriors.ps}
\caption{Continued.}
\end{figure*}

\twocolumn
\raggedright

\listofobjects
\end{document}